\documentclass[12pt]{article}
\usepackage{amsmath,amsthm,amsfonts,amssymb,graphicx}
\usepackage{cite,faktor}
\usepackage{xcolor}			% for coloring texts
\usepackage{subfig}          	% for \subfloat commands

\newlength{\xtrawidth}
\newlength{\xtraheight}
\numberwithin{equation}{section}
\numberwithin{table}{section}
\newdimen\tableauside\tableauside=1.0ex
\newdimen\tableaurule\tableaurule=0.4pt
\newdimen\tableaustep
\def\phantomhrule#1{\hbox{\vbox to0pt{\hrule height\tableaurule width#1\vss}}}
\def\phantomvrule#1{\vbox{\hbox to0pt{\vrule width\tableaurule height#1\hss}}}
\def\sqr{\vbox{%
  \phantomhrule\tableaustep
  \hbox{\phantomvrule\tableaustep\kern\tableaustep\phantomvrule\tableaustep}%
  \hbox{\vbox{\phantomhrule\tableauside}\kern-\tableaurule}}}
\def\squares#1{\hbox{\count0=#1\noindent\loop\sqr
  \advance\count0 by-1 \ifnum\count0>0\repeat}}
\def\tableau#1{\vcenter{\offinterlineskip
  \tableaustep=\tableauside\advance\tableaustep by-\tableaurule
  \kern\normallineskip\hbox
    {\kern\normallineskip\vbox
      {\gettableau#1 0 }%
     \kern\normallineskip\kern\tableaurule}%
  \kern\normallineskip\kern\tableaurule}}
\def\gettableau#1{\ifnum#1=0\let\next=\null\else
\squares{#1}\let\next=\gettableau\fi\next}

\DeclareMathOperator{\Res}{Res}
\DeclareMathOperator{\Disc}{Disc}

\theoremstyle{plain}% default
\newtheorem{thm}{Theorem}[section]
\newtheorem{lem}[thm]{Lemma}
\newtheorem{prop}[thm]{Proposition}

\graphicspath{{figs/}}	% Please put all the figures, plots, and graphics in this subdirectory.
\begin{document}

\begin{titlepage}
\begin{center}
\hfill BONN-TH-2014-03\\
\hfill BRXTH-674\\
\vskip 0.55in

{\Large \bf Knot Invariants from Topological Recursion on}\\[2ex]
{\Large \bf Augmentation Varieties}\\

\vskip 0.3in

{ Jie Gu${}^{a}$, Hans Jockers${}^{a}$, Albrecht Klemm${}^{a}$, Masoud Soroush${}^{b,c}$}\\

\vskip 0.2in
{\footnotesize
\begin{tabular}{p{7cm}p{7cm}}
${}^{\, a}${\em Bethe Center for Theoretical Physics,} &
${}^{\, b}${\em Martin A. Fisher School of Physics,} \\
$\phantom{{}^{\, b}}${\em Physikalisches Institut, Universit\"at Bonn} &
$\phantom{{}^{\, a}}${\em Brandeis University} \\
$\phantom{{}^{\, b}}${\em 53115 Bonn, Germany} &
$\phantom{{}^{\, a}}${\em Waltham, MA 02453, USA}\\[4ex]
\multicolumn{2}{c}{\hbox to 5cm {${}^{\, c}${\em Department of Mathematics,}\hfil}} \\
\multicolumn{2}{c}{\hbox to 5cm {$\phantom{{}^{\, c}}${\em Brandeis University}\hfil}} \\
\multicolumn{2}{c}{\hbox to 5cm {$\phantom{{}^{\, c}}${\em Waltham, MA 02453, USA}\hfil}}
\end{tabular}
}
\end{center}
\vskip0.5ex

\begin{center} {\bf Abstract} \end{center}

Using the duality between Wilson loop expectation values of $SU(N)$ Chern--Simons theory on $S^3$ and topological open-string amplitudes on the local mirror of the resolved conifold, we study knots on $S^3$ and their invariants encoded in colored HOMFLY polynomials by means of topological recursion. In the context of the local mirror Calabi--Yau threefold of the resolved conifold, we generalize the topological recursion of the remodelled B-model in order to study branes beyond the class of toric Harvey--Lawson special Lagrangians --- as required for analyzing non-trivial knots on $S^3$. The basic ingredients for the proposed recursion are the spectral curve, given by the augmentation variety of the knot, and the calibrated annulus kernel, encoding the topological annulus amplitudes associated to the knot. We present an explicit construction of the calibrated annulus kernel for torus knots and demonstrate the validity of the topological recursion. We further argue that --- if an explicit form of the calibrated annulus kernel is provided for any other knot --- the proposed topological recursion should still be applicable. We study the implications of our proposal for knot theory, which exhibit interesting consequences for colored HOMFLY polynomials of mutant knots. 

\vfill

\noindent January, 2014

{
\let\thefootnote\relax
\footnotetext{jiegu@th.physik.uni-bonn.de, jockers@uni-bonn.de, aklemm@th.physik.uni-bonn.de, soroush@brandeis.edu}
}

\end{titlepage}
%%%%%%%%%%%%%%%%%%%%%%%%%%%%%%%%%%%%%%%%%%%%%%%%%%%%%%%%%%%%%%%%%%%%

\tableofcontents
\newpage

%%%%%%%%%%%%%%%%%%%%%%%%%%%%%%
\section{Introduction}
%%%%%%%%%%%%%%%%%%%%%%%%%%%%%%

About twenty five years ago, Witten showed that Wilson loops in Chern--Simons theory furnish a suitable framework to study knots and their invariants on three manifolds \cite{Witten:1988hf}. Since then, this momentous work has led to many important insights, both in quantum field theory and in knot theory.\footnote{For a review, see for instance~\cite{Labastida:1998ud,MR2177747,Gukov:2012jx} and references therein.} As $SU(N)$ Chern--Simons theory on the three sphere $S^3$ is equivalent the A-twisted topological string theory on the total space $T^*S^3$ of the cotangent bundle of the three sphere \cite{Witten:1992fb}, knot invariants can also be studied in terms of topological strings~\cite{Ooguri:1999bv}. 

A topological version of the holographic principle gives rise to a large $N$ transition~\cite{Gopakumar:1998ki}, which maps the topological open string theory on $T^*S^3$ to a dual topological closed string theory on the resolved conifold $\mathcal{O}(-1)\oplus\mathcal{O}(-1)\rightarrow\mathbb{P}^1$. Furthermore, the mirror curve that defines the local mirror geometry of the resolved conifold, has been interpreted as the spectral curve of a matrix model \cite{Dijkgraaf:2002fc,Marino:2006hs,Bouchard:2007ys}, which provides for a powerful method to compute closed and certain open string correlation functions using the topological recursion introduced by Eynard and Orantin \cite{Eynard:2007kz}. In these duality correspondences expectation values of Wilson loops of finite-dimensional representations of $SU(N)$ Chern--Simons theory are mapped to open topological string amplitudes in the presence of (non-compact) Lagrangian probe branes, which model the Wilson loop and hence the knot under consideration \cite{Ooguri:1999bv}. From a knot theory perspective the Wilson loop expectation values yield the HOMFLY polynomials of knots colored with finite representations of $SU(N)$ \cite{Witten:1988hf}.

The aim of this note is to calculate knot invariants from open-string correlation functions in the topological B-model of the local mirror geometry of the resolved conifold by means of topological recursion. For the unknot on $S^3$ such calculations have been carried out a long time ago \cite{Ooguri:1999bv}. In this case a tremendous simplification occurs because the mirror curve of the resolved conifold coincides with the moduli space of the probe brane for the unknot. This is a general feature of the Harvey--Lawson special Lagrangian branes~\cite{MR666108,Aganagic:2000gs,Aganagic:2001nx} in local toric Chern--Simons manifolds and is responsible for the relative simplicity of the matrix model formulation in the remodelling approach~\cite{Marino:2006hs,Bouchard:2007ys}. However, in general the moduli space of a probe brane associated to a knot does not coincide with the mirror curve, which makes the calculation of topological correlation functions more challenging.

Using the topological recursion approach \cite{Eynard:2007kz,Bouchard:2007ys}, an interesting construction to calculate topological correlators for torus knots has been put forward in ref.~\cite{Brini:2011wi}. However, the used spectral curve of the matrix model does not solely describe the moduli space of the relevant probe branes, and therefore it contains (at each level of the topological recursion) redundant contributions that --- at least to our knowledge --- do not enjoy a physical interpretation directly linked to the analyzed torus knots.

Another interesting recent development links the augmentation variety of the differential graded algebra of knot contact homology \cite{MR2116316,MR2175153,MR2376818,MR2807087,MR3070519} to the moduli space of the associated probe brane in the resolved conifold geometry~\cite{Ng:2012qq,Aganagic:2013jpa}. While the connection to the moduli space of the probe brane admits an immediate extraction of topological disk invariants \cite{Aganagic:2012jb,Fuji:2012nx,Jockers:2012pz,Aganagic:2013jpa}, the implementation of the topological recursion along the lines of ref.~\cite{Eynard:2007kz} requires a careful reexamination of the second ingredient of the topological recursion, namely the annulus kernel.

In this work we provide a modified definition for this kernel, which enables us to perform the topological recursion directly based upon the moduli space of the probe brane --- that is to say upon the augmentation variety of the differential graded algebra in knot contact homology. We construct these annulus kernels in particular for torus knots and demonstrate the viability of the proposed recursion explicitly by showing that all our results conform with the existing literature. Our construction is inspired by the approach of ref.~\cite{Brini:2011wi}, but let us emphasize that by applying the topological recursion directly to the augmentation variety, we obtain at each step in the recursion only correlators specific to the analyzed knot without any redundancies as in ref.~\cite{Brini:2011wi}. In ref.~\cite{Aganagic:2012jb}, it is argued that the moduli space of the probe brane for any knot may also enjoy an interpretation as an unconventional mirror curve to the resolved conifold geometry and proposed that a drastic simplification of the periods of these degenerate curves of generally high genus occurs, which is necessary to give them a closed string interpretation in the conifold background. These simplifications are not found, but we do obtain the expected conifold closed string amplitudes. In particular, we check that a generalization of the variational principle of ref.~\cite{Borot:2013lpa} gives the conifold planar free energy from the augmentation varieties for general knots, and we use the modified kernels to check the genus one free energy for torus knots.

Our explicit construction of the kernels for the augmentation varieties clarifies the relation between the approaches in ref.~\cite{Brini:2011wi} and in ref.~\cite{Aganagic:2012jb}. So far it requires rather detailed knowledge of that part of the information in the colored HOMLFY polynomial that determines the annulus amplitude and is currently only available for torus knots. However, we expect that this is a technical restriction and the topological recursion based on the modified kernel works for more general knots as well. Based on general properties of colored HOMFLY polynomials of mutant knots \cite{MR1395780}, we argue that the knowledge of the augmentation variety itself is not sufficient to deduce the kernel required for carrying out the proposed topological recursion. Nevertheless, assuming only the existence of a kernel that renders the topological recursion possible, we are able to conjecture some intriguing consequences for a general pair of mutant knots. For instance, if two mutant knots are distinguishable by some HOMFLY polynomials colored with a finite representation of $SU(N)$, then --- for sufficiently large $N$ --- this pair of mutant knots can certainly be distinguished by the colored HOMFLY polynomials of a representation of $SU(N)$ given by a Young tableau with two rows.

The outline of this paper is as follows: In Section~\ref{sec:curves}, starting from the spectral curve and the matrix model formulation introduced in ref.~\cite{Brini:2011wi}, we construct the moduli space varieties of the probe branes associated to torus knots on $S^3$, which --- as we demonstrate --- are equivalent to the augmentation varieties of the corresponding differential graded algebra in knot contact homology. This correspondence allows us to derive the annulus kernels, generating the annulus amplitudes and required for carrying out the topological recursion on the level of the augmentation varieties. In particular, the sample calculations in Section~\ref{sec:Simpleexamples} explain the relation between the curve of~\cite{Brini:2011wi}, whose construction is reviewed in Section~\ref{sec:BEM}, and the augmentation variety discussed in Section~\ref{sec:aug}. In Section~\ref{sec:stretched}, we provide an independent evidence in favor of the validity of the annulus kernel found in Section~\ref{sec:kernel}. Section~\ref{sec:toprec} is the main part of this paper. Building on the results of the previous section, we propose a topological recursion for knot invariants in the spirit of Eynard and  Orantin \cite{Eynard:2007kz}. We check our proposal by applying the topological recursion to torus knots explicitly. In particular, for the trefoil knot, we deduce from the topological recursion the three-point amplitude in genus zero, the one-point amplitude in genus one, and the closed string free energy $F^{(1)}$ for the associated probe brane in the resolved conifold. We discuss our findings, and show that all our results are in agreement with the existing literature. We also demonstrate that it is possible to extract the planar free energy $F^{(0)}$ of the conifold resolution from the augmentation variety of an arbitrary knot. In Section~\ref{sec:knotinv}, we discuss general features of the proposed topological recursion and its implications for HOMFLY polynomials colored with $SU(N)$ representations. Assuming generality of our proposal we conjecture some consequences for colored HOMFLY polynomials of pairs of mutant knots. In Section~\ref{sec:conc}, we present our conclusions, discuss the implications of our results and point out some future directions. We present the technical details of our calculations in a series of appendices. In Appendix~\ref{app:B23}, we explicitly present the physical annulus kernel of the trefoil knot. In Appendix~\ref{app:Localization}, using equivariant localization techniques, we calculate the (stretched) annulus instanton numbers for the first few windings in the A-model for several torus knots. In Appendix~\ref{app:composite}, we calculate the full stretched annulus amplitudes from the quantum group invariants of composite representations. Appendix~\ref{sec:SymVariance} is concerned with the symplectic transformation properties of $F^{(1)}$ in the remodelled B-model. Appendix~\ref{sec:Props} provides detailed proofs for some technical statements made in Section~\ref{sec:toprec}. Finally, Appendix~\ref{sec:AugPoly} summarizes the augmentation polynomials of those non-torus knots for which the corresponding prepotentials, $F^{(0)}$, have been calculated.

%%%%%%%%%%%%%%%%%%%%%%%%%%%%%%
\section{Spectral curves and annulus kernels of knots} \label{sec:curves}
%%%%%%%%%%%%%%%%%%%%%%%%%%%%%%
Following ref.~\cite{Witten:1992fb,Ooguri:1999bv}, a knot $\mathcal{K}$ on $S^3$ is modeled in string theory on the deformed conifold geometry $T^*S^3$ as the intersection locus of a stack $N$ compact special Lagrangian branes on the compact three-cycle $S^3$ with a stack $M$ non-compact special Lagrangian probe branes $\mathcal{L}_\mathcal{K}$ of topology $\mathcal{K}\times \mathbb{R}^2$ with the non-compact directions suitably embedded in the cotangent directions of $T^*S^3$. The extremal transition to the resolved conifold geometry $\mathcal{O}(-1)\oplus\mathcal{O}(-1)\to\mathbb{P}^1$ realizes a large $N$ transition \cite{Gopakumar:1998ki}. In this transition the stack of branes on $S^3$ are replaced by a background flux supported on $\mathbb{P}^1$, while the $M$ non-compact probe branes still describe the knot $\mathcal{K}$ \cite{Diaconescu:2011xr}. In this transition the volume of $\mathbb{P}^1$ of the resolved conifold is given by the 't~Hooft coupling
$$
  \operatorname{vol}(\mathbb{P}^1) \,=\, g_s N \ ,
$$
with the string coupling $g_s$, which in turn gets complexified by the expectation value of the B-field to the complexified K\"ahler parameter
$$
  t\,=\, B + i \operatorname{vol}(\mathbb{P}^1) \ .
$$  

The described string theory on the conifold is dual to $SU(N)$ Chern--Simons theory on $S^3$ \cite{Witten:1988hf}, where the string coupling $g_s$ is identified with the coupling constant $k$ of the $SU(N)$ Chern--Simons theory as\footnote{For a review of these discussed dualities, see for instance ref.~\cite{Marino:2004uf} and references therein.}
$$
  g_s\,=\,\frac{2\pi}{k+N} \ .
$$

Furthermore, the Wilson loop expectation value of the knot $\mathcal{K}$ in some finite representation of $SU(N)$ --- which yields in knot theory the HOMFLY polynomial of $\mathcal{K}$ colored with this representation of $SU(N)$ \cite{Witten:1992fb,Ooguri:1999bv} --- is given to leading order in $g_s$ by the disk instanton generated superpotential of the stack of probe branes. Starting from the annulus kernel --- generating the annulus instanton numbers of the probe branes in the resolved conifold --- the higher order corrections in $g_s$ arise from topological recursion to be discussed in detail in Section~\ref{sec:toprec}. 

The aim of this section is to establish the local mirror symmetry geometry for probe branes on the resolved conifold describing the knot $\mathcal{K}$ on $S^3$. To set the stage and to establish our notation, we first review the local mirror symmetry picture for unknot, which naturally leads us to the matrix model formulation of torus knots in terms of fractional unknots as introduced by Brini--Eynard--Mari\~no \cite{Brini:2011wi}. For torus knots, we establish the precise relationship between the picture of fractional unknots and the probe brane moduli space, as arising from the augmentation varieties of the differential graded algebras in knot contact homology\cite{MR2116316,MR2175153,MR2376818,MR2807087,MR3070519}.\footnote{The fractional unknots approach is a B-model point of view, in which the augmentation variety plays the role of the classical moduli space of the B-brane mirror to the Lagrangian brane $\mathcal{L}_\mathcal{K}$. The differential graded algebra perspective is an A-model picture, and the augmentation variety furnishes the quantum moduli space of the Lagrangian brane $\mathcal{L}_\mathcal{K}$.} This allows us to derive both the disk instanton generated superpotential and the kernel for the annulus instantons of the augmentation varieties for torus knots. We hope that in the future fractional unknots can be used more generally to deduce the superpotentials and the annulus kernels for knots beyond the class of torus knots, and we plan to get back to this issue elsewhere.

\subsection{The spectral curve for unknot}
The local mirror geometry of the resolved conifold $\mathcal{O}(-1)\oplus\mathcal{O}(-1)\to\mathbb{P}^1$ is given by the genus zero spectral curve \cite{Hori:2000kt}
\begin{equation} \label{eq:unknot}
   (1- Q \beta) - \alpha\beta^f\,(1-\beta) \,=\, 0  \ , \qquad \alpha,\beta \in \mathbb{C}^* \ ,
\end{equation}
with the integral parameter $f$ and the complex structure parameter $Q$, which is mapped by local mirror symmetry to the complexified volume of the compact one-cycle $\mathbb{P}^1$ of the resolved conifold
\begin{equation} \label{eq:vol}
  t\,=\,\frac{1}{2\pi i} \log Q \ .
\end{equation}

Viewed as the mirror spectral curves of the resolved conifold, all choices of the integral parameter $f$ are equivalent due to simple coordinate redefinitions of the $\mathbb{C}^*$ variable $\alpha$. However, the spectral curve \eqref{eq:unknot} also enjoys the interpretation as the open-string moduli space of non-compact toric brane of the conifold \cite{Leung:1997tw,Aganagic:2000gs}. In particular, $\alpha$ is the algebraic coordinate in the semi-classical regime of the toric special Lagrangian brane~$\mathcal{L}$ of topology $S^1\times\mathbb{R}^2$ residing on an exterior leg of the toric skeleton of the resolved conifold. Then the parameter $f$ describes the framing of this non-compact brane~$\mathcal{L}$, and its (framing dependent) quantum superpotential --- generated by disk instantons corrections --- becomes \cite{Aganagic:2000gs,Aganagic:2001nx}
\begin{equation} \label{eq:UnknotSuper}
    W_{\mathcal{L},f}(\alpha;Q) \,=\, \int \frac{d\alpha}{\alpha} \log \beta(\alpha;Q) \ ,
\end{equation}
where $\beta(\alpha;Q)$ is the function defined implicitly by the spectral curve \eqref{eq:unknot}. The probe brane $\mathcal{L}$ describes the unknot on $S^3$ \cite{Ooguri:1999bv}, and the framing parameter $f$ becomes the framing of the unknot in $S^3$, while the disk instanton generated superpotential \eqref{eq:UnknotSuper} captures the leading order contribution in $g_s$ of appropriate linear combinations of colored HOMFLY polynomials of unknot. 

Moreover, the annulus instantons of the probe brane $\mathcal{L}$ are generated by the (framing dependent) annulus kernel \cite{Eynard:2007kz,Bouchard:2007ys}
\begin{equation} \label{eq:Kunknot}
  B_{\mathcal{L},f}(\alpha_1,\alpha_2;Q)d\alpha_1 d\alpha_2\,=\, \frac{\beta'(\alpha_1) \beta'(\alpha_2)}{\left(\beta(\alpha_1)-\beta(\alpha_2)\right)^2}\,d\alpha_1 d\alpha_2 \ ,
\end{equation}
defined on the (symmetric) product of the spectral curve  \eqref{eq:unknot}. Since the moduli space of the probe brane $\mathcal{L}$ for unknot coincides with bulk spectral curve, the annulus kernel is given by the Bergman kernel. For genus zero spectral curves, the Bergman kernel is uniquely defined by the property that the double pole on the diagonal spectral curve $\alpha_1=\alpha_2$ (normalized to one) is the only pole of the kernel. As we will see for more general knots, the moduli space of the associated probe branes do not coincide with the genus zero curve \eqref{eq:unknot} of the mirror conifold geometry any more, and as a consequence the annulus kernel does not exhibit the simple pole structure of the Bergman kernel anymore.

\subsection{Brini--Eynard--Mari\~no approach to torus knots}\label{sec:BEM}
Using $SL(2,\mathbb{Z})$ transformations acting on the spectral curve of the unknot \eqref{eq:unknot}, Brini, Eynard and Mari\~no show that the probe branes $\mathcal{L}_{r,s}$ associated to torus knots $\mathcal{K}_{r,s}$ (with the co-prime integers $r$ and $s$) are encoded in the spectral curve \cite{Brini:2011wi}
\begin{equation} \label{eq:fracunknot}
   h_{r,s}(\zeta,\rho;Q)\,=\,(1- Q \rho^r) - \zeta \rho^s\,(1-\rho^r) \,=\, 0  \ .
\end{equation}
Identifying $(\zeta,\rho^r)$ with the $\mathbb{C}^*$ coordinate $(\alpha,\beta)$ in \eqref{eq:unknot}, this spectral curve can be viewed as the unknot spectral curve with fractional framing
\begin{equation} \label{eq:fracframing}
  f \,=\, \frac{s}{r} \,\in\, \mathbb{Q} \ .
\end{equation}
A priori the fractional framing of the unknot does not have an immediate physical interpretation.  Therefore, in the following we sometimes refer to the above spectral curve~\eqref{eq:fracunknot} as the auxiliary spectral curve. 

Furthermore, expanded in the vicinity $\zeta=0$, $\rho^r=Q^{-1}$ of this auxiliary curve, the authors show that the superpotentials
\begin{equation} \label{eq:fracsuper}
  W_{r,s}^{(k)}(\zeta;Q) \,=\, r \int \frac{d\zeta}{\zeta} \log \rho^{(k)}(\zeta;Q) \ , \qquad k=1,\ldots,r \ ,
\end{equation}
with the superscript $k$ labeling the distinct expansion points $(\zeta,\rho)=(0,e^{2\pi i k /r} Q^{-1/r})$, encode the physical superpotential $W_{r,s}$ of the Lagrangian probe brane $\mathcal{L}_{r,s}$ associated to the torus knot $\mathcal{K}_{r,s}$ in the following way. First, we occasionally note that the superpotentials $W_{r,s}^{(k)}$ --- which we denote accordingly as the auxiliary superpotentials in the following --- enjoy the series expansion
\begin{equation}
  W_{r,s}^{(k)}(\zeta;Q) \,=\, \log \left( e^{2\pi i k/r} Q^{-1/r} \right) \log\zeta^r + \sum_{n=1}^{\infty} \widetilde P^{(r,s)}_n(Q) \zeta^n \ ,
\end{equation}  
where the classical term that is linear in the flat open-string modulus $(\log\zeta)$ is a space-time Fayet--Iliopolous term \cite{Aganagic:2001ug}. The non-perturbative disk instanton corrections from the probe branes $\mathcal{L}_{r,s}$ are encoded in the coefficient polynomials $\widetilde P^{(r,s)}_n(Q)$ of the series expansion in the algebraic open-string coordinate $\zeta$, and the exponent $n$ yields the winding number of the disk instantons enumerated in $\widetilde P^{(r,s)}_n(Q)$. 

We observe that the phase transformation $\zeta\to e^{2\pi i/r}\zeta$ cyclicly permutes the solutions $\rho^{(k)}(\zeta)$.\footnote{As $r$ and $s$ are co-prime integers, B\'ezout's Lemma ensures that we can always find integers $p$ and $q$ with $rp-sq=1$. As a consequence, if $(\zeta,\rho)$ is a point on the auxiliary spectral curve~\eqref{eq:fracunknot}, then one readily checks that the point $(e^{2\pi i/r} \zeta,e^{2\pi i q/r}\rho)$ resides on the curve, too. Finally, since the solutions $\rho^{(k)}$ obey $\lim_{\zeta\to0}\rho^{(k)}(\zeta)=e^{2\pi ik/r} Q^{1/r}$, the phase transformation $\zeta\to e^{2\pi i/r}\zeta$ results in the cyclic permutation $\rho^{(k)}(\zeta) \to \rho^{(k+q \mod r)}(\zeta)$.} This phase shift acts trivially on the algebraic coordinate~$\alpha=\zeta^r$, which in ref.~\cite{Brini:2011wi} is identified with the physical open-string coordinate of the probe brane~$\mathcal{L}_{r,s}$. The sum of all superpotentials~$W_{r,s}^{(k)}$ is invariant with respect to this phase transformation, and therefore yields the physical superpotential of the brane~$\mathcal{L}_{r,s}$
\begin{equation} \label{eq:supersum}
   W_{r,s} \,=\, \sum_{k=1}^r W_{r,s}^{(k)} \,=\,r \int \frac{d\zeta}{\zeta} \log \left(\rho^{(1)}(\zeta;Q)\cdot\ldots\cdot\rho^{(r)}(\zeta;Q)  \right) \ .
\end{equation}
In this way only disk instanton corrections at winding numbers that are integral in the variable $\alpha$ contribute, which justifies the identification of $W_{r,s}$ with the superpotential of the brane $\mathcal{L}_{r,s}$. Thus in terms of the algebraic open-string coordinate $\alpha$ the physical superpotential becomes\footnote{In ref.~\cite{Brini:2011wi} already the superpotentials~\eqref{eq:fracsuper} have been expressed in terms of the open-string coordinates $\alpha=\zeta^r$. Then it is argued that only the analytic terms --- that is to say only contributions $\alpha^q$ with $q\in\mathbb{Z}$ --- give rise to disk instantons relevant for Wilson loop expectation values of the torus knots $\mathcal{K}_{r,s}$.}
\begin{equation} \label{eq:physsuper}
     W_{r,s}(\alpha;Q) \,=\, -\log Q \log \alpha + r \sum_{n=1}^{\infty} \widetilde P^{(r,s)}_{rn}(Q) \alpha^n \ , \qquad \alpha=\zeta^r \ .
\end{equation}

We can also infer that the framing of the described torus knot $\mathcal{K}_{r,s}$ is given by $r\cdot s = \frac{s}{r}\cdot r^2$. Here the fractional framing \eqref{eq:fracframing} is multiplied by one factor of $r$ due to the restriction to winding contributions $\zeta^r$, while the second factor of $r$ arises due to concatenation of all solutions $\rho^{(k)}(\zeta)$ in eq.~\eqref{eq:supersum}. Note that --- even though the superpotentials $W^{(k)}_{r,s}$ and $W^{(k)}_{s,r}$ are inequivalent --- the physical superpotential $W_{r,s}$ and $W_{s,r}$ (for $r\ne s$) are the same \cite{Brini:2011wi}, which is consistent with the fact that the framed torus knots $\mathcal{K}_{r,s}$ and $\mathcal{K}_{s,r}$ are also identical. Note that similarly this symmetry appears in the Rosso--Jones formula for torus knots as well \cite{MR1209320}. 

One of the beautiful results of the work by Brini, Eynard and Mari\~no is that the auxiliary spectral curve \eqref{eq:fracunknot} also gives rise to auxiliary annulus kernels
\begin{equation} \label{eq:auxAnn}
   B_{r,s}^{(k,\ell)}(\zeta_1,\zeta_2;Q)d\zeta_1 d\zeta_2 = 
   \frac{\left(r\rho^{(k)}(\zeta_1)^{r-1}{\rho^{(k)}}{}'(\zeta_1)\right)\left(r\rho^{(\ell)}(\zeta_2)^{r-1}\rho^{(\ell)}{}'(\zeta_2)\right)}{\left(\rho^{(k)}(\zeta_1)^r-\rho^{(\ell)}(\zeta_2)^r\right)^2} 
   d\zeta_1 d\zeta_2\ ,
\end{equation}
for $k,\ell=1,\ldots,r$. They can be interpreted as Bergman kernels of the unknot spectral curve with fractional windings \cite{Brini:2011wi}. These Bergman kernels contain also auxiliary fractional windings. Nevertheless, they arise from a matrix model such that the topological recursion of Eynard and Orantin is applicable as shown in ref.~\cite{Borot:2013lpa}. As before the physical amplitudes relevant for the torus knots $\mathcal{K}_{r,s}$ are extracted from the integral winding contributions of these amplitudes in terms of the algebraic open-string coordinates $\alpha_\mu = (\zeta_\mu)^r$.

The physical annulus kernel --- in which all fractional windings are removed --- can again be realized by summing up the auxiliary kernels according to
\begin{equation} \label{eq:Bphys}
   B_{r,s}(\alpha_1,\alpha_2;Q)d\alpha_1d\alpha_2 \,=\, \sum_{k,\ell=1}^{r} B_{r,s}^{(k,\ell)}(\zeta_1,\zeta_2;Q)d\zeta_1 d\zeta_2 \ , \quad \alpha_\mu = (\zeta_\mu)^r \ , \quad \mu=1,2 \ .
\end{equation}   
 In the Section~\ref{sec:toprec}, we demonstrate that the topological recursion of Eynard and Orantin applies directly to the spectral curve arising from the physical superpotential~\eqref{eq:physsuper} together with the physical annulus kernel~\eqref{eq:Bphys}, without the need to remove any auxiliary fractional winding in the calculated amplitudes.

\subsection{Spectral curves and augmentation varieties for torus knots}\label{sec:kernel}
The construction of the physical superpotential~\eqref{eq:supersum} of torus knots allows us to also construct a physical spectral curve from the auxiliary spectral curve~\eqref{eq:fracunknot}. In ref.~\cite{Jockers:2012pz} we have observed that the physical superpotential~\eqref{eq:physsuper} for torus knots is given by
\begin{equation} \label{eq:Wphys2}
   W_{r,s}(\alpha;Q) \,=\, \int \frac{d\alpha}{\alpha} \log \beta(\alpha) \ ,
\end{equation}
where the series expansion $\beta(\alpha)$ again arises from a (unique) curve $\mathcal{C}_{r,s}$ given by the zero locus of the polynomial $F_{r,s}(\alpha,\beta;Q)$ in the coordinates $(\alpha,\beta)\in{\mathbb{C}^*}^2$. As the physical spectral curve $\mathcal{C}_{r,s}$ generates the disk instantons of the physical superpotential \cite{Aganagic:2012jb,Jockers:2012pz,Aganagic:2013jpa}, it is the moduli space of the Lagrangian probe branes $\mathcal{L}_{r,s}$ \cite{Aganagic:2000gs}. Furthermore, the curve $\mathcal{C}_{r,s}$ is shown in various examples to coincide with the augmentation variety of the differential graded algebra in knot contact homology \cite{MR2116316,MR2175153,MR2376818,MR2807087,MR3070519}. This correspondence --- namely the identification of the probe brane moduli space with the augmentation variety of the differential graded algebra in knot contact homology --- is expected to hold for  knots in general, i.e., even beyond the class of torus knots \cite{Aganagic:2013jpa}.

As the physical superpotential~\eqref{eq:Wphys2} of torus knots is encoded in the auxiliary superpotentials \eqref{eq:fracsuper}, we can calculate the physical spectral curve $\mathcal{C}_{r,s}$ from the auxiliary spectral curve \eqref{eq:fracunknot}. In order to remove the unphysical fractional contributions $W_{r,s}^{(k)}$, we first consider the ideal generated by $r$ copies of the auxiliary spectral curves $h_{r,s}(\zeta,\rho^{(k)};Q)$, $k=1,\ldots, r$, in terms of the $\mathbb{C}^*$~variables $\zeta, \rho^{(1)},\ldots, \rho^{(r)}$, together with the relations
\begin{equation}\label{equ:RelatingBEMandAV}
   \alpha\,=\,\zeta^r \ , \qquad \beta\,=\,(-1)^{r+1} \rho^{(1)}\cdot\ldots\cdot \rho^{(r)} \ ,
\end{equation} 
for the additional $\mathbb{C}^*$-variables $\alpha, \beta$. We have encountered the first relation already in the previous section, while the second relation arises from matching the two expressions of the physical superpotential~\eqref{eq:supersum} and \eqref{eq:Wphys2}.\footnote{The sign $(-1)^{r+1}$ is chosen to match the conventions of ref.~\cite{Jockers:2012pz}, and it only introduces a trivial constant shift in the physical superpotential~$W_{r,s}$.} 

To arrive at the physical superpotential, it is crucial that all the $r$ distinct implicit solutions $\rho^{(k)}(\zeta)$ appear in the sum~\eqref{eq:supersum}. This ensures that only integral windings in the physical open-string coordinate $\alpha$ appear in \eqref{eq:physsuper}. This can be achieved by enlarging the ideal such that the zero locus of the ideal consists only of those points, which correspond to distinct implicit solutions $\rho^{(k)}(\zeta)$. The resulting enlarged ideal then becomes
\begin{equation} \label{eq:Ihat}
     \widehat{\mathcal{I}}_{r,s} \, =\,\big\langle \alpha-\zeta^r,\,\beta+(-1)^r \rho^{(1)}\cdots\rho^{(r)}\big\rangle 
       \cup \left(\bigcup_{1\le k_1<\ldots<k_i\le r} \!\!\!\!\big\langle h_{r,s}(\zeta,\rho^{(k_1)},\ldots,\rho^{(k_i)};Q) \big\rangle \right) \ ,
\end{equation} 
where the polynomials $h_{r,s}(\zeta,\rho^{(k_1)},\ldots,\rho^{(k_i)};Q)$ are symmetric in the variables $\rho^{(k)}$. Starting from the auxiliary unknot equations $h_{r,s}(\zeta,\rho^{(k)};Q)$, $k=1,\ldots,r$, they are recursively defined by
\begin{equation} \label{eq:hdef}
\begin{aligned}
     &h_{r,s}(\zeta,\rho^{(k_1)},\ldots,\rho^{(k_i)};Q)\,\equiv\, \\
     &\quad\sum_{1\le n<m\le i} \frac{h_{r,s}(\zeta,\rho^{(k_1)},\ldots,\widehat{\rho^{(k_n)}},\ldots,\rho^{(k_i)};Q)-h_{r,s}(\zeta,\rho^{(k_1)},\ldots,\widehat{\rho^{(k_m)}},\ldots,\rho^{(k_i)};Q)}{\rho^{(k_n)}-\rho^{(k_m)}} \ ,
\end{aligned}     
\end{equation}
where $\widehat{\rho^{(k_n)}}$ indicates the omission of the variable $\rho^{(k_n)}$. Due to taking differences of the defining unknot equations and by dividing out factors $\rho^{(k_{i})}-\rho^{(k_{j})}$, $i\ne j$, the ideal $\widehat{\mathcal{I}}_{r,s}$, that is to say the zero locus of $\widehat{\mathcal{I}}_{r,s}$, does not contain any of the unwanted points that correspond to non-distinct implicit solutions $\rho^{(k)}(\zeta)$. Thus the physical spectral curve $\mathcal{C}_{r,s}$, furnishing the moduli space of the probe brane $\mathcal{L}_{r,s}$ is given by the elimination ideal 
\begin{equation}
    \mathcal{I}_{r,s} \,=\, \widehat{\mathcal{I}}_{r,s} \cap \mathbb{C}(Q)[\alpha,\beta]  \,\subset \, \mathbb{C}(Q)[\alpha,\beta] \ ,
\end{equation}    
in the polynomial ring $\mathbb{C}(Q)[\alpha,\beta]$ over the extension field $\mathbb{C}(Q)$ of $\mathbb{C}$. It is generated by the augmentation polynomials $F_{r,s}(\alpha,\beta;Q)$ of the differential graded algebra of knot contact homology
\begin{equation} \label{eq:AugCurve}
    \mathcal{I}_{r,s} \,=\, \big\langle F_{r,s}(\alpha,\beta;Q) \big\rangle \ .
\end{equation}

Alternatively, the ideal $\mathcal{I}_{r,s}$ can be thought of as the elimination ideal arising from the ideal generated by all symmetric sums of the generators $h_{r,s}(\zeta,\rho^{(k_1)},\ldots,\rho^{(k_i)};Q)$. These generators then provide for relations
\begin{equation} \label{eq:SymRels}
     R_{r,s}^{(k)}(\zeta,S^{(1)},\ldots,S^{(r)},\beta)\,=\,0 \ , \qquad k=1,\ldots,r \ , 
\end{equation}     
in terms of the elementary symmetric functions $S^{(j)}$ in the variables $\rho^{(k)}$. These relation allows us to eliminate the variables $\rho^{(k)}$ in favor of $\beta$. By additionally replacing $\zeta$ in favor of $\alpha$, we arrive at the physical spectral curve ideal $\mathcal{I}_{r,s}$. 

The latter point of view also allows us to get a handle on the physical annulus kernel~\eqref{eq:Bphys} in terms of the variables $(\alpha_1,\beta_1,\alpha_2,\beta_2)$. By construction the kernel~\eqref{eq:Bphys} can be written as a rational function in the elementary symmetric polynomials $S^{(k)}_\mu$, $\mu=1,2$, where the index $\mu$ labels the two sets of variables $\rho^{(k)}_\mu$, $\mu=1,2$. Then using the symmetric relations \eqref{eq:SymRels} together with the relations for $\alpha_\mu$, $\mu=1,2$, for both sets of variables, we obtain the physical annulus kernel~\eqref{eq:Bphys} as a rational function of~$(\alpha_1,\beta_1,\alpha_2,\beta_2)$, i.e., 
\begin{equation}\label{eq:Annulus-kernel}
     B_{r,s}(\alpha_1,\beta_1,\alpha_2,\beta_2;Q)d\alpha_1d\alpha_2 \,=\, 
     \frac{N_{r,s}(\alpha_1,\beta_1,\alpha_2,\beta_2;Q)}{D_{r,s}(\alpha_1,\beta_1,\alpha_2,\beta_2;Q)}d\alpha_1d\alpha_2 \ ,
\end{equation}
in terms of the polynomials $N_{r,s}(\alpha_1,\beta_1,\alpha_2,\beta_2;Q)$ and $D_{r,s}(\alpha_1,\beta_1,\alpha_2,\beta_2;Q)$. Note that since the physical annulus kernel is defined for points $(\alpha_\mu,\beta_\mu)$ on the curve $\mathcal{C}_{r,s}$, a given rational function $N_{r,s}/D_{r,s}$ is not a unique representative of this kernel, but instead two representatives $N_{r,s}/D_{r,s}$ and $\widetilde{N}_{r,s}/\widetilde{D}_{r,s}$ give rise to the same kernel, if they are related by
\begin{equation}
     \frac{r\cdot N_{r,s}+\sum_\mu p_\mu\cdot F_{r,s}(\alpha_\mu,\beta_\mu;Q)}{r\cdot D_{r,s}+\sum_\mu q_\mu\cdot F_{r,s}(\alpha_\mu,\beta_\mu;Q)}\,=\,
     \frac{\tilde r\cdot \widetilde{N}_{r,s}+\sum_\mu \tilde{p}_\mu\cdot F_{r,s}(\alpha_\mu,\beta_\mu;Q)}{\tilde{r} \cdot \widetilde{D}_{r,s}+\sum_\mu \tilde{q}_\mu\cdot F_{r,s}(\alpha_\mu,\beta_\mu;Q)}
\end{equation}
with some Q-dependent polynomials $p_\mu, \tilde p_\mu, q_\mu, \tilde q_\mu, r, \tilde r$ of $(\alpha_1,\beta_1,\alpha_2,\beta_2)$. Finding a nice representative in practice --- i.e., finding a representative of low degree in variables $\alpha_\mu$ and $\beta_\mu$ --- is often a cumbersome and challenging task.

Geometrically, the physical annulus kernel~\eqref{eq:Annulus-kernel} is a meromorphic bi-differential on the product of two physical spectral curves $\mathcal{C}_{r,s}\times\mathcal{C}_{r,s}$. Along the diagonal curve $\mathcal{C}^\text{diag}_{r,s} \subset \mathcal{C}_{r,s}\times\mathcal{C}_{r,s}$, it develops a double pole, which --- expressed in the local coordinates $(\alpha_1,\alpha_2)$ --- takes in the vicinity of $\mathcal{C}^\text{diag}_{r,s}$ the characteristic form
\begin{equation}
    B_{r,s}(\alpha_1,\beta(\alpha_1),\alpha_2,\beta(\alpha_2)) d\alpha_1d\alpha_2 \,=\, \frac{r\, d\alpha_1 d\alpha_2}{(\alpha_1 - \alpha_2)^2} \,+\, \ldots \ ,
\end{equation}
where `$\ldots$' refers to regular terms. This can readily be seen because each diagonal summand $B^{(k,k)}_{r,s}$ in the sum~\eqref{eq:Bphys} contributes a double pole $(\zeta_1-\zeta_2)^{-2} d\zeta_1d\zeta_2$, which (up to regular terms) corresponds to a double pole $(\alpha_1-\alpha_2)^{-2} d\alpha_1d\alpha_2$ in the local coordinates $(\alpha_1,\alpha_2)$. This observation allows us to define the calibrated annulus kernel
\begin{equation} \label{eq:Bcal}
  \widehat{B}_{r,s}(\alpha_1,\beta_1,\alpha_2,\beta_2)d\alpha_1d\alpha_2 \,=\, \left( B_{r,s}(\alpha_1,\beta_1,\alpha_2,\beta_2;Q) - \frac{(r-1)}{(\alpha_1-\alpha_2)^2}\right) d\alpha_1d\alpha_2 \ .
\end{equation}  
with the leading order behavior 
\begin{equation}
   \widehat{B}_{r,s}(\alpha_1,\beta(\alpha_1),\alpha_2,\beta(\alpha_2)) d\alpha_1d\alpha_2 \,=\, \frac{d\alpha_1 d\alpha_2}{(\alpha_1 - \alpha_2)^2} \,+\, \ldots \ ,
\end{equation}
along the diagonal curve $\mathcal{C}^\text{diag}_{r,s}$.

There are several reasons for introducing the calibrated annulus kernel~\eqref{eq:Bcal}. First, we note that the calibration of the double pole along the diagonal curve does not modify the open-string invariants of the brane $\mathcal{L}_{r,s}$. But in addition, the calibration renders the kernel $\widehat{B}_{r,s}$ symmetric under the exchange of the integers $r$ and $s$, which is natural as both branes $\mathcal{L}_{r,s}$ and $\mathcal{L}_{s,r}$ describe the same torus knot. Moreover, this calibration choice coincides with the calibration of the Klein bi-differential associated to the spectral curve $\mathcal{C}_{r,s}$, which is a symmetric meromorphic bi-differential on the product $\mathcal{C}_{r,s}\times\mathcal{C}_{r,s}$ that is regular expect for a double pole with coefficient one along the diagonal curve $\mathcal{C}_{r,s}^\text{diag}$ \cite{MR1510386,MR1510518}. Finally, it is only the calibrated kernel $\widehat{B}_{r,s}$ that transforms covariantly under framing transformations $\alpha \to \alpha\beta^f$. While these arguments are of a somewhat heuristic nature, we will show in Section~\ref{sec:toprec} that the calibrated annulus kernel~\eqref{eq:Bcal} is indeed the correct choice to define a consistent topological recursion.

Let us also mention that the Bergman kernel --- usually to be used in the context of the topological recursion --- is a particular Klein bi-differential with the additional property that all its integrals over $A$-cycles with respect to a symplectic basis of one-cycles of $H_1(\mathcal{C}_{r,s},\mathbb{Z})$ vanish \cite{MR0054140,Eynard:2007kz}. Therefore, it is tempting to identify the calibrated annulus kernel~\eqref{eq:Bcal} with a Klein bi-differential or even the Bergman kernel. However, analyzing in detail the pole structure of the calibrated annulus kernel --- as we will do in Section~\ref{sec:toprec} --- reveals that the calibrated annulus kernel has additional poles apart from the double pole along the diagonal curve $\mathcal{C}_{r,s}^\text{diag}\subset \mathcal{C}_{r,s} \times \mathcal{C}_{r,s}$. 

We believe that the emergence of additional poles indicates non-trivial short distance interactions among branes located at different points on the physical spectral curves $\mathcal{C}_{r,s}$. Such phenomena do not occur if the mirror curve coincides with the physical spectral curve of the brane, because then distinct points on the physical spectral curve of the brane --- that is to say distinct points on the mirror curve --- are automatically spatially separated. The latter situation occurs in particular for the mirror-symmetric description of toric non-compact Lagrangian branes in local toric Calabi--Yau threefolds as introduced in refs.~\cite{Aganagic:2000gs,Aganagic:2001nx}. In this case the Bergman kernel is indeed a natural candidate for the generating function of the annulus instantons \cite{Bouchard:2007ys}.

\subsection{Simple examples}\label{sec:Simpleexamples}
Let us illustrate the relationship between the the auxiliary spectral curves \eqref{eq:fracunknot} and the physical spectral curve~\eqref{eq:AugCurve} by a few examples. To this end, we first illustrate the construction with the torus knot $\mathcal{K}_{2,1}$, which is just the unknot at framing $f=2$. Then the ideal $\widehat{\mathcal{I}}_{2,1}$ is generated by
\begin{equation}
    \widehat{\mathcal{I}}_{2,1} \,=\, \left\langle \alpha-\zeta^2,\, \beta+\rho^{(1)}\rho^{(2)},\, h_{2,1}(\zeta,\rho^{(1)};Q),\, h_{2,1}(\zeta,\rho^{(2)};Q),\, h_{2,1}(\zeta,\rho^{(1)},\rho^{(2)};Q)\right\rangle \ ,
\end{equation}
with
\begin{equation}
\begin{aligned}
     h_{2,1}(\zeta,\rho^{(1)},\rho^{(2)};Q)\,&=\,\frac{h_{2,1}(\zeta,\rho^{(1)};Q) - h_{2,1}(\zeta,\rho^{(2)};Q)}{\rho^{(1)}-\rho^{(2)}} \\ 
     \,&=\,-Q(\rho^{(1)}+\rho^{(2)})-\zeta\left( 1 + \rho^{(1)}\rho^{(2)}-(\rho^{(1)}+\rho^{(2)})^2 \right) \ .
\end{aligned}     
\end{equation}
The last polynomial together with the definition of $\alpha$ and $\beta$ allows us to eliminate from the sum $h_{2,1}(\zeta,\rho_1;Q) + h_{2,1}(\zeta,\rho_2;Q)$ the variables $\rho_1$ and $\rho_2$ so as to arrive at the elimination ideal 
\begin{equation}
    \mathcal{I}_{2,1} \,=\, \big\langle  (1- Q \beta) - \alpha\beta^2\,(1-\beta)  \big\rangle \ ,
\end{equation}
which reproduces the spectral curve of unknot~\eqref{eq:unknot} at framing $f=2$. Furthermore, the physical annulus kernel is generated by
\begin{equation}
   B_{2,1}(\zeta_\mu,\rho^{(1)}_\mu,\rho^{(2)}_\mu;Q) d\alpha_1 d\alpha_2 \,=\, \sum_{k,\ell=1}^2 \tfrac{\rho^{(k)}_1\rho^{(\ell)}_2}{\zeta_1\zeta_2 (\rho_1^{(k)}-\rho_2^{(\ell)})^2} \cdot
     \tfrac{h^{(1,0)}_{2,1}(\zeta_1,\rho_1^{(k)};Q)h^{(1,0)}_{2,1}(\zeta_2,\rho_2^{(\ell)};Q)}{h^{(0,1)}_{2,1}(\zeta_1,\rho^{(k)}_1;Q)h^{(0,1)}_{2,1}(\zeta_2,\rho_2^{(\ell)};Q)}\,d\alpha_1 d\alpha_2 \ ,
\end{equation}
where $h^{(1,0)}_{2,1}$ and $h^{(0,1)}_{2,1}$ are the derivatives of the auxiliary unknot curves \eqref{eq:fracunknot} with respect to the first and second argument, which appear to express the derivatives ${\rho^{(k)}_\mu}{}'(\zeta)$ in eq.~\eqref{eq:auxAnn} algebraically in terms of the variables $\zeta_k$ and $\rho_\mu^{(k)}$.  By construction the kernel $B_{2,1}(\zeta_\mu,\rho_\mu^{(1)},\rho_\mu^{(2)}) d\alpha_1 d\alpha_2$ is a symmetric rational function in the variables $\rho_\mu^{(1)}$ and $\rho_\mu^{(2)}$ for both $\mu=1,2$, respectively. Therefore, with the help of the relations
\begin{equation}
   \rho_\mu^{(1)}+\rho_\mu^{(2)} \,=\, - \frac{1-Q\beta_\mu}{\zeta_\mu\,\beta_\mu} \ , \quad \rho_\mu^{(1)}\rho_\mu^{(2)}\,=\,-\beta_\mu \ , \quad 
   \zeta^2_\mu\,=\,\alpha_\mu \ , \quad \mu=1,2 \ ,
\end{equation}
for the elementary symmetric polynomials in $\rho_\mu^{(k)}$ and for $\zeta_\mu$ (arising from the ideal $\widehat{\mathcal{I}}_{2,1}$), we eliminate the variables $\rho_\mu^{(k)}$ and $\zeta_\mu$, and obtain the physical annulus kernel as a rational function in $\alpha_\mu$ and $\beta_\mu$. Keeping in mind that the variables reside on the unknot spectral curve $F_{2,1}(\alpha_\mu,\beta_\mu;Q)=0$, we arrive with a few steps of algebra at the final expression for the calibrated annulus kernel
\begin{equation}
    \widehat{B}_{2,1}(\alpha_\mu,\beta_\mu;Q)d\alpha_1 d\alpha_2 \,=\,  
    \frac{1}{(\beta_1-\beta_2)^2}\frac{F_{2,1}^{(1,0)}(\alpha_1,\beta_1;Q)F_{2,1}^{(1,0)}(\alpha_2,\beta_2;Q)}{F_{2,1}^{(0,1)}(\alpha_1,\beta_1;Q)F_{2,1}^{(0,1)}(\alpha_2,\beta_2;Q)} d\alpha_1 d\alpha_2 \ ,
\end{equation} 
which --- as expected --- coincides with the Bergman kernel of the unknot in the framing $f=2$.

As our second example, we analyze the torus knot $\mathcal{K}_{2,3}$, which is the trefoil knot represented in framing $f=6$ by the ideal $\widehat{I}_{2,3}$, which is generated by
\begin{equation}
    \widehat{\mathcal{I}}_{2,3} \,=\, \left\langle \alpha-\zeta^2,\, \beta+\rho_1\rho_2,\, h_{2,3}(\zeta,\rho_1;Q),\, h_{2,3}(\zeta,\rho_2;Q),\, h_{2,3}(\zeta,\rho_1,\rho_2;Q)\right\rangle \ ,
\end{equation}
with polynomials $h_{2,3}(\zeta,\rho_1,\rho_2;Q)$ defined in eq.~\eqref{eq:hdef}. We calculate the elimination ideal $\mathcal{I}_{2,3}$ with the help of the computer algreba system {\sc Singular} \cite{DGPS}, and we find that it is generated by the polynomial
\begin{equation} \label{eq:AugK23}
   F_{2,3}(\alpha,\beta;Q)=1-Q \beta +
       \alpha\beta^3 \left(1-\beta+2 \beta^2 -2 Q\beta^2- Q \beta^3+Q^2\beta^4 \right) -
       \alpha^2\beta^9(1-\beta) \ .
\end{equation}

Note that the curve $F_{2,3}(\alpha,\beta;Q)=0$ describes the open-string moduli space of the Lagrangian probe brane $\mathcal{L}_{2,3}$ in framing $f=6$ \cite{Aganagic:2012jb,Jockers:2012pz}. Furthermore, after changing to zero framing, it coincides with the augmentation variety of the differential graded algebra of the knot contact homology of the trefoil knot $\mathcal{K}_{2,3}$ \cite{MR2376818}. 

%%%%%%%%%%%%%%%%%%%%%%
\begin{figure}
\begin{center}
\subfloat[~\hskip18ex~]{\includegraphics[width=0.25\textwidth]{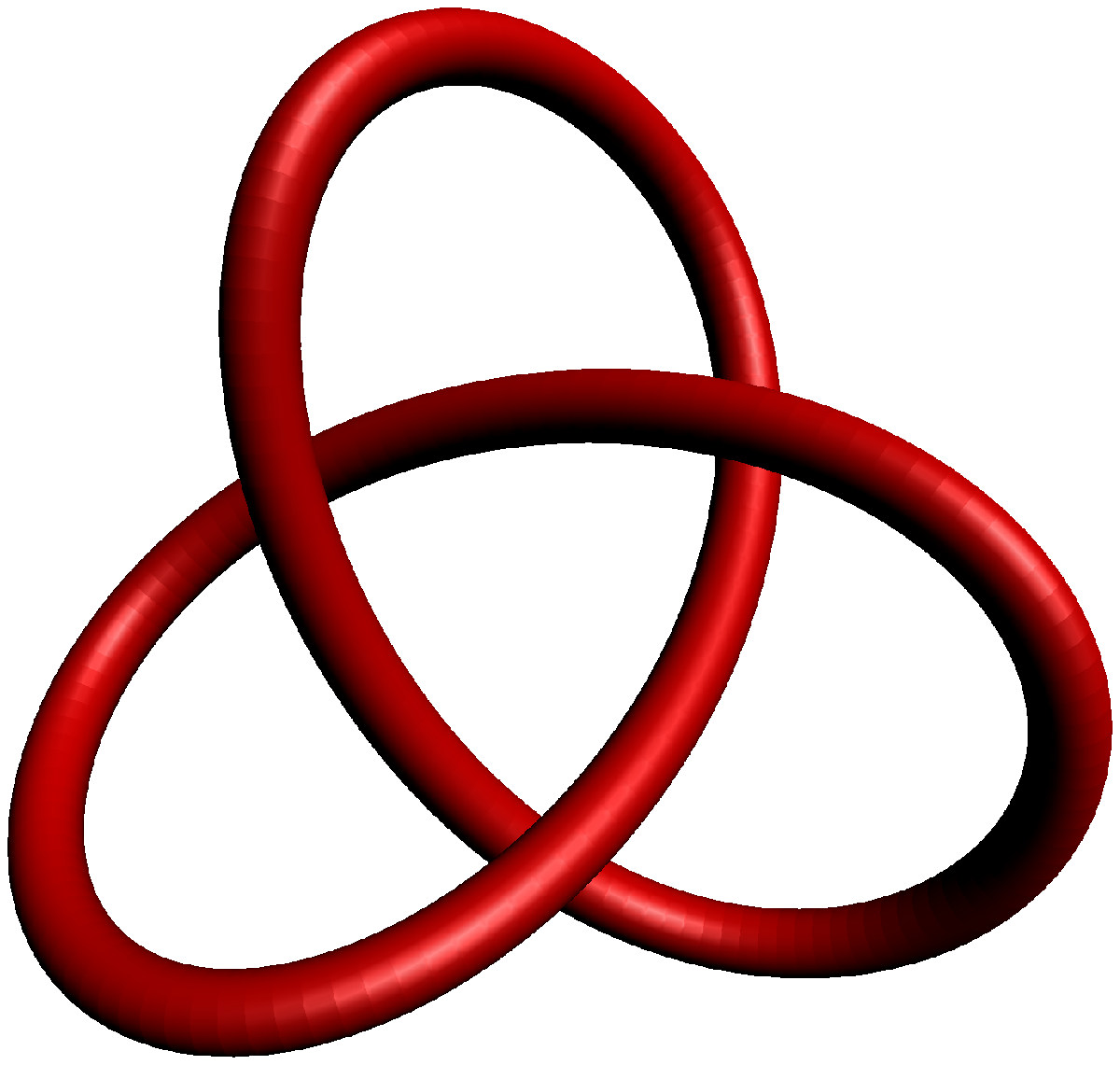}}\hskip6ex
\subfloat[~\hskip18ex~]{\includegraphics[width=0.25\textwidth]{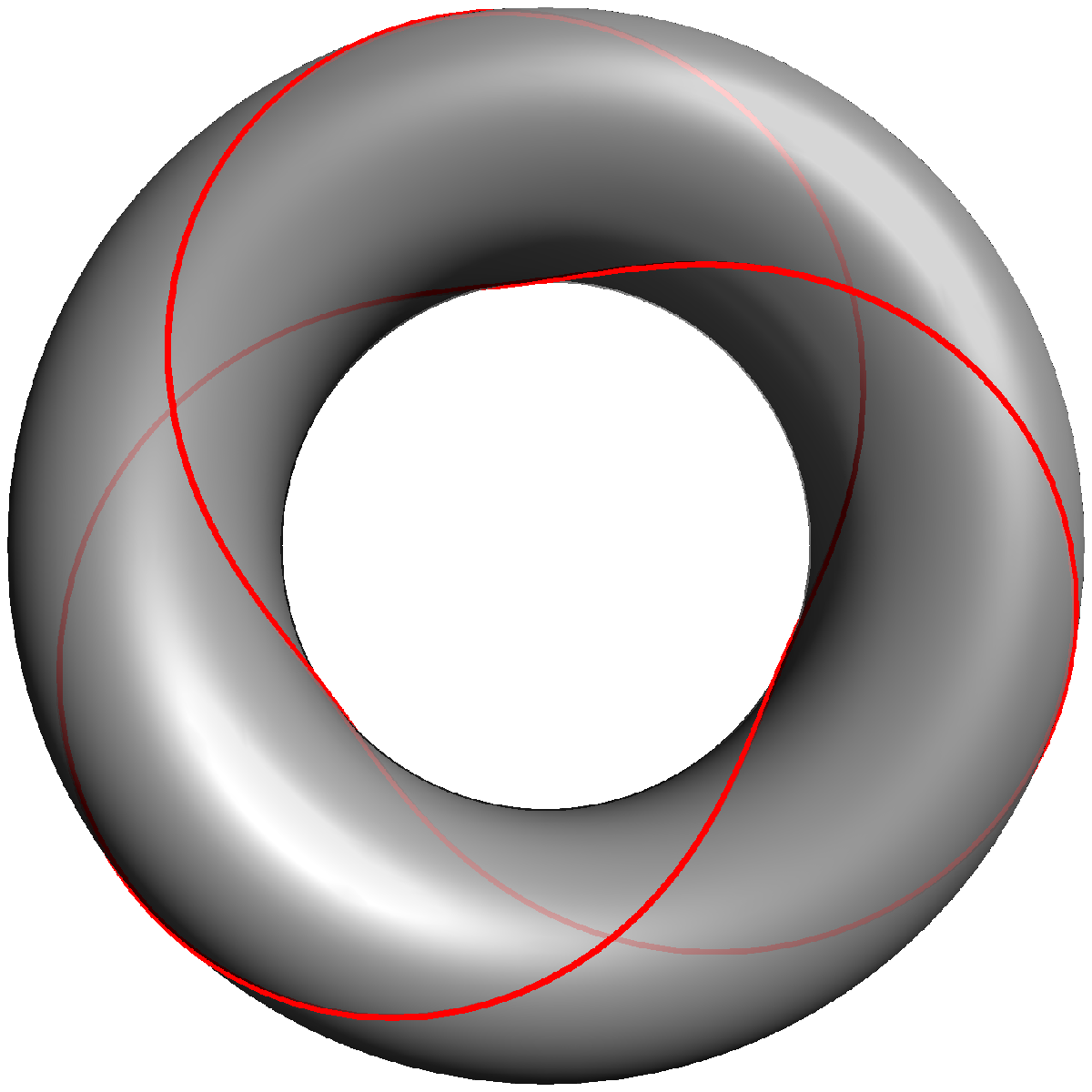}}\hskip6ex
\subfloat[~\hskip18ex~]{\includegraphics[width=0.25\textwidth]{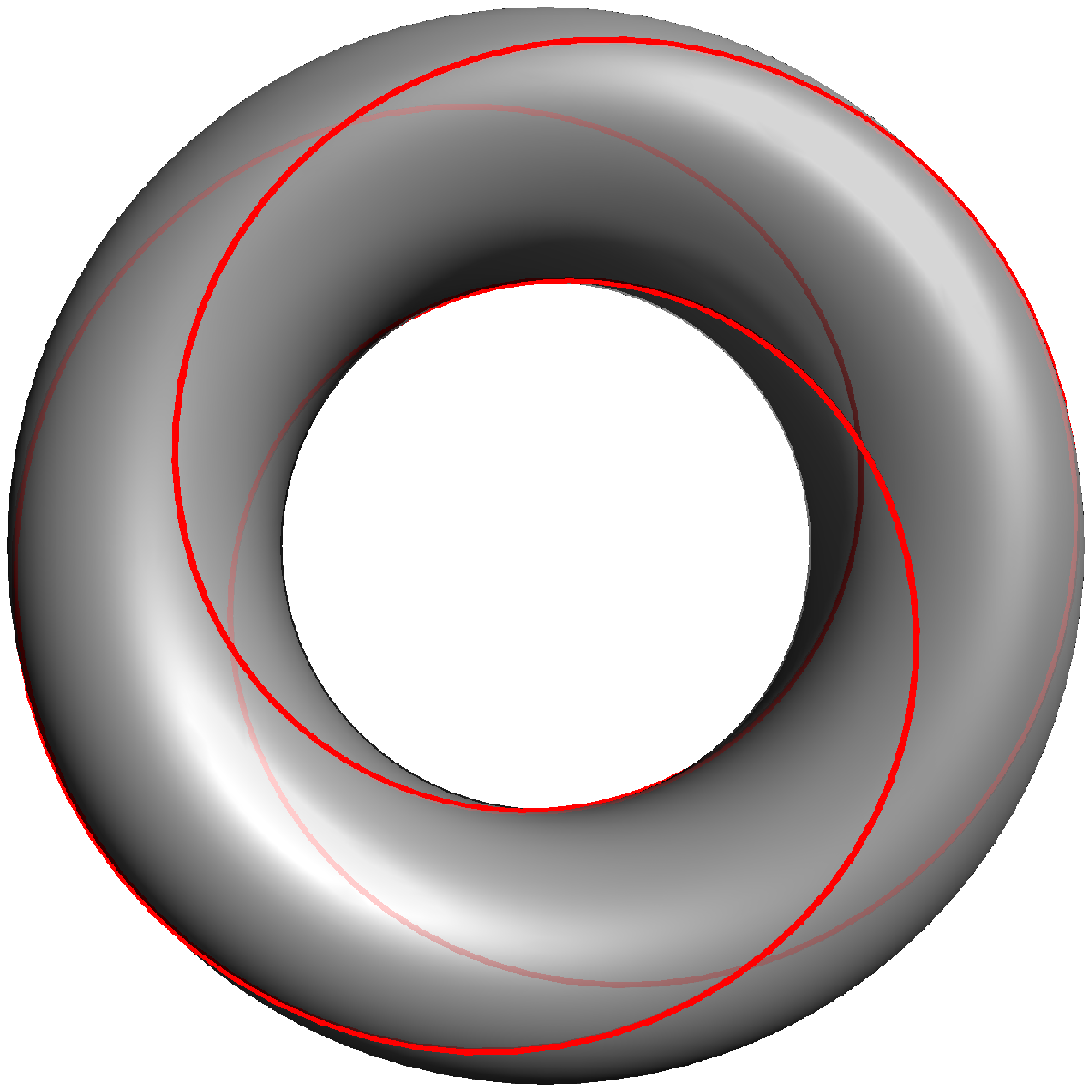}}
\end{center}
\caption{In figure (a) we depict the trefoil knot. As the trefoil is a torus knot, it can be drawn on the surface of a two-dimensional torus embedded in $S^3$. Either the trefoil winds around the small circle twice --- as in figure (b) --- or three times --- as in figure (c). These two choices map to the (equivalent) representations $\mathcal{K}_{2,3}$ and $\mathcal{K}_{3,2}$ of the trefoil, respectively.} \label{fig:trefoil}
\end{figure}
%%%%%%%%%%%%%%%%%%%%%%

Alternatively, we can model the trefoil knot by exchanging the integers $(r,s)=(2,3)$ to $(r,s)=(3,2)$. Then the ideal $\widehat{I}_{3,2}$ is generated by
\begin{equation}
\begin{aligned}
    \widehat{\mathcal{I}}_{3,2} \,=\, &\left\langle \ \alpha-\zeta^3,\, \beta-\rho_1\rho_2\rho_3,\, h_{3,2}(\zeta,\rho_1;Q),\, h_{3,2}(\zeta,\rho_2;Q),\,h_{3,2}(\zeta,\rho_3;Q),\,  \right.\\
    & \quad \left. h_{3,2}(\zeta,\rho_1,\rho_2;Q),\, h_{3,2}(\zeta,\rho_1,\rho_3;Q),\, h_{3,2}(\zeta,\rho_2,\rho_3;Q),  \right. \\
    & \quad \left. h_{3,2}(\zeta,\rho_1,\rho_2,\rho_3;Q) \ \right\rangle \ ,
\end{aligned}     
\end{equation}
with its generators recursively defined in eq.~\eqref{eq:hdef}. While the ideals $\widehat{\mathcal{I}}_{2,3}$ and $\widehat{\mathcal{I}}_{3,2}$ are rather distinct, by employing again the computer algebra system {\sc Singular} \cite{DGPS}, we find that the elimination ideal $\mathcal{I}_{3,2}$ is generated by the same polynomial \eqref{eq:AugK23}, i.e., 
\begin{equation} \label{eq:I23}
   \mathcal{I}_{2,3} \,=\, \mathcal{I}_{3,2} \ .
\end{equation}
This is in agreement with the fact that both torus knots $\mathcal{K}_{2,3}$ and $\mathcal{K}_{3,2}$ represent the trefoil (in framing $f=6$) as depicted in Figure~\ref{fig:trefoil}. Hence, also the moduli spaces of their Lagrangian probe branes $\mathcal{L}_{2,3}$ and $\mathcal{L}_{3,2}$ ought to be equivalent. The resulting agreement \eqref{eq:I23} --- which holds more generally for the two realizations $\mathcal{K}_{r,s}$ and $\mathcal{K}_{s,r}$ of torus knots --- serves as a non-trivial check of our proposal for the physical spectral curve of torus knots. This issue is discussed further in the next subsection.

Using the general formula~\eqref{eq:Bphys}, we can also determine the physical and the calibrated annulus kernel for the trefoil torus knot. The resulting kernels are rather long and not very illuminating. However, as we demonstrate the topological recursion, which we introduce in Section~\ref{sec:toprec}, at the example of the trefoil knot explicitly, the calibrated annulus kernel is required for carrying out our example computations. Therefore, for reference and for completeness the annulus kernels of trefoil are listed in Appendix~\ref{app:B23}.

Expanding the constructed calibrated annulus kernel in $\alpha_1$ and $\alpha_2$, after rescaling $\alpha_\mu \mapsto Q^5 \alpha_\mu,\mu=1,2,$ we read off the annulus instanton numbers of the trefoil knot $\mathcal{K}_{2,3}$ in framing $6$ from the expansion
\begin{equation} \label{eq:Kexp}
\begin{aligned}
 \widehat{B}&{}_{2,3}(\alpha_1,\alpha_2) d\alpha_1 d\alpha_2 \,=\, \frac{d\alpha_1 d\alpha_2}{(\alpha_1 -\alpha_2)^2}
 +d\alpha_1 d\alpha_2 \Big[\,60 - 144 Q + 117 Q^2 - 36 Q^3 + 3 Q^4\\
 &+ (1680 - 6048 Q + 8568 Q^2 - 6020 Q^3+ 2160 Q^4 - 360 Q^5 + 20 Q^6) (\alpha_1+\alpha_2) \\
 &+ (45045 - 216216 Q + 436293 Q^2- 479952 Q^3 + 311850 Q^4 \\
 &\qquad\qquad\qquad - 120960 Q^5 + 26838 Q^6 - 3024 Q^7 + 126 Q^8) (\alpha_1^2+\alpha_2^2) \\
 &+ (52920 - 254016 Q + 512736 Q^2 - 564480 Q^3 + 367290 Q^4 \\
 &\qquad\qquad\qquad -142800 Q^5 + 31800 Q^6 - 3600 Q^7 + 150 Q^8) \alpha_1 \alpha_2 +\ldots \Big] \ ,
\end{aligned}
\end{equation}
which agrees with the results for the trefoil knot presented in ref.~\cite{Brini:2011wi}. Guided by ref.~\cite{Diaconescu:2011xr}, we also calculated for some low degrees and windings the annulus instanton numbers in Appendix~\ref{app:Localization} explicitly. Comparing eq.~\eqref{eq:Kexp} with numbers listed in~\eqref{eq:locK23} we observe the expected agreement.

\subsection{Knot spectral curves and augmentation varieties}\label{sec:aug}
Using the results of Brini--Eynard--Mari\~no \cite{Brini:2011wi}, we have constructed the augmentation polynomial $F_{r,s}(\alpha,\beta;Q)$, which describes the physical spectral curve $\mathcal{C}_{r,s}$ of the brane $\mathcal{L}_{r,s}$ associated to the torus knot $\mathcal{K}_{r,s}$. The concept of knot augmentation polynomials actually arises from the differential graded algebra of the knot contact homology of a knot $\mathcal{K}$ in $\mathbb{R}^3$ \cite{MR2116316,MR2175153,MR2376818,MR2807087,MR3070519}, which amounts to calculating the (quantum) moduli space of the Lagrangian brane $\mathcal{L}$ of the knot directly in the A-model \cite{Aganagic:2013jpa}. 

For later reference, we briefly summarize a few aspects of augmentation varieties from differential graded algebra in knot contact homology. Each knot $\mathcal{K}$ embedded in $\mathbb{R}^3$ can be extended to a Legendrian submanifold $\Lambda_\mathcal{K}$ in the (five-dimensional) unit conormal bundle $ST^*\mathbb{R}^3$. For the knot $\mathcal{K}$ the Legendrian submanifold $\Lambda_\mathcal{K}$ has the topology of a two-torus, and therefore the relative homology $H_2(\Lambda_\mathcal{K},ST^*\mathbb{R}^3)\simeq H_1(\Lambda_\mathcal{K})\oplus H_2(ST^*\mathbb{R}^3)$ is generated by the one-cycles $\alpha, \beta$ of $H_1(\Lambda_\mathcal{K})$ and the two-cycle~$Q$ of $H_2(ST^*\mathbb{R}^3)$ generating the $S^2$-fiber of the conormal bundle $ST^*\mathbb{R}^3$.\footnote{Here, $\alpha, \beta$ and $Q$ are viewed as multiplicative generators of the homology group~$H_2(\Lambda_\mathcal{K},ST^*\mathbb{R}^3)$. These generators are related to the generators $\lambda,\mu$ and $U$ of ref.~\cite{MR2807087} according to~\eqref{eq:Conversion}.}  Furthermore, the Legendrian submanifold $\Lambda_\mathcal{K}$ associates to the knot $\mathcal{K}$ a (non-commutative) differential graded algebra $(\mathcal{A}_\mathcal{K},\partial)$ over the ring $\mathbb{Z}[\alpha^{\pm 1},\beta^{\pm 1},Q^{\pm 1}]$. The homology $HC_*(\mathcal{K})$ of this differential graded algebra refers to the knot contact homology of $\mathcal{K}$. An augmentation of the differential graded algebra over the field $\mathbb{C}$ is a graded ring homomorphism $\epsilon: \mathcal{A} \rightarrow \mathbb{C}$ with $\epsilon(1)=1$ that maps all boundaries  to zero, i.e., $\epsilon \circ \partial=0$. Finally, the knot augmentation variety $\operatorname{Aug}_\mathcal{K}(\alpha,\beta;Q)$ is the variety of all augmentations, which describes a variety in $\operatorname{Hom}(\mathbb{Z}[\alpha^{\pm 1},\beta^{\pm 1},Q^{\pm 1}],\mathbb{C})\simeq (\mathbb{C}^*)^3$. For the precise definitions and for further details on aspect of augmentation varieties from contact knot homology, we refer the interested reader to refs.~\cite{MR2116316,MR2175153,MR2376818,MR2807087,MR3070519,Aganagic:2013jpa,Ekholm:2013xoa}.

It follows that augmentation varieties $\operatorname{Aug}_\mathcal{K}(\alpha,\beta;Q)$ in codimension one in $({\mathbb{C}^*})^3$ can be represented by an augmentation polynomial $\widetilde{F}_\mathcal{K}(\alpha,\beta;Q)$,\footnote{The augmentation polynomial $\widetilde{F}_\mathcal{K}(\alpha,\beta;Q)$ is actually only defined up to multiplications by $c\,\alpha^n\beta^mQ^p$ with $c\in\mathbb{C}, n,m,p\in\mathbb{Z}$, that is to say up to multiplications by units. We use this ambiguity to represent the augmentation polynomial as a Q-dependent polynomial in the variables $\alpha$ and $\beta$ with a constant term normalized to one.} which we view here as the generator of an ideal $\mathcal{I}_\mathcal{K}$ in the polynomial ring $\mathbb{C}(Q)[\alpha,\beta]$. Furthermore, Aganagic, Ekholm, Ng and Vafa \cite{Aganagic:2013jpa} propose that the augmentation polynomial $\widetilde{F}_\mathcal{K}(\alpha,\beta;Q)$ describes the physical spectral curve of the moduli space of the Lagrangian brane $\mathcal{L}_\mathcal{K}$. 

Using refs.~\cite{Brini:2011wi,Jockers:2012pz}, we have proposed in this work an independent construction for the physical spectral curve of torus knots, which implies that the augmentation polynomial~$F_{r,s}(\alpha,\beta;Q)$ in eq.~\eqref{eq:AugCurve} ought to be identical to the augmentation polynomial~$\widetilde{F}_{r,s}(\alpha,\beta;Q)$ of torus knots $\mathcal{K}_{r,s}$ (calculated in the framing $r\cdot s$). This is indeed confirmed by checking many explicit examples of torus knots (see also ref.~\cite{Jockers:2012pz}). Thus we conjecture for all torus knots $\mathcal{K}_{r,s}$ the equivalence
\begin{equation} \label{eq:equiv}
  F_{r,s}(\alpha,\beta;Q) \,\equiv \, \widetilde{F}_{r,s}(\alpha,\beta;Q) \ ,
\end{equation}
This agreement furnishes a non-trivial check both on the physical interpretation of augmentation varieties put forward in refs.~\cite{Aganagic:2012jb,Aganagic:2013jpa} and on the construction of the physical spectral curve of this work, and it sheds light on both approaches. From the differential graded algebra point of view the calculation of the augmentation variety $\operatorname{Aug}_\mathcal{K}(\alpha,\beta;Q)$ --- i.e., the computation of the physical spectral curve --- is not limited to torus knots. The braid representation of any knot~$\mathcal{K}$ directly results in the associated differential graded algebra $(\mathcal{A}_\mathcal{K},\partial)$ and hence the augmentation variety $\operatorname{Aug}_\mathcal{K}(\alpha,\beta;Q)$ \cite{Ng:2012qq}. On the other hand, the concatenation of fractional unknot spectral curves canonically yields the calibrated annulus kernel generating the annulus instantons. Therefore, it would be interesting to generalize this method beyond the class of torus knots. We hope to come back to this issue elsewhere.

\subsection{Phase transitions and stretched annulus instantons}\label{sec:stretched}

As shown in refs.~\cite{MR2807087,Ng:2012qq}, the augmentation varieties $\operatorname{Aug}_\mathcal{K}(\alpha,\beta;Q)$ and hence the physical spectral curves have an involutive symmetry $\iota$, which maps the algebraic $\mathbb{C}^*$ coordinates according to\footnote{Note that for the choice of the algebraic $\mathbb{C}^*$~coordinates used for the unknot spectral curve~\eqref{eq:unknot}, this involutive symmetry is a bit modified, i.e., $\alpha\mapsto Q^{f+1}/\alpha$ and $\beta\mapsto 1/(Q\beta)$. However, the other augmentation polynomials stated in this work conform with the involutive symmetry of eq.~\eqref{eq:invol}. \label{ft:unknot}}
\begin{equation} \label{eq:invol}
   \iota: \ \alpha \mapsto \frac{Q^{f-1}}{\alpha} \ , \ \beta\mapsto \frac{1}{Q \beta} \ ,
\end{equation}
where $f$ refers to the framing of the knot $\mathcal{K}$.

The existence of this involution has interesting physical consequences for the phase structure of the probe branes $\mathcal{L}_\mathcal{K}$. Since, the vicinity $(\alpha,\beta)=(0,Q^{-1})$ of the physical spectral curve describes the probe brane $\mathcal{L}_\mathcal{K}$ in the large volume regime of the resolved conifold, it implies that also the image brane $\iota_*\mathcal{L}_\mathcal{K}$ in the vicinity $(\alpha,\beta)=(\infty,1)$ is a large volume brane of the resolved conifold. Note that these two points in the brane moduli space are not continuously connected within a single large volume phase. Hence, moving the large volume brane $\mathcal{L}_\mathcal{K}$ continuously to the large volume brane $\iota_*\mathcal{L}_\mathcal{K}$ indicates a phase transition in the open-string moduli space along such a path.

As both branes $\mathcal{L}_\mathcal{K}$ and $\iota_*\mathcal{L}_\mathcal{K}$ are large volume branes, the annulus amplitude between these two branes is generated by annulus instantons stretching between them. We also refer to these instantons as the stretched annuli. In practice, the stretched annulus instanton numbers are obtained by expanding the calibrated annulus kernel $\widehat{B}_\mathcal{L}(\alpha_1,\beta_1,\alpha_2,\beta_2) d\alpha_1 d\alpha_2$ (which is a bi-differential on the product $\mathcal{C} \times \mathcal{C}$ of the physical spectral curve) about the large volume points of the brane $\mathcal{L}_\mathcal{K}$ and $\iota_*\mathcal{L}_\mathcal{K}$ in the two spectral curve factors~$\mathcal{C}$, respectively. That is to say, the expansion of the calibrated annulus kernel
\begin{equation}
  \widehat{B}_\mathcal{L}(\alpha_1,\beta_1,\alpha_2,\beta_2)d\alpha_1d\alpha_2 \,=\, \widehat{B}_\mathcal{L}( \alpha_1,\beta(\alpha_1),\tilde\alpha_2,\beta(\tilde\alpha_2) ) d\alpha_1 d\tilde\alpha_2 \ ,
\end{equation}
about the coordinates $\alpha_1$ and $\tilde\alpha_2$ with $\tilde\alpha_2=\iota(\alpha_2)=\frac{Q^{f-1}}{\alpha_2}$ (and $d\alpha_2=\iota^*d\alpha_2$) yields the stretched annulus numbers.

In particular for the trefoil torus knots, the explicit stretched annulus numbers (in framing $f=6$) after rescaling $\alpha_1 \mapsto Q^5 \alpha_1, \tilde{\alpha}_2 \mapsto Q^5 \tilde{\alpha}_2$ read
\begin{equation} \label{eq:B23stretch}
\begin{aligned}
   \widehat{B}&{}_{2,3}( \alpha_1,\beta(\alpha_1),\tilde\alpha_2,\beta(\tilde\alpha_2) ) d\alpha_1 d\tilde\alpha_2 \,=\, 
   d\alpha_1d\tilde{\alpha}_2 \Big[   -9 Q + 16 Q^2 - 9 Q^3 + Q^4\\
   &+ (-168 Q + 504 Q^2 - 576 Q^3 + 300 Q^4- 60 Q^5)(\alpha_1+\tilde\alpha_2)\\
   &+ (-3861 Q + 16236 Q^2 - 28215 Q^3 + 25920 Q^4 \\
   &\qquad\qquad\qquad\qquad\qquad\qquad  - 13230 Q^5 + 3528 Q^6 -  378 Q^7) (\alpha_1^2+\tilde\alpha_2^2)\\
   & + (-3136 Q + 13230 Q^2 - 23040 Q^3 + 21280 Q^4 - 11088 Q^5 \\
   &\qquad\qquad\qquad\qquad\qquad\qquad + 3150 Q^6 - 400 Q^7 + 2 Q^8) \alpha_1 \tilde{\alpha}_2 +\ldots  \Big]   \ .
\end{aligned}
\end{equation}
We observe that there is no leading classical term and all stretched annuli contain at least one power of $Q$, which is the exponentiated complexified volume \eqref{eq:vol} of the one-cycle~$\mathbb{P}^1$ of the resolved conifold. Thus, the tension of all annulus instantons scales with the volume of the compact cycle of the resolved conifold, which confirms that the two branes $\mathcal{L}_{2,3}$ and $\iota_*\mathcal{L}_{2,3}$ are indeed spatially separated and reside in two distinct large volume phases.

%%%%%%%%%%%%%%%%%%%%%%%
\begin{figure}[t]
\begin{center}
\includegraphics[width=0.78\textwidth]{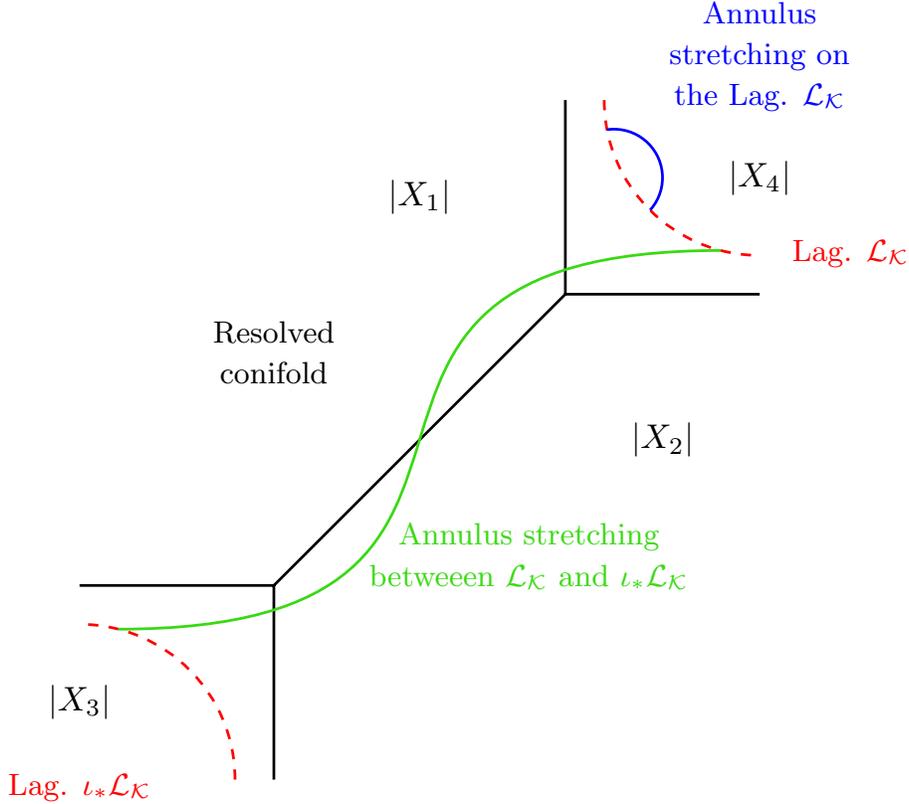}\hskip6ex
\end{center}
\caption{The dashed red lines schematically indicate two sets of branes $\mathcal{L}_\mathcal{K}$ and $\iota_*\mathcal{L}_\mathcal{K}$ in the resolved conifold. The blue line symbolizes an annulus instanton with both boundaries on the brane $\mathcal{L}_\mathcal{K}$. The green segment shows an annulus amplitude stretching between the branes $\mathcal{L}_\mathcal{K}$ and $\iota_*\mathcal{L}_\mathcal{K}$.} 
\label{fig:Stretched-Annulus}
\end{figure}
%%%%%%%%%%%%%%%%%%%%%%

In order to shed more light on the geometry of the two branes $\mathcal{L}_\mathcal{K}$ and $\iota_*\mathcal{L}_\mathcal{K}$, we trace the involution \eqref{eq:invol} on the physical spectral curve to the resolved conifold. Recall that a priori the physical spectral curve is a concept in the topological B-model, while the description of the Lagrangian brane $\mathcal{L}_\mathcal{K}$ arises in the topological A-model. Thus tracing the involution of the physical spectral curve to the resolved conifold amounts to tracing the involution through the local mirror map. The mirror geometry of the resolved conifold is described by \cite{Aganagic:2000gs,Aganagic:2001nx}
\begin{equation}\label{eq:mirrorCY}
   uv\,=\,y_{1}+y_{2}+y_{3}+y_{4}\ , \qquad y_{1}y_{2}\,=\,Q\,y_{3}y_{4} \ ,
\end{equation}
where $u,v\in{\mathbb{C}}$, and $y_{i}$, $i=1,\ldots,4,$ are the homogeneous coordinates subject to the constraint in \eqref{eq:mirrorCY}. The latter coordinates relate to the toric coordinates $X_i$, $i=1,\ldots,4$, of the toric skeleton of the resolved conifold according to $y_i = \log |X_i|^2$ as depicted in Figure~\ref{fig:Stretched-Annulus}. To describe the augmentation polynomial of the brane $\mathcal{L}_\mathcal{K}$ in the large volume regime, we need to trace the corresponding large volume point to the resolved conifold. For instance, we choose the patch $y_{4}\neq0$ and define the ${\mathbb{C}}^{*}$ variables $\alpha=-\frac{y_{1}}{y_{4}}$ and $\beta=-\frac{y_{2}}{Q\,y_{4}}$. We notice that with this definition of local coordinates, we find the mirror spectral curve \eqref{eq:unknot} (with $f=0$). In order to geometrically interpret the stretched annuli numbers, we also need to identify the image of the large volume point of the brane $\mathcal{L}_\mathcal{K}$ with respect to the involution $\iota$. To this end we notice that the involution $\iota$ maps the local patch $(\alpha,\beta)$ to the local patch $(\tilde\alpha,\tilde\beta)$ given by $\tilde{\alpha}=-\frac{y_{2}}{y_{3}}$ and $\tilde{\beta}=-\frac{y_{1}}{Q\,y_{3}}$.\footnote{Recall that the involution is rescaled for the curve~\eqref{eq:unknot} according to footnote~\ref{ft:unknot}.} This justifies the location of the branes $\mathcal{L}_{\mathcal{K}}$ and $\iota_*\mathcal{L}_{\mathcal{K}}$ as schematically summarized in Figure~\ref{fig:Stretched-Annulus}. Note that we have used here that the involution~$\iota$ is both a symmetry of the mirror spectral curve and the physical spectral curve of $\mathcal{L}_\mathcal{K}$ as described by the augmentation polynomial. For the unknot these two notions become the same.

Such stretched amplitudes play a decisive role for topological open-string amplitudes of orientifolds, where they become covering contributions to the orientifold amplitudes. In the context of local toric Calabi--Yau orientifolds with Harvey--Lawson branes such covering amplitudes are computed in refs.~\cite{Bouchard:2004iu,Bouchard:2004ri,Krefl:2009mw}. For orientifolds of the conifold the composite representations of quantum groups encode these covering amplitudes \cite{Marino:2009mw}. 

With this geometric understanding, we can now check the proposed stretched annulus numbers~\eqref{eq:B23stretch} for the trefoil knot~$\mathcal{K}_{2,3}$ (and in principal also for other torus knots $\mathcal{K}_{r,s}$), explicitly. Implementing the brane $\mathcal{L}_{2,3}$ and the image brane $\iota_*\mathcal{L}_{2,3}$ into the localization calculation for torus knots as discussed in Appendix~\ref{app:Localization} along the lines of ref.~\cite{Diaconescu:2011xr}, we calculated the stretched annulus numbers for the trefoil knot for some low windings explicitly as listed in~\eqref{eq:locK23}, and we find perfect agreement with~\eqref{eq:B23stretch}. Furthermore, we also extracted the stretched annulus amplitude numbers in Appendix~\ref{app:composite} from quantum groups in composite representations. Again, by comparing~\eqref{eq:B23stretch} with the end result~\eqref{StrAnn} for the trefoil knot (with framing $f=6$), we find perfect agreement.

Apart from demonstrating that the calibrated annulus kernel also encodes the stretched annuli numbers, which as shown in ref.~\cite{Marino:2009mw} and in Appendix~\ref{app:composite} interestingly relate to composite representations of quantum groups, the presented comparison serves as a non-trivial check on the validity of the calibrated annulus kernel. Since one of the expansion points is mapped by the involution $\iota$ to different point on the physical spectral curve, the extracted numbers \eqref{eq:B23stretch} probe the global analytic structure of the calibrated annulus kernel in Appendix~\ref{app:B23}. In particular, it confirms the calibration of the annulus kernel, which has been introduced somewhat ad hoc in eq.~\eqref{eq:Bcal}. The calibration condition is indeed crucial to extract the stretched annulus instanton numbers correctly.

%%%%%%%%%%%%%%%%%%%%%%%%%%%%%%
\section{Topological recursion on augmentation curves} \label{sec:toprec}
%%%%%%%%%%%%%%%%%%%%%%%%%%%%%%
Guided by the Rosso--Jones identity for quantum groups of torus knots \cite{MR1209320} and based on the remodelled B-model \cite{Bouchard:2007ys, Marino:2006hs}, Brini, Eynard and Mari\~no successfully apply a topological recursion to the spectral curve~\eqref{eq:fracunknot} associated to probe branes of torus knots \cite{Brini:2011wi}. As already indicated in Section~\ref{sec:curves}, at each stage in this recursion the construction yields fractional winding contributions in the algebraic brane modulus. Furthermore, it seems difficult to generalize this recursion --- which is intimately based on the spectral curve~\eqref{eq:fracunknot} --- beyond the class of torus knots. To address these two points, we discuss in this section the methods to apply the topological recursion based on the remodelled B-model directly to the augmentation variety of knot contact homology \cite{MR2116316,MR2175153,MR2376818,MR2807087,MR3070519}, which can be constructed for any given knot. Although currently our techniques can only be carried out explicitly for torus knots as well, we believe that in principal the computational scheme based on augmentation varieties is applicable beyond the class of torus knots.

The topological recursion of the remodelled B-model is suitable to describe toric branes in local Calabi--Yau geometries, for which its validity has been shown in ref.~\cite{Eynard:2012nj}. However, the branes associated to knots and described by augmentation varieties, generically generalize the class of toric branes in the context of the resolved conifold. Therefore, it is legitimate to further modify the recursion of the remodelled B-model, as we do here. A crucial step in this modification is to replace the Bergman kernel --- a basic ingredient in the ordinary topological recursion --- by the calibrated annulus kernel of the knot, as constructed for torus knots in Section~\ref{sec:BEM}. We argue that the proposed adaption of the recursion to the setting of knot augmentation varieties is consistent, and more importantly it produces the correct correlation functions.

Since the work of Aganagic and Vafa \cite{Aganagic:2012jb}, a puzzle is whether or in what sense the augmentation variety describes the closed-string sector of the resolved conifold. We address this question by computing explicitly the first two orders in the string coupling constant of the free energy. This is achieved by adopting the new viewpoint towards free energies in the context of topological recursion \cite{Borot:2013lpa} to our proposal of the modified recursion. In particular, we show that (the third derivative of) the genus zero contribution to the free energy of the resolved conifold can be computed directly from the augmentation variety of the knot without the knowledge of the annulus kernel, which at present we only know how to construct for torus knots. In this way, we correctly reproduce the genus zero part of the free energy from the augmentation varieties of both torus and non-torus knots. 

This section is structured as follows. In Section~\ref{sec:RemodelReview} we start with a brief review on the key ingredients of the topological recursion of the remodelled B-model, and we give computational definitions of $F^{(0)}$ and $F^{(1)}$ in the context of remodelled B-model, which have not been spelt out in ref.~\cite{Bouchard:2007ys}.

In Section~\ref{sec:RemodellingChanged} we consider the application of (modified) B-model remodelling to knot augmentation varieties. Section~\ref{sec:RemodellingChanged} begins with the argument of why one cannot use the standard Bergman kernel of knot augmentation varieties in order to compute the correlation differentials, and the rationale behind the construction of the calibrated annulus kernel. The next two subsections are concerned with the consistency of the modified B-model modelling. In Section~\ref{sec:PoleStructure} we analyze the physical annulus kernel and justify its calibration, and Section~\ref{sec:KnotFreeEnergy} shows that the definition of free energies based on the variational formula is still valid.

Section~\ref{sec:Results} contains our main results. After explaining in Section~\ref{sec:Techniques} the techniques we use to facilitate the computation of correlation functions, we present in Section~\ref{sec:Trefoil} the planar three-point function, the genus one one-point function, as well as the genus one free energy computed from the augmentation variety of the trefoil knot. We show that our results are consistent with the knot invariants arising from colored HOMFLY polynomials of trefoil. In Section~\ref{sec:F0} we show the remarkable fact that (the derivative of) the planar free energy computed from the augmentation variety of any knot is consistent with the free energy of the conifold resolution.

\subsection{The remodelled B-model in a nutshell}\label{sec:RemodelReview}
%%%%
We first review the key ingredients of the topological recursion of the remodelled B-model \cite{Bouchard:2007ys}, which realizes a local mirror B-model description of toric branes in non-compact toric Calabi--Yau threefolds. The remodelling is based on the original topological recursion developed by Eynard and Orantin, and we refer the reader for further details to refs.~\cite{Eynard:2007kz, Eynard:2008we}.

Given a spectral curve $\mathcal{C}$ in $\mathbb{C}^*\times\mathbb{C}^*$ defined as the zero locus of a polynomial $F(\alpha,\beta)$, one can extract three sets of information
\begin{itemize}
\item the ramification points $a_i$ with respect to the projection on the $\alpha$-plane given in terms of the zeros of $d\alpha(p)/\alpha(p)$ (which form a subset of the zeros of $d\alpha$);
\item the canonical meromorphic one-form $ \omega^{(0)}(p) = \Theta(p) = \log \beta(p)d\alpha(p)/\alpha(p)$;
\item and the Bergman kernel $\omega^{(0)}_2(p_1,p_2) = B(p_1,p_2)$ associated to the Riemann surface \cite{MR0054140},\footnote{For ease of notation, in the following we often absorb the differentials of forms into their name, e.g., for the Bergman kernel we write $B(p_1,p_2)$ instead of $B(p_1,p_2)d\alpha(p_1)d\alpha(p_2)$.} which is a suitable compactification of the spectral curve $\mathcal{C}$;
\end{itemize}
where $p$, $p_1$ and $p_2$ denote points on the spectral curve $\mathcal{C}$.

One can then compute an infinite series of \emph{stable correlation differentials} $\omega^{(g)}_n(p_1,\cdots,p_n)$ with $2g-2+n>0$ by the recursive equation\footnote{This formalism is valid if the spectral curve has only simple ramification points. Suitable formalisms for arbitrary ramification points have been developed in \cite{Bouchard:2012an, Bouchard:2012yg}. However, since we will only encounter simple ramification points in our computations, we do not need such extension here.}
\begin{equation} \label{equ:TopRecursion}
\begin{aligned}
	\omega^{(g)}_{n+1}(p,p_1,\cdots,p_n) \,=\,&
	 \sum_{i} \mathop{\Res}_{q\rightarrow a_i} K(p,q)\Big(\omega^{(g-1)}_{n+2}(q,\bar{q},J)  \\
	&\qquad +\sum_{h=0}^g \sum_{I\subset J}^{'}\omega^{(h)}_{|I|+1} (q,I)
	 \omega^{(g-h)}_{n-|I|+1}(\bar{q},J\backslash I) \Big) \ .	
\end{aligned}
\end{equation}
Here $J$ is the set of points $p_1,\cdots,p_n$ and $(I, J\backslash I)$ is a two-part partition of the set $J$.  The symbol $\mathop{\sum}^{'}$ means the exclusion of $(h,I) = (0,\varnothing)$ as well as  $(g,J)$. The point $\bar{q}$ is the conjugate point of $q$ near the simple ramification point $a_i$.  The recursion kernel $K(p,q)$ is constructed from $B(p,q)$  and $\Theta(q)$ as 
\begin{equation}
		K(p,q) = -\frac{1}{2}\frac{\int_{q'=\bar{q}}^q B(p,q' )}{\Theta(q) - \Theta(\bar{q})}\ .
\end{equation}
For instance, the genus one one-point correlation differential is computed by
\begin{equation}\label{equ:omega11}
	\omega^{(1)}_1(p) = \sum_i \mathop{\Res}_{q\rightarrow a_i} K(p,q)B(q,\bar{q}) \ ,
\end{equation}
and the genus zero three-point correlation differential reads
\begin{equation} \label{equ:omega3}
\begin{aligned}
	\omega^{(0)}_3(p_1,p_2,p_3)  \,=&\, \sum_{i} \mathop{\Res}_{q \rightarrow a_i} \frac{B(p_1,q) B(p_2,q) B(p_3,q) \alpha(q) \beta(q)}{d\alpha(q) d\beta(q)} \\
	\,=&\, \sum_{i} \frac{B(p_1,a_i) B(p_2,a_i) B(p_3,a_i) \alpha(a_i) \beta(a_i)}{d^2\alpha(a_i)/d\beta^2} \ ,
\end{aligned}
\end{equation}
where $B(p_1,a_i) = \frac{B(p_1,q)}{d\beta(q)}|_{q\rightarrow a_i}$. Note the only unstable correlation differentials are $\omega^{(0)}_1(p)$ and $\omega^{(0)}_2(p_1,p_2)$, which are given by $\Theta(p)$ and $B(p_1,p_2)$, respectively.

As proved in ref.~\cite{Bouchard:2007ys}, the integrated correlation functions $A^{(g)}_n = \int \omega^{(g)}_n(p_1,\cdots,p_n)$ expressed in flat open and closed coordinates calculate open $n$-point functions of the mirror topological A-model that are generated by open instantons at genus $g$. In particular, the disk instanton amplitude $A^{(0)}_1$, namely the disk-generated superpotential $W(p)$, and the annulus instanton amplitude $A^{(0)}_2$ are given by
\begin{gather}
	A^{(0)}_1(p) = \int \Theta(p) = \int \log \beta(p) \frac{ d \alpha(p)}{ \alpha(p)}\ , \\
	A^{(0)}_2 (p_1,p_2)= \int  \left(  B(p_1,p_2) - \frac{dp_1 dp_2}{(p_1-p_2)^2}  \right)\ .	\label{equ:AnnulusAmplitude}
\end{gather}

The free energies $F^{(g)}$, however, cannot directly be computed from \eqref{equ:TopRecursion}. Instead we advocate the viewpoint towards free energies proposed in \cite{Borot:2013lpa} based on the generalization of the variational formula for the correlation differentials, which in turn is the generalization of the loop insertion operator in the matrix model \cite{Eynard:2007kz,Eynard:2008we}. In this way we can give definitions of (derivatives of) $F^{(0)}$ and $F^{(1)}$ in the remodelled B-model. The variation of the spectral curve caused by the variation of some complex structure parameters of the spectral curve induces the variation of the canonical 1-form $\Theta(q)$
\begin{equation}
	\Omega (q) := \delta_{\Omega} \log(\beta(q))|_{\alpha(q)} \frac{d\alpha(q)}{\alpha(q)} = \frac{\delta_{\Omega}\; \beta(q)|_{\alpha(q)} d\alpha(q) }{\alpha(q) \beta(q) } \ .
\end{equation}
Then the meromorphic one-form $\Omega(q)$ can serve equally well as the yet unspecified variation parameters to characterize the variation of the spectral curve. Any meromorphic one-form can be decomposed into three pieces: a holomorphic differential, a meromorphic differential with only simple poles, and a meromorphic differential with only high order poles.  Each piece can be obtained by the integration of the Bergman kernel over some path with a suitable multiplier function (see for instance Section~4.3 of \cite{Eynard:2008we}). As a consequence, one can always find an integration path $\partial\Omega$ and a multiplier function $\Lambda(p)$ associated to the variation~$\Omega(q)$ such that\footnote{For different meromorphic one-forms $\Omega(q)$, examples of paths~$\partial\Omega$ and multipliers~$\Lambda(p)$ can be found in ref.~\cite{Eynard:2008we}.}
\begin{equation} \label{equ:OmegaLambda}
	\Omega(q) = \int_{\partial\Omega} B(p,q) \Lambda(p)\ .			
\end{equation}
Then the stable correlation differentials $\omega^{(g)}_n(p_1,\ldots,p_n)$ satisfy the following variational formulas \cite{Eynard:2007kz,Eynard:2008we}
\begin{equation}\label{equ:CorrelatorVariation}
	\delta_\Omega \omega^{(g)}_n(p_1,\ldots,p_n) = \int_{\partial \Omega} \omega^{(g)}_{n+1}(p,p_1,\ldots,p_n) \Lambda(p) \ ,
\end{equation}
as shown by simply adapting the arguments in ref.~\cite{Eynard:2007kz} to the remodelled scenarios.

In the same spirit as in ref.~\cite{Borot:2013lpa}, one can view the free energies~$F^{(g)}$ as correlation zero-forms~$\omega^{(g)}_{n=0}$ and define the free energies $F^{(g)}$ in the remodelled B-model by\begin{equation}\label{equ:FreeEnergyVariation}
	\delta_\Omega F^{(g)} =  \int_{\partial \Omega} \omega^{(g)}_{1}(p) \Lambda(p)	
	\quad \text{for}\quad g\geq 1\ .	
\end{equation}
Indeed this relation is consistent with the definition of the free energies~$F^{(g)}$ for $g\geq 2$ introduced in ref.~\cite{Bouchard:2007ys}.

The variational principle allows us to define the genus zero and genus one free energies~$F^{(0)}$ and $F^{(1)}$ as well. The (third derivative of the) planar free energy $F^{(0)}$ becomes
\begin{equation}\label{equ:F0Variation}
	\delta_{\Omega_i} \delta_{\Omega_j} \delta_{\Omega_k} F^{(0)} \,=\, \int_{\partial \Omega_i}\Lambda_i(p_1) \int_{\partial \Omega_j}\Lambda_j(p_2)  \int_{\partial \Omega_k}\Lambda_k(p_3) \omega^{(0)}_3(p_1,p_2,p_3) \ ,
\end{equation}
and using the identities \eqref{equ:OmegaLambda} and \eqref{equ:omega3} we arrive at 
\begin{equation}\label{equ:d3F0NoB}  
    \delta_{\Omega_i} \delta_{\Omega_j} \delta_{\Omega_k} F^{(0)} 
    \,=\, \sum_{\ell} \frac{\Omega_i(a_\ell)\Omega_j(a_\ell)\Omega_k(a_\ell) \alpha(a_\ell) \beta(a_\ell)}
    {d^2\alpha(a_\ell)/d\beta^2} \ .		
\end{equation}
Note that the last expression does not depend on the Bergman kernel any more. This is an important observation for us, as it allows us to compute the planar free energy of the resolved conifold from any knot augmentation variety (without the knowledge of the Bergman kernel).

The genus one free energy $F^{(1)}$ in the remodelled B-model is computed by inserting $\omega^{(1)}_1(p)$ of \eqref{equ:omega11} into the variational formula~\eqref{equ:FreeEnergyVariation}. Then integrating with respect to the variational parameters we arrive at the expression
\begin{equation}\label{equ:RemodelledF1}
	F^{(1)} = \frac{1}{24} \ln \left( \tau_B^{12} \prod_{i} \frac{\beta'(a_i)}{ \beta(a_i) \alpha(a_i) } \right) \ ,
\end{equation}
where the product is over all the zeros of the meromorphic form $d\alpha/\alpha$. The Bergman tau function $\tau_B$ is a function over the moduli space of the spectral curve $\mathcal{C}. $\footnote{To be precise, $\tau_B$ is a function on the Hurwitz space $H_{g,N}$ which is the moduli space of branched covering $\alpha: \mathcal{C} \rightarrow \mathbb{P}^1$. The curve $\mathcal{C}$ has genus $g$ and the meromorphic function $\alpha$ has degree $N$. $H_{g,N}$ is stratified according to poles and critical points of $\alpha$. In a generic stratum, the Hurwitz space is locally parametrized by the branch points $\alpha(a_i)$. See for instance \cite{MR2211143}.} It is characterized by the following property 
\begin{equation}
	\frac{\partial \ln  \tau_B}{\partial \alpha(a_i)}  = \mathop{\Res}_{q\rightarrow a_i} \frac{B(q,\bar{q})}{d\alpha} \ .
\end{equation}
$\beta'(a_i)$ is the derivative with respect to the local coordinate $z_i$ in the neighborhood of $a_i$ defined by
\begin{equation}
		\alpha = \alpha(a_i) + z_i^2 \ .
\end{equation}
Note that the formula~\eqref{equ:RemodelledF1} is very similar to but slightly different from the $F^{(1)}$ given in the original topological recursion \cite{Eynard:2007kz,Eynard:2008we}.

Inspired by the derivation of the planar free energy~$F^{(0)}$, it turns out to be computationally more feasible to insert the expression~\eqref{equ:omega11} into the variational formula~\eqref{equ:FreeEnergyVariation} before carrying out the integral. Then $\delta_\Omega F^{(1)}$ becomes an expression  only in terms of $\alpha(q), \beta(q), B(p,q)$, and $\Omega(q)$:
\begin{equation}\label{equ:d1F1Expanded}
\begin{aligned}
	\delta_\Omega F^{(1)} \,=\,& \frac{1}{2}\sum_{i}\delta_\Omega\, \alpha(a_i) \cdot \mathop{\Res}_{q\rightarrow a_i} \frac{B(q,\bar{q})}{d \alpha(q)}\\
	&+\sum_{i} \frac{ \Omega'' \beta' \alpha \beta^2 + \Omega(-\beta^{(3)} \alpha \beta^2 + 3\beta'' \beta' \alpha \beta- 2(\beta')^3 \alpha + 6 \beta' \beta^2 )  }{ 96 (\beta')^2 \beta } \Big|_{q=a_i}		
\end{aligned}
\end{equation}
Here we often use local coordinate $z_i$ in the neighborhood of the ramification points~$a_i$, and then derivatives $'$ are taken with respect to $z_i$; in particular 
$\Omega^{(k)} = \frac{d^k}{dz_i^k}\left( \frac{\Omega(q)}{dz_i}\right)$. 
Furthermore, the variation of the branch points $\delta_\Omega\, \alpha(a_i)$ can be computed from the definition of $\Omega(q)$ in the remodelled scenario as 
\begin{equation}
	\delta_\Omega\, \alpha(a_i) = -\frac{ \alpha(q) \beta(q) \Omega(q) }{ d\beta(q) } \Big|_{q=a_i}\ .
\end{equation}

We should also mention that in the remodelled B-model the genus one free energy~$F^{(1)}$ is often \emph{not} invariant under the exchange of the algebraic coordinates $\alpha$ and $\beta$, and as a consequence the projection coordinate has to be chosen properly. We relegate the discussion of this technical issue to Appendix~\ref{sec:SymVariance}.

\subsection{Remodeled B-model for torus knot augmentation curves}\label{sec:RemodellingChanged}
%%%%
Attempting to apply the topological recursion of the remodelled B-model on knot augmentation curves, one immediately faces difficulties with the Bergman kernel. The augmentation curves of nontrivial knots are usually of higher genus, and the Bergman kernel of a curve of genus greater than zero is of a transcendental nature. For instance, the augmentation polynomial~$F_{2,3}(\alpha,\beta;Q)$ of the trefoil knot describes a Q-depend genus one curve. Its Bergman kernel is given by the Akemann kernel~\cite{Akemann:1996zr}, which involves elliptic functions of the parameter $Q$. On the other hand, the Bergman kernel is supposed to be the generating function of the annulus instanton numbers. Its expansion in terms of brane modulus $\alpha$ reads
\begin{equation}\label{equ:BergmanExpansion}
B(p_1,p_2)=\frac{d\alpha_1 d\alpha_2}{(\alpha_1-\alpha_2)^2}
+d\alpha_1d\alpha_2\left(p_{(2)}+p_{(1,1)}(\alpha_1+\alpha_2)+p_{(0,2)}\alpha_1\alpha_2+\ldots\right) \ .	\end{equation}
In this notation subscript vector $\vec{k}=(k_1,k_2,\ldots)$ in $p_{\vec{k}}$ means that $k_j$ boundary components of the instanton have winding number $j$. In particular, for the annulus instanton numbers we have $|\vec{k}| = \sum_i k_i=2$. The coefficients $p_{\vec k}$ are proportional to the free energies $F_{0,\vec{k}}(Q)$ with $|\vec k|=2$ according to
\begin{equation}
	p_{\vec{k}} \,=\, \prod_j j^{k_j} F_{0,\vec{k}}(Q) \quad \text{for} \quad |\vec k|=2 \ . 
\end{equation}

\noindent Note that the prefactor $\prod_j j^{k_j}$ drops out after integrating the Bergman kernel to the annulus amplitude as in (\ref{equ:AnnulusAmplitude}). Furthermore, the large $N$ duality relates the free energy $F_{g,\vec{k}}(Q)$ to the connected Wilson loop expectation values $W^{(c)}_{\vec{k}}(q,Q)$ \cite{Gopakumar:1998ki, Ooguri:1999bv,MR2177747}
\begin{equation}\label{equ:StringGaugeDuality}
	\sum_{g=0} (-1)^g g_s^{2g-2+|\vec{k}|} F_{g,\vec{k}}(Q)  = \frac{1}{\prod_j j^{k_j}} W^{(c)}_{\vec{k}}(q,Q) \Big|_{q=e^{g_s}}	\ .
\end{equation}
In particular this implies that $F_{0,\vec{k}}(Q)  =  W^{(c)}_{\vec{k}}(q,Q)/\prod_j j^{k_j}|_{q=1}$ for $|\vec k|=2$. So the coefficients of the Bergman kernel in the expansion of $\alpha_1,\alpha_2$ are the same as the expectation values $W^{(c)}_{\vec{k}}(q,Q)$ in the planar limit. Since $W^{(c)}_{\vec{k}}(q,Q)$ can be written as polynomials in the HOMFLY polynomials $W^{\mathcal{K}}_R(q,Q)$ of the knot describing the Wilson loop\cite{Witten:1988hf}, one is led to the contradictory conclusion that the Bergman kernel itself at most can be a rational function of $Q$.

Nonetheless, one can assume that a non-trivial open-string mirror map exists among the open moduli of the A-branes on the resolved conifold and their mirror symmetric B-branes, so that secretly spectral curve associated to the bane is in fact of genus zero. This, however, is precluded under the assumption that the brane modulus $\alpha$ is a good affine coordinate in the neighborhood of $\alpha=0$.\footnote{This assumption in principle can be violated because to be more accurate $u=\frac{1}{2\pi i}\log\alpha$ instead of $\alpha$ itself is the brane modulus. The construction in \cite{Brini:2011wi} circumvents the following argument precisely because it violates this assumption. But this is also the reason in the instanton generating functions of the fractional power terms, which defy explanation and which we want to avoid.} Indeed on a Riemann sphere one can always find a coordinate $y$ whose coordinate chart covers almost the whole Riemann sphere except for one point, such that its zero coincides with the zero of $\alpha$. Then, $\alpha$ being locally regular, we can Taylor expand $\alpha$ in terms of $y$ with the leading coefficient normalized to one, i.e.,
\begin{equation}
	\alpha(y) = y+c_1 y^2 + c_2 y^3 + \ldots \ .
\end{equation}
On the other hand, in terms of the coordinate $y$ the Bergman kernel of the Riemann sphere has the simple form
\begin{equation}
	B(y_1,y_2)dy_1dy_2 \,=\, \frac{dy_1 dy_2}{(y_1-y_2)^2} \ .
\end{equation}
Now, we expand the coordinate $\alpha_1,\alpha_2$ in (\ref{equ:BergmanExpansion}) and compare the coefficients with the formula above to obtain the identities
\begin{equation}\label{equ:pcIdentities}
\begin{aligned}
	0 &= p_{(2)}-c_1^2 + c_2,\\
	0 &= p_{(1,1)}+2c_1^3 +2p_{(2)}c_1-4c_1c_2,\\
	0 &= p_{(0,2)} - 6 c_1^4 + 4 p_{(2)} c_1^2 + 16 c_1^2 c_2 + 4 p_{(1,1)}c_1 - 6 c_2^2,\\
	0 &= p_{(1,0,1)}- 3 c_1^4 + 9 c_1^2c_2 + 3 p_{(1,1)} c_1 + 3 p_{(2)} c_2 - 3 c_2^2,\\
	&\cdots \;\; \cdots
\end{aligned}
\end{equation}
After eliminating the coefficients~$c_i$ in \eqref{equ:pcIdentities} a new list of identities $\{J_i=0\}$ involving only the annulus instanton numbers $p_{\vec{k}}$ can be constructed. The simplest identity reads
\begin{equation}
	J_1 = 6 p_{(2)}^2 - 4 p_{(1,0,1)} + 3 p_{(0,2)} \,=\, 0 \ .
\end{equation}
However, this identity does not hold for a generic torus knots $\mathcal{K}_{r,s}$.\footnote{By construction this identity holds for unknot. Hence, it implies a non-linear relation among the quantum dimensions of $SU(N)$.} For instance, inserting the known annulus numbers in arbitrary framing $f$ the identities $J_1(\mathcal{K}_{2,3})$ and $J_1(\mathcal{K}_{2,5})$ for the torus knots $\mathcal{K}_{2,3}$ and $\mathcal{K}_{2,5}$ are respectively given by\footnote{One may be surprised to find that $J_1$ does not depend on the framing $f$. This, however, does not hold for the other identities $J_i$ starting at $i\ge2$. It would be interesting to know if the quantity $J_1(\mathcal{K})$ has a geometric meaning in knot theory.} 
\begin{equation}
\begin{aligned}
	J_1(\mathcal{K}_{2,3})\,&=\,36 (-1 + Q)^4 (5 - 4 Q + Q^2) \ , \\
	J_1(\mathcal{K}_{2,5})\,&=\,60 (-1 + Q)^4 (98 - 168 Q + 105 Q^2 - 28 Q^3 + 3 Q^4) \ . \\
\end{aligned}
\end{equation}
which are clearly non-vanishing. 

Therefore, we conclude that that the Bergman kernel of neither a proper high genus curve nor a Riemann sphere can be the generating function of annulus instanton numbers of a non-trivial knot (under the assumption that the coordinate $\alpha$ is locally affine in the neighborhood of $\alpha=0$).

Inspired by the engineering of an instanton enumeration problem out of the topological recursion computation \cite{Eynard:2009qr}, we reverse the line of thought and conjecture that the topological recursion might still work as long as the canonical form $\Theta(p)$ and the ``modified'' kernel $\widetilde{B}(p_1,p_2)$ are still generating functions of disk instanton numbers and annulus instanton numbers, respectively. The latter  might not be a Bergman kernel of the spectral curve anymore, but nevertheless the correlation differentials are the results of the discussed topological recursion computation and give rise to the generating functions of instanton numbers with the corresponding topology.

In Section~\ref{sec:BEM} we provide for a method to construct for the knot augmentation curves of torus knots the desired modified kernel, which we call the calibrated annulus kernel $\widehat{B}_{r,s}(p_1,p_2):= \widehat{B}_{r,s}(\alpha(p_1), \alpha(p_2);Q)d\alpha(p_1) d\alpha(p_2)$. It is the generating function of the annulus instanton numbers of the associated probe brane, and it is a rational function in $Q$ as expect from the large $N$ duality. We propose that the remodelled topological recursion can be applied to the knot augmentation curves together with the calibrated annulus kernel, which replaces the role of the Bergman kernel.\footnote{For torus knots, a similar idea has been put forward in ref.~\cite{Borot:2013lpa}.} First, however, we need to discuss some issues concerning the physical annulus kernel $B_{r,s}(p_1,p_2)$ introduced in Section~\ref{sec:BEM}, which may immediately jeopardize the validity of the proposed topological recursion.

\subsubsection{Pole structure of the physical annulus kernel}\label{sec:PoleStructure}
Since the calibrated annulus kernel~$\widehat{B}_{r,s}(p_1,p_2)$ for torus knots is a more general bi-differential than the Bergman kernel, it exhibits a more complicated pole structure. Let us first have a look at the uncalibrated kernel $B_{r,s}(p_1,p_2)$ constructed in \eqref{eq:Bphys} from the Bergman kernels of the auxiliary spectral curves $h_{r,s}(\zeta,\rho;Q)$ in eq.~\eqref{eq:fracunknot}. 

The curve $h_{r,s}(\zeta,\rho;Q)$ is of degree $(r+s)$ and hence it is a $(r+s)$-sheeted cover of the $\zeta$-plane. This implies that for a generic value $\zeta$, we find $\rho^{(k)}(\zeta), k=1,\ldots,r+s$, solutions that give rise to $r+s$ distinct points $(\zeta,\rho^{(k)})$ on the curve $h_{r,s}(\zeta,\rho;Q)$. Among all the $(r+s)$ solutions --- analytically continued to $\zeta=0$ --- there are exactly $r$ solutions that are regular at $\zeta=0$ --- corresponding to the $r$ expansion centers mentioned immediately below \eqref{eq:fracsuper} --- and that enter into the superpotential~\eqref{eq:fracsuper}. On the hand, when one uses the ideal~$\widehat{I}_{r,s}$ in \eqref{eq:Ihat} to construct the physical spectral curve $\mathcal{C}_{r,s}$, each $\rho^{(k)}$ is in fact a complete copy of the auxiliary curve $h_{r,s}(\zeta,\rho;Q)$, capable of moving to any of the $(r+s)$ covering sheets. As a result, the $r$ distinct $\rho$-components of $\beta$ can be chosen arbitrarily from the $(r+s)$ solutions in total, different selections giving rise to different covering sheets of the $\alpha$-plane, whose entirety is the physical spectral curve. The expansion of the physical superpotential $W_{r,s}(\alpha;Q)$ picks up only one covering sheet of the $\alpha$-plane, which contains the point $(\alpha,\beta)=(0,1/Q)$. This covering sheet automatically corresponds to the selection of the $r$ solutions of $\rho$ which are regular at $\zeta=0$.

Since from now on we want to stress that the choice of the $r$ distinct $\rho$-components is arbitrary, we rewrite the physical algebraic coordinates $\alpha$ and $\beta$ as given by 
\begin{equation}\label{equ:RelatingBEMandAVNew}
   \alpha\,=\,\zeta^r \ , \qquad \beta=(-1)^{r+1}\rho^{(\ell_1)}\cdot\ldots\cdot \rho^{(\ell_r)} \ ,
\end{equation} 
for any $1\le \ell_1 < \ldots < \ell_r\le r+s$. Also the construction of the uncalibrated kernel \eqref{eq:auxAnn} and \eqref{eq:Bphys} is rewritten to stress this arbitrary choice,
\begin{align}\label{eq:BphysNew}
   B_{r,s}(\alpha_1,\alpha_2;Q)d\alpha_1d\alpha_2 \,&=\, \sum_{m,n=1}^{r} B_{r,s}^{(m,n)}(\zeta_1,\zeta_2;Q)d\zeta_1 d\zeta_2 \nonumber\\
   &= \, \sum_{m,n=1}^{r}
   \frac{\left(r\rho^{(\ell_m)}(\zeta_1)^{r-1}{\rho^{(\ell_m)}}{}'(\zeta_1)\right)\left(r\rho^{(\ell_n)}(\zeta_2)^{r-1}\rho^{(\ell_n)}{}'(\zeta_2)\right)}{\left(\rho^{(\ell_m)}(\zeta_1)^r-\rho^{(\ell_n)}(\zeta_2)^r\right)^2} 
   d\zeta_1 d\zeta_2 \ , \nonumber\\
   &\qquad\qquad\qquad\qquad\quad\qquad\quad \alpha_\mu = (\zeta_\mu)^r \ , \quad \mu=1,2 \ .
\end{align}

The kernel construction~\eqref{eq:BphysNew} shows that the physical kernel $B_{r,s}(p_1,p_2)$ develops a double pole whenever $B_{r,s}^{(m,n)}(\zeta_1,\zeta_2;Q)d\zeta_1 d\zeta_2$ has a double pole. Thus according to eq.~\eqref{eq:auxAnn}, we have a double pole whenever $\rho^{(\ell_m)}(\zeta_1)=\eta \rho^{(\ell_n)}(\zeta_2)$ with $\eta$ an $r$-th root of unity. Due to $h_{r,s}(\zeta,\eta \rho) = h_{r,s}(\eta^{-s}\zeta,\rho)$ and since $\alpha$ and $\beta$ are invariant under $(\zeta,\eta\rho^{(\ell_m)})\to(\eta^{-s}\zeta,\rho^{(\ell_m)})$, we absorb the phase $\eta^{-s}$ into $\zeta$ and assume in the following that $\rho^{(\ell_m)}(\zeta_1)$ and $\rho^{(\ell_n)}(\zeta_2)$ are equal at the location of a double pole. This implies that at the location of a double pole indeed $\alpha_1$ and $\alpha_2$ have to be the same but $\beta_1$ and $\beta_2$ may be distinct, as long as some $\rho$ components coincide. This leads to a proliferation of double poles, which may induce two major problems for the topological recursion.

Let us turn to the first potential problem. The topological recursion often involves components of the form $B_{r,s}(q,\bar{q})$ where $\bar{q}$ is the conjugate point of $q$ near a ramification point. Due to the enhanced pole structure, the kernel $B_{r,s}(q,\bar{q})$ could be ill-defined at the point $(q,\bar q)$ on $\mathcal{C}_{r,s}\times\mathcal{C}_{r,s}$. To investigate this problem, we need an understanding of the ramification points of the augmentation curve $F_{r,s}(\alpha,\beta;Q)$ in terms of points on the auxiliary curve $h_{r,s}(\zeta,\rho;Q)$.

A ramification point $a_i=(\alpha_i,\beta_i)$ occurs on the augmentation curve, if there are two distinct local solutions $\beta(\alpha)$ and $\hat\beta(\alpha)$ that coincide at the point $a_i$, i.e., $\beta_i=\beta(\alpha_i)=\hat\beta(\alpha_i)$. This can happens when $\beta(\alpha)$ and $\hat\beta(\alpha)$ differ only by one $\rho$-component while the other $(r-1)$ $\rho$-components are already identical. Without loss of generality let the differing $\rho$-components be $\rho^{(\ell_1)}$ and $\rho^{(\hat{\ell}_1)}$ and the identical $\rho$-components be $\rho^{(\ell_k)}, k=2,\ldots,r$. Then we have a ramification point $a_i=(\alpha_i,\beta_i)$, if the two distinct $\rho$ components coincide at $\zeta_i$, i.e.,
\begin{equation}\label{equ:AugPolyRam}
\begin{aligned}
  (-1)^{r+1} \rho^{(\hat\ell_1)} \rho^{(\ell_2)}\cdots \rho^{(\ell_r)}&=\hat\beta \, \to \,
  \beta=(-1)^{r+1} \rho^{(\ell_1)} \rho^{(\ell_2)}\cdots \rho^{(\ell_r)} \ , \quad \hat\ell_1\ne \ell_1 \ ,\\ 
  &\text{if} \ \rho^{(\hat\ell_1)} \to \rho^{(\ell_1)} \ .
\end{aligned}
\end{equation}
Here all the $\rho$'s are evaluated for the same $\zeta$. This also shows that a ramification point of the augmentation curve corresponds to a ramification point of the auxiliary curve, namely when $\rho^{(\hat\ell_1)}$ and $\rho^{(\ell_1)}$ coincide at a ramification point of the auxiliary curve. Moreover, by construction in the vicinity of such a ramification point, $\hat\beta(\alpha)$ actually describes the conjugate point $\bar q$ of the point $q$ given by $\beta(\alpha)$. But due to the $(r-1)$ pairs of identical $\rho$ components in $\beta$ and $\hat\beta$, the annulus kernel $B_{r,s}(p_1,p_2)$ has a double pole at $(p_1,p_2)=(q,\bar q)$. This makes the expression $B_{r,s}(q,\bar{q})$ and hence the topological recursion ill-defined.

Calibrating the kernel can cure this problem! We observe that the principal part of the auxiliary kernel $B_{r,s}^{(m, n)}(\zeta_1,\zeta_2;Q)d\zeta_1d\zeta_2$ (when $\rho^{(\ell_m)}(\zeta_1)$ approaches $\rho^{(\ell_n)}(\zeta_2)$) is 
$$
 \frac{  d\rho^{(\ell_m)}(\zeta_1) d\rho^{(\ell_n)}(\zeta_2)}{(\rho^{(\ell_m)}(\zeta_1)-\rho^{(\ell_n)}(\zeta_2))^2} \ .
$$
This is in fact the Bergman kernel of the auxiliary curve, and expanding in $\zeta_1, \zeta_2$ yields the principal part $\frac{d\zeta_1 d\zeta_2}{(\zeta_1 - \zeta_2)^2}$. Thus, the kernel $B_{r,s}(q,\bar{q})$ becomes well-defined, if we remove the $(r-1)$ double poles caused by the $(r-1)$ identical components $\rho^{(\ell_2)},\ldots,\rho^{(\ell_r)}$ in $\beta$ and $\hat\beta$. This can be achieved by subtracting
\[
		\frac{(r-1)d\alpha_1 d\alpha_2}{(\alpha_1 - \alpha_2)^2} = (r-1) \frac{r^2 \zeta_1^{r-1}\zeta_2^{r-1}d\zeta_1d\zeta_2}{(\zeta_1^r - \zeta_2^r)^2} = (r-1) \frac{d\zeta_1 d\zeta_2}{(\zeta_1 - \zeta_2)^2} + \textrm{reg.}
\]
which is precisely how we calibrate the physical annulus kernel in (\ref{eq:Bcal}). From now on the annulus kernel always refers to the calibrated one unless otherwise specified.

Before we move on to the second problem, let us briefly summarize the pole structure of the calibrated kernel $\widehat{B}_{r,s}(p_1,p_2)$ has. Recall that for a generic given value $\zeta$ on the auxiliary curve $h_{r,s}(\zeta,\rho;Q)$ there are $(r+s)$ different values $\rho^{(k)}$. According to \eqref{equ:RelatingBEMandAV}, a point on the augmentation curve corresponds to the combination of arbitrary $r$ distinct $\rho$ components. And there are $r+s\choose r$ ways of selecting them. So a generic value of $\alpha$ value is shared by $r+s\choose r$ different points on the augmentation curve (with canonical framing). Fix a value for $\beta = (-1)^{r+1}\rho^{(\ell_1)}\cdot\ldots\cdot\rho^{(\ell_r)}$ describing the point~$p_1$ in $\widehat{B}_{r,s}(p_1,p_2)$. Then a double pole arises if $\hat\beta$ of the second point $p_2$ shares $i$ (with $i\ne r-1$) different $\rho$-components with $\beta$ of the first point $p_1$ and is evaluated at the same value of $\alpha$, i.e., $p_1=(\alpha,\beta)$ and $p_2=(\alpha,\hat\beta)$. There are ${r\choose i}{s \choose r-i}$ such possibilities. For these ${r\choose i}{s \choose r-i}$ double poles, the coefficient of the pole is $i-(r-1) = i+1-r$, as described in terms of the local coordinates $\alpha_1$ and $\alpha_2$ by  
$$
	\frac{(i+1-r )d\alpha_1 d\alpha_2}{(\alpha_1 - \alpha_2)^2}\ .
$$
(When $i=r-1$, the calibration of $\widehat{B}_{r,s}(p_1,p_2)$ cancels all the double poles and the calibrated kernel becomes regular at these points.) Using the Chu--Vandermonde identity we find in total
$\sum_{i=0,i\neq r-1}^{r} {r\choose i}{s \choose r-i} = {r+s\choose r}- r\cdot s$
points for a given generic value of $\alpha$. Note the above description can be made $r,s$ symmetric if one uses index $j=r-i$ to classify the values of $\beta$ for a given $\alpha$. Then the number of poles $\mathcal{N}_j$ in a given class $j$ and the associated principal part $\mathcal{P}_j$ of the calibrated kernel $\widehat{B}_{r,s}(p_1,p_2)$ in the local coordinates $\alpha_1$ and $\alpha_2$ are given by
\begin{equation}
	\mathcal{N}_j\,=\,{r\choose j}{s \choose j} \ , \quad \mathcal{P}_j\,=\,\frac{(1-j)d\alpha_1 d\alpha_2}{(\alpha_1 - \alpha_2)^2}\ . 
\end{equation}

\begin{figure}[t]
\centering
\subfloat[With Bergman kernel]{\includegraphics[width=0.45\textwidth]{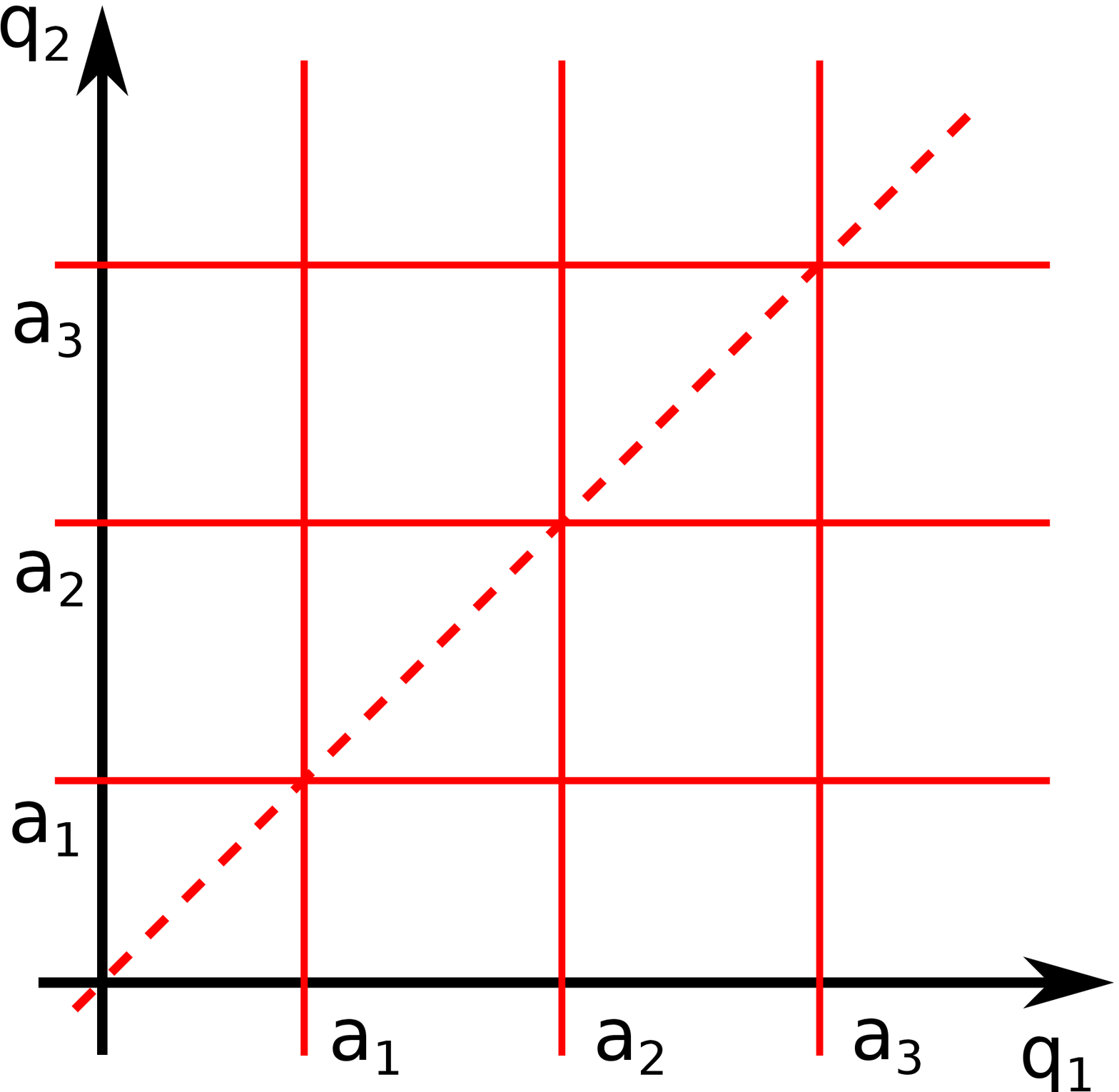}\label{fig:ZeroDivisor1}}
\subfloat[With annulus kernel $\widehat{B}_{r,s}(q_1,q_2)$]{\includegraphics[width=0.45\textwidth]{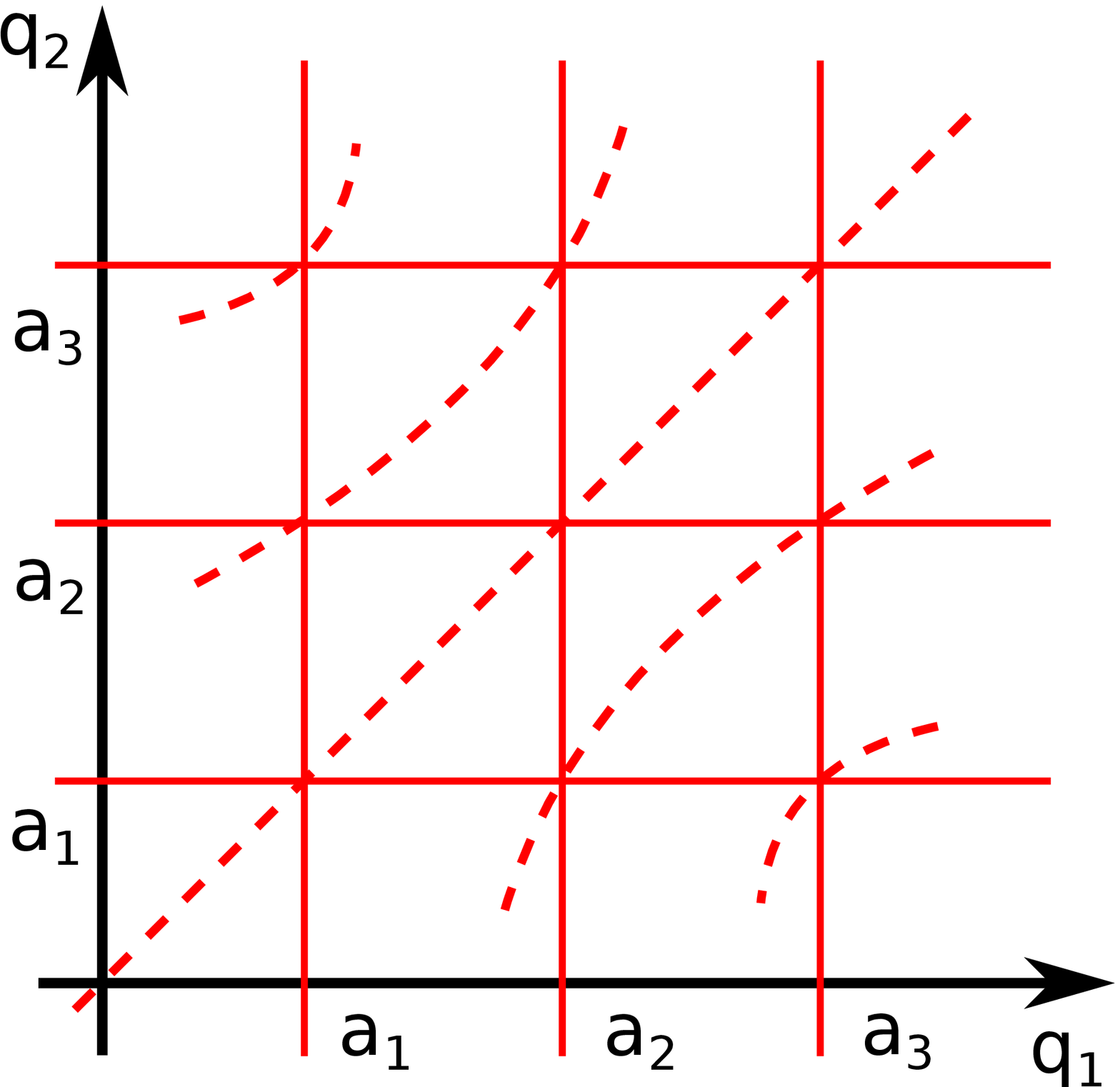}\label{fig:ZeroDivisor2}}
\caption{Here is an schematic illustration of the zero divisors of the denominators of some $I^{(g)}_n$, which has two internal variables $q_1,q_2$ parametrizing the variety $\mathcal{C}\times\mathcal{C}$. Figure (a) is the case of (remodelled) topological recursion. The horizontal and vertical red lines are the divisors $q_i -a_i$ while the diagonal dashed red line is the divisor $q_1- q_2$ from the pole of $B(q_1,q_2)$. The intersection points are $(a_i,a_j)$. In figure (b) the Bergman kernel is replaced by the annulus kernel $B_{r,s}(q_1,q_2)$. Although the zero divisor from the poles of $B_{r,s}(q_1,q_2)$ (red dashed curves) is more complicated, the intersection points are still $(a_i,a_j)$.} 
\end{figure}

Let us know consider the second problem, which is more subtle. In the recursive equation of the unmodified B-model remodelling (\ref{equ:TopRecursion}), the residues are only taken at the ramification points because the integrand on the right hand side consisting of the recursion kernel and other correlation differentials have poles only at the ramification points. The latter feature is due to the underlying matrix model construction, or can be argued from the zero loci of the canonical one-form $\Theta(p)$ and the pole loci of the Bergman kernel. In the modified B-model remodelling adapted to the knot augmentation curves, the underlying matrix model is yet missing. With the Bergman kernel replaced by the annulus kernel $\widehat{B}_{r,s}(p_1,p_2)$, one has to study whether new poles can arise in the integrand on the right hand of (\ref{equ:TopRecursion}). If new poles do arise, their residues should be included in the recursion computation.

Let us look at the computation of the differential $\omega^{(g)}_n(p_1,\ldots,p_n)$ in the unmodified B-model remodelling more closely. Apply the recursive formula (\ref{equ:TopRecursion}) repeatedly until all the correlation differentials on the right hand side are removed, we get
\begin{equation}\label{equ:IntegrandI}
	\omega^{(g)}_n(p_1,\ldots,p_n)\,=\,\sum_{i_1,\ldots,i_h}\mathop{\Res}_{q_1\rightarrow a_{i_1}}\cdots\mathop{\Res}_{q_h\rightarrow a_{i_h}}I^{(g)}_n(p_1,\ldots,p_n,q_1,\ldots,q_h) \ .
\end{equation}
The integrand $I^{(g)}_n$ on the right hand side now consists of only copies of the canonical form $\Theta(p)$ and the Bergman kernel $B(p_1,p_2)$.
The residues are taken over the \emph{internal variables} $q_i$. The number of these internal variables is $h=2g-2+n$ \cite{Eynard:2007kz,Eynard:2008we}. Then the statement in the previous paragraph that the integrand in (\ref{equ:TopRecursion}) only has poles at the ramification points is equivalent to the observation that the integrand $I^{(g)}_n$ only has residues at the positions $(a_{i_1},\ldots,a_{i_h})$. If we think of the $h$ internal variables $q_i$ as parametrizing the $h$-dimensional complex space $\mathcal{C}\times\ldots\times\mathcal{C}$, the differential $I^{(g)}_n$ only has residues at the intersection points of the zero divisors of the denominator of $I^{(g)}_n$. The zero divisors are either $q_i -a_i$ which are the zero loci of $\Theta(q)-\Theta(\bar{q})$, or $q_i - q_j$ which are the pole loci of $B(q_i,q_j)$, and their intersection points are inevitably tuples $(a_{i_1},\ldots,a_{i_h})$. The case when $h=2$ is illustrated schematically in Figure~\ref{fig:ZeroDivisor1}.

In the case of modified B-model remodelling adapted to knot augmentation curves, one can show that although the annulus kernel $\widehat{B}_{r,s}(q_1,q_2)$ has a more involved pole structure, the zero loci of $\Theta(q_i)-\Theta(\bar{q_i})$ and the pole loci of $\widehat{B}_{r,s}(q_i,q_j)$ still only intersect at tuples of ramification points. The proof is given in Appendix~\ref{sec:PolePositions}. Therefore the differential $I^{(g)}_n$ has still only residues at $(a_{i_1},\ldots,a_{i_h})$, and no new residues need to be added.

\subsubsection{Variational formula and free energies}	\label{sec:KnotFreeEnergy}
In order to compute from the knot augmentation curves not only the correlation differentials $\omega^{(g)}_n(p_1,\ldots,p_n)$ but also the free energies $F^{(g)}$, we need to check whether the definitions of the free energies are still consistent with the proposed modification to the remodelled B-model. We now demonstrate that the definitions~\eqref{equ:FreeEnergyVariation} given in Section~\ref{sec:RemodelReview} readily generalize by just including an appropriate normalization factor, namely 
\begin{align}
\delta_{\Omega_i} \delta_{\Omega_j} \delta_{\Omega_k} F^{(0)} \,&=\, \frac{1}{N_{r,s}}\int_{\partial \Omega_i}\Lambda_i(p_1) \int_{\partial \Omega_j}\Lambda_j(p_2)  \int_{\partial \Omega_k}\Lambda_k(p_3) \omega^{(0)}_3(p_1,p_2,p_3)	\ ,	\label{equ:ModifiedF0}\\
	\delta_\Omega F^{(g)} \,&=\,  \frac{1}{N_{r,s}}\int_{\partial \Omega} \omega^{(g)}_{1}(p) \Lambda(p)		\ , \qquad	g\geq 1 \ ,	 \label{equ:ModifiedFg}
\end{align}
where the normalization factor reads
\begin{equation}
	N_{r,s} = {r+s-2 \choose r-1} \ .
\end{equation}
Furthermore, the basis of the definition (\ref{equ:FreeEnergyVariation}) --- namely the variational formula for the stable correlation differentials (\ref{equ:CorrelatorVariation}) --- is still valid in the modified B-model remodelling up to the normalization $N_{r,s}$, i.e., 
\begin{equation}\label{equ:VaritationTorus}
	\delta_\Omega \; \omega^{(g)}_n(p_1,\ldots,p_n) \,=\, \frac{1}{N_{r,s}}\int_{\partial \Omega} \omega^{(g)}_{n+1}(p,p_1,\ldots,p_n) \Lambda(p) \ ,
\end{equation}
where the integration path $\partial \Omega$ and the multiplier $\Lambda(p)$ are given by
\begin{equation}\label{equ:Lambda}
	\Omega (q) \,=\, \int_{\partial \Omega} \widehat{B}_{r,s}(p,q) \Lambda(p) \ .
\end{equation}
We should stress that, as the annulus kernel~$\widehat{B}_{r,s}(p,q)$ generalizes the Bergman kernel, the definition of $\partial \Omega$  and $\Lambda(p)$ are just formal because given an arbitrary $\Omega(q)$ we may not be able to find $\partial \Omega$  and $\Lambda(p)$ such that they satisfy the relation~\eqref{equ:Lambda}. This, however, is of no concern since the first point $p$ in correlation differential~$\omega^{(g)}_{n+1}(p,p_1,\ldots,p_n)$ only appears in the recursion kernel $K(p,q_1)$ containing $\widehat{B}_{r,s}(p,q_1)$. So $\partial\Omega$  and $\Lambda(p)$ are always combined with $\widehat{B}_{r,s}(p,q_1)$ and thus can immediately be replaced by $\Omega(q_1)$.
 
The new variational formula for correlation differentials can be easily proven by adapting the arguments in Section~5 of ref.~\cite{Eynard:2007kz} together with the observation that the annulus kernel $\widehat{B}_{r,s}(p_1,p_2)$ satisfies the Rauch variational formula including the normalization factor $N_{r,s}$
\begin{equation}
	\delta_\Omega \widehat{B}_{r,s}(p_1,p_2)\Big|_{\alpha(p_1),\alpha(p_2)} \,=\, \frac{1}{N_{r,s}}\sum_i\mathop{\Res}_{q\rightarrow a_i} \frac{\Omega(q) \widehat{B}_{r,s}(p_1,q) \widehat{B}_{r,s}(p_2,q) \alpha(q)\beta(q)}{ d\alpha(q)d\beta(q)} \ .
\end{equation}
We verify this normalized Rauch variational formula in Appendix~\ref{sec:RauchCalibratedKernel}.

One caveat is that the appearance of the normalization factor $N_{r,s}$ in the variational formula indicates the correlation differentials may be greater than what we expect by several powers of $N_{r,s}$. A careful study shows that when one computes $\omega^{(g)}_n(p_1,\ldots,p_n)$ using modified B-model remodelling on an augmentation curve, every time the sum over residues at ramification points in terms of an internal variable $q$ is performed, a copy of $N_{r,s}$ appears. So a stable correlation differential $\omega^{(g)}_n(p_1,\ldots,p_n)$ is enhanced by a factor of $N_{r,s}^{2g-2+n}$, where the exponent is the number of internal variables. We can remove the normalization factor by defining the \emph{normalized stable correlation differential}
\begin{equation}
	\widehat{\omega}^{(g)}_n(p_1,\ldots,p_n) \,=\, \frac{1}{N_{r,s}^{2g-2+n}} \omega^{(g)}_n(p_1,\ldots,p_n) \ ,
\end{equation}
which generates instead the correct instanton numbers. Using this notation, one can also remove the normalization factor in (\ref{equ:ModifiedF0}), (\ref{equ:ModifiedFg}), and (\ref{equ:VaritationTorus}).
\begin{gather}
\delta_{\Omega_i} \delta_{\Omega_j} \delta_{\Omega_k} F^{(0)} \,=\, \int_{\partial \Omega_i}\Lambda_j(p_1) \int_{\partial \Omega_j}\Lambda_j(p_2)  \int_{\partial \Omega_k}\Lambda_k(p_3) \widehat{\omega}^{(0)}_3(p_1,p_2,p_3) \ ,		\label{equ:NormalizedF0}\\
	\delta_\Omega F^{(g)} \,=\,  \int_{\partial \Omega} \widehat{\omega}^{(g)}_{1}(p) \Lambda(p)\ ,		\qquad	g\geq 1 \ ,	 \label{equ:NormalizedFg}\\
	\delta_\Omega \; \widehat{\omega}^{(g)}_n(p_1,\cdots,p_n) \,=\, \int_{\partial \Omega} \widehat{\omega}^{(g)}_{n+1}(p,p_1,\cdots,p_n) \Lambda(p).		\label{equ:VariationTorusKnotNormalized}
\end{gather}
Once the variational formula for correlation differentials is established, the definitions of the free energies (\ref{equ:NormalizedF0}) and (\ref{equ:NormalizedFg}) are justified. In practice it is more convenient to use (\ref{equ:d3F0NoB}) and (\ref{equ:d1F1Expanded}), with proper normalization, to compute the derivatives of $F^{(0)}$ and $F^{(1)}$ with respective to the curve parameters.

In the next section we demonstrate that we can indeed extract the free energies of conifold resolution from knot augmentation curves. In particular, since (the third derivative of) the planar free energy $F^{(0)}$ does not require the knowledge of the annulus kernel (which at present we only know how to construct for torus knots), we verify that --- up to the normalization factor $N_\mathcal{K}$ --- the planar free energy can indeed be computed for knots beyond the class of torus knots. Therefore, we conjecture that the variational formula \eqref{equ:VaritationTorus} with the appropriate normalization factor $N_\mathcal{K}$ is valid any knots $\mathcal{K}$.

\subsection{Computational results from the topological recursion}\label{sec:Results}
%%%%%
In this section we present our computational results in favor of our proposal that the remodelled B-model can be applied to knot augmentation curves equipped with the annulus kernel $\widehat{B}_{r,s}(p_1,p_2)$ constructed in (\ref{eq:Bphys}) and (\ref{eq:Bcal}). Before spelling out the details, we first explain the  techniques we use to facilitate our computation. This is especially useful if only one internal variable $q$ arises in the computation, e.g. in the cases of of $W^{(0)}_3(p_1,p_2,p_3)$, $W^{(1)}_1(p)$ as well as in the derivatives of the free energies $F^{(0)}$ and $F^{(1)}$.

\subsubsection{Interlude: Discriminants, symmetrization and inversion}\label{sec:Techniques}
The augmentation curve of a nontrivial knot has many ramification points. As shown in Appendix~\ref{sec:RamPoints}, for the augmentation curve $F_{r,s}(\alpha,\beta;Q)$ of a torus knot $\mathcal{K}_{r,s}$ --- in a generic framing and for a generic value of $Q$ --- the number $\#(a_i)$ of ramification points $a_i$ is 
\begin{equation}
	\#(a_i)\,=\, 2 {r+s-2\choose r-1}\ .
\end{equation}
So already the torus knot $\mathcal{K}_{2,3}$ has six ramification points, and in general it is not possible to find those ramification points analytically. Instead, one expects the final result of computing a correlation differential to be a symmetric polynomial in terms of the ramification points. Therefore, one can reduce the correlation differential to elementary symmetric polynomials, which are in turn expressed by the coefficients of some character polynomial, which is a polynomial whose zeros include all the ramification points. To make such a structure plausible, however, we should be able to parametrize all the ramification points with just one discriminating coordinate.

This discriminating coordinate can usually be found. For example on a hyperelliptic curve given by
$$
		y^2 = \sigma(x) \ ,
$$		
the ramification points are completely distinguishable by their values of $x$. Even if neither of the original coordinates $x$ and $y$ can completely distinguish all the ramification points, one can always tilt the coordinate planes slightly to break the degeneracy. In practice, this means one can find a new coordinate $z$ being a linear combination of $x$ and $y$ which can make all the distinction. Note here we are not changing the projection plane with respect to which the ramification points are defined. 

In the case of knot augmentation curves, $\beta$ is a suitable discriminating coordinate, while $\alpha$ is not (at least for all the knots that we have checked so far). Thus we need to find the character polynomial $\Disc(\beta)$ termed the \emph{discriminant}, whose zeros should give us all $\beta(a_i)$. However, the correlation differentials, nonetheless, are expressed in terms of both coordinates $\alpha$ and $\beta$ evaluated at the ramification points. The question is then how to eliminate the $\alpha(a_i)$'s in favor of the $\beta(a_i)$'s.

We notice that the ramification points are the simple zeros of the zero-dimensional ideal $\mathcal{J}_{r,s}$ generated by $F_{r,s}(\alpha,\beta)$ and $\partial_\beta F_{r,s}(\alpha,\beta)$ in the ring $\mathbb{Q}(Q)[\alpha,\beta]$, and to obtain $\Disc(\beta)$ we essentially face an elimination problem. To this end, we perform the Euclidean division on polynomials
$$
  F_{r,s}(\alpha,\beta) \,=\, \partial_\beta F_{r,s}(\alpha,\beta) P_1(\alpha,\beta) + S_1(\alpha,\beta )\ ,
$$  
so that $S_1(\alpha,\beta )$ belongs to the ideal $\mathcal{J}_{r,s}$ and has degree of $\alpha$ strictly lower than that in $\partial_\beta F_{r,s}(\alpha,\beta)$. In fact since $F_{r,s}(\alpha,\beta)$ and $\partial_\beta F_{r,s}(\alpha,\beta)$ have the same degree in $\alpha$ their roles can be exchanged. Then we continue this procedure to find $S_2(\alpha,\beta ), \ldots,$ all being in the ideal $\mathcal{J}_{r,s}$
$$
\begin{aligned}
	\partial_\beta F_{r,s}(\alpha,\beta) \,&=\, S_1(\alpha,\beta) P_2(\alpha,\beta) + S_2(\alpha,\beta )\ ,\\
	\cdots \;& \; \cdots \;\; \cdots\\
	S_{n-2}(\alpha,\beta) &= S_{n-1}(\alpha,\beta) P_n(\alpha,\beta) + S_n(\beta )\ ,
\end{aligned}
$$
until $S_n(\beta)$ is only a polynomial in $\beta$. Then the square-free part of $S_n(\beta)$ is the sought discriminant $\Disc(\beta)$. Note that the $\alpha(a_i)$ to $\beta(a_i)$ conversion problem is also solved along the way. If $\beta$ is the discriminating coordinate, then the relation $S_{n-1}(\alpha,\beta)=0$ --- linear in $\alpha$ --- can be solved for $\alpha$, and we obtain
\begin{equation}\label{equ:RemoveAlpha}
	\alpha(a_i) \,=\, \alpha(\beta(a_i)) \ ,
\end{equation}
as a rational function of $\beta$, which converts $\alpha(a_i)$ into a rational function of $\beta(a_i)$.

After using (\ref{equ:RemoveAlpha}) to remove all the $\alpha(a_i)$ in the result of correlation differential computation, one finds another problem. The result is presented as the sum of rational function $R(\beta(a_i))$ in all the different $\beta(a_i)$'s. Adding them together yields a horrendously lengthy rational function, the symmetric reduction of whose numerator and denominator would take enormous computer time. However since now the base ring $\mathbb{Q}(Q)[\beta]/\Disc(\beta)$ where the ramification points live is actually a field, one can find the inverse of the denominator of $R(\beta(a_i))$ and convert the rational function $R(\beta(a_i))$ to a polynomial $P(\beta(a_i))$. The way to do it is solve the equation
\begin{equation}\label{equ:PolyGCD}
	\Disc(\beta) M(\beta)+ D(\beta) N(\beta) = L\ ,		
\end{equation}
where $\Disc(\beta)$ and $D(\beta)$ are the discriminant and the denominator of $R(\beta(a_i))$ we want to invert respectively. The aim is to find polynomials $M(\beta)$ and $N(\beta)$ such that $L$ is independent of $\beta$ (although it is still a polynomial in $Q$). This is similar to the B\'ezout's problem in number theory, which is to find integral solutions $p$ and $q$ to the equation
\begin{equation}
	r \; p+s \; q \,=\, \gcd(r,s)	\ ,\label{equ:GCD}
\end{equation}
with $\gcd(r,s)$ the greatest common divisor of the integers $r$ and $s$. Similarly, the equation \eqref{equ:PolyGCD} has a solution as long as $\Disc(\beta) $ and $D(\beta)$ do not have any factors in common, which is the case since $D(\beta)$ does not vanish at a ramification point. Analogously, as for B\'ezout's problem repetitive Euclidean divisions allow us to determine $N(\beta)$, and then $N(\beta)/L$ is the inverse of $D(\beta)$ in the quotient ring $\mathbb{Q}(Q)[\beta]/\Disc(\beta)$.

Once the rational function $R(\beta(a_i))$ is converted to a polynomial $P(\beta(a_i))$, the sum $\sum_i P(\beta(a_i))$, which is a symmetric polynomial in $\beta(a_i)$ and which has power at most the degree of $\Disc(\beta)$ minus one, is very easy to reduce to elementary symmetric functions in $\beta(a_i)$. The latter in turn can be replaced by coefficients of $\Disc(\beta)$.

Finally, let us remark that although the algebraic method introduced here is efficient and powerful, sometimes it may still take a lot of time to complete the computation in practice due to the complexity of the annulus kernel~$\widehat{B}_{r,s}(p_1,p_2)$ (see for instance the expression for the trefoil knot in Appendix~\ref{app:B23}). Then one can resort to the numerical computations for several specific numerical values of $Q$, and attempt to reconstruct the final result as a polynomial/rational function in $Q$.
%%%%%%%%%%%%%%%%%%%%%%%%%%%%%%

\subsubsection{The trefoil knot} \label{sec:Trefoil}
The augmentation curve of the trefoil knot in the canonical framing six is given in (\ref{eq:AugK23}). The calibrated annulus kernel $\widehat{B}_{r,s}(p_1,p_2)$ is a lengthy rational function in $\alpha_1,\beta_1,\alpha_2,\beta_2$ stated in its full glory in Appendix~\ref{app:B23}. We have checked its validity as the annulus instanton generating function in Section~\ref{sec:curves}.

The normalized planar three-point correlation differential $\widehat{\omega}^{(0)}_3(p_1,p_2,p_3)$ is computed by specializing the formula (\ref{equ:omega3}) to the annulus kernel and taking into account the proper normalization,
\begin{equation}
	\widehat{\omega}^{(0)}_3(p_1,p_2,p_3)\,=\,\frac{1}{N_{2,3}}\sum_{i} \mathop{\Res}_{q\rightarrow a_i} \frac{\widehat{B}_{r,s}(p_1,q) \widehat{B}_{r,s}(p_2,q) \widehat{B}_{r,s}(p_3,q) \alpha(q)\beta(q) }{d\alpha(q) d\beta(q)} \ ,
\end{equation}
where $N_{2,3}=3$. We give the first few terms of the expansion of $\omega^{(0)}_3(p_1,p_2,p_3)$ corresponding to the first three winding vectors $\vec{k}=\{(3),(2,1),(1,2)\}$. After rescaling $\alpha_\mu \mapsto Q^5 \alpha_\mu, \mu=1,2,3$,
\begin{equation}
\begin{aligned}
\widehat{\omega}&{}^{(0)}_3(p_1,p_2,p_3) \, = \, d\alpha_1 d\alpha_2 d\alpha_3 \Big[ -7200 + 24192 Q - 31536 Q^2 + 19980 Q^3 - 6264 Q^4   \\ 
&+ 864 Q^5 - 36 Q^6+ ( -302400 + 1378944 Q - 2624832 Q^2 + 2699424 Q^3 \\
& - 1620216 Q^4 +570960 Q^5 - 112320 Q^6 + 10800 Q^7 - 360 Q^8 )	(\alpha_1+\alpha_2+\alpha_3)	\\
& + (-12700800 + 73156608 Q-183145536 Q^2 + 261128448 Q^3 - 233372160 Q^4 \\
&  + 135520560 Q^5 - 
 51246720 Q^6+12283200 Q^7 - 1749600 Q^8 + 129600 Q^9 \\
& - 3600 Q^{10})(\alpha_1\alpha_2+\alpha_2\alpha_3+\alpha_3\alpha_1)  + \ldots \Big] \ .
\end{aligned}
\end{equation}
This result reproduces the correct instanton numbers as also computed from the Chern--Simons theory.

As a further check, we also expand one boundary component --- say corresponding to the point $p_3$ on $\mathcal{C}_{2,3}$ --- about the large volume point $\tilde p_3=\iota(p_3)$ associated to the image brane $\iota_*\mathcal{L}_{2,3}$. According to \eqref{eq:invol} this amounts to carrying out an expansion in terms of $\tilde{\alpha}_3=Q^5/\alpha_3$. The coefficients of the expansion of $\widehat{\omega}^{(0)}_3(p_1,p_2,\tilde p_3)$ now count the numbers of stretched planar three-holed instantons with two boundaries on the brane $\mathcal{L}_{2,3}$ and one boundary on the brane $\iota_*\mathcal{L}_{2,3}$. This expansion becomes
\begin{equation}
\begin{aligned}\label{equ:Stretched3pt}
	\widehat{\omega}&{}^{(0)}_3(p_1,p_2,\tilde{p}_3)\,=\, d\alpha_1 d\alpha_2 d\tilde{\alpha}_3 \Big[  432 Q - 1224 Q^2 + 1296 Q^3 - 612 Q^4 + 108 Q^5\\
	&+(18144 Q - 72576 Q^2 + 118944 Q^3 - 101952 Q^4 + 47880 Q^5 - 
 11520 Q^6 \\
 	&+ 1080 Q^7)(\alpha_1+\alpha_2)+(8064 Q - 33264 Q^2 + 56160 Q^3 - 49824 Q^4 + 24624 Q^5 \\
 	&- 6480 Q^6 +  720 Q^7) \tilde{\alpha}_3+\ldots \Big] \ .
\end{aligned}
\end{equation}
Here the constant terms, the coefficients of $\alpha_1,\alpha_2$ and the coefficients of $\tilde{\alpha}_3$ correspond to the windings $(1,1,\underline{1}), (1,2,\underline{1})$ and $(1,1,\underline{2})$, respectively, where the underlined entry refers to the distinguished boundary component mapped to the image brane $\iota_*\mathcal{L}_{2,3}$. Comparing with~\eqref{eq:Str3pta}, \eqref{eq:Str3ptb}, and \eqref{eq:Str3ptc} for $f=6$, we find agreement with the results from the quantum groups of composite representations discussed in Appendix \ref{app:composite}  (for $f=6$). We should stress that this is non-trivial check on the global structure of the correlation differential $\widehat{\omega}^{(0)}_3(p_1,p_2,p_3)$, and thus a non-trivial check on the prosed topological recursion.

The one-point function at genus one is computed by specializing the formula (\ref{equ:omega11}) to the calibrated annulus kernel with the proper normalization. After the usual rescaling $\alpha \mapsto Q^5\alpha$ its expansion becomes
\begin{equation}
\begin{aligned}
\widehat{\omega}&{}^{(1)}_1(p) \,=\,d\alpha\Big[ \frac{1}{24}\left(22 - 21 Q - Q^2\right) +\frac{1}{12} \left(
1722 - 3752 Q + 2625 Q^2 - 620 Q^3 + 25 Q^4\right)\alpha \\
&+\frac{1}{24}\left( 
213213 - 719433 Q + 940500 Q^2 - 595980 Q^3 + 185850 Q^4  \right.\\
 & \left. - 25074 Q^5+ 924 Q^6\right)\alpha^2 + \ldots \Big]   \ ,
\end{aligned}
\end{equation}
which is in agreement with the corresponding expected instanton numbers.

The next task is to compute the free energies $F^{(0)}$ and $F^{(1)}$ from the augmentation curve of the trefoil knot, which reproduces the expected result of the closed-string sector of the resolved conifold. For this purpose, we use formulas~(\ref{equ:d3F0NoB}) and (\ref{equ:d1F1Expanded}), respectively. Here we vary the complex structure of the curve with respect to the flat closed-string modulus $t$ of eq.~\eqref{eq:vol} in order to derive (the third derivative) of $F^{(0)}$. For arbitrary framing we find 
\begin{equation}\label{yukawa}
	\frac{\partial^3}{\partial t^3} F^{(0)} \,=\, (2\pi i)^3 \left(\frac{Q}{ 1-Q}+ \frac{f^3+f^2-5 f+3}{3 f (f-3) }\right) \ . 
\end{equation}
The first term in (\ref{yukawa}) is in accordance with the Yukawa coupling of the resolved conifold, and hence reproduces the prepotential of the resolved conifold correctly. The framing dependence $f$ that appears in the second term of (\ref{yukawa}) only contributes to the classical term of $F^{(0)}$. Due to the non-compactness of the resolved conifold, the classical piece depends on a regularization recipe. It would interesting to understand the resulting framing dependence from this point of view.

Similarly, we find for the the derivative of $F^{(1)}$ for an arbitrary framing $f$ 
\begin{equation}\label{f1t}
	\frac{\partial}{\partial t} F^{(1)} \,=\, 2\pi i\left(\frac{Q}{12(1-Q)} + \frac{f^3-4 f^2+28 f-69}{72 f (f-3)}\right) \ .
\end{equation}
As for the prepotential, the first term of (\ref{f1t}) coincides with the $F^{(1)}$ of the resolved conifold, while the second term in (\ref{f1t}) is again a classical contribution, which depends on the used regularization scheme. 

To calculate the free energies in various framings $f$, we remind the reader that the augmentation curve~(\ref{eq:AugK23}) and the annulus kernel~$\widehat{B}_{r,s}(p_1,p_2)$ constructed in Section~\ref{sec:curves} of the trefoil knot $\mathcal{K}_{2,3}$ are given in framing six, but can easily be transformed to arbitrary framings with the transformation
\begin{equation}\label{equ:FramingTransformation}
	(\alpha,\beta) \mapsto (\alpha \beta^{f-6}, \beta) \ .
\end{equation}
In particular, it is straight forward to check that the annulus kernel $\widehat{B}_{r,s}(p_1,p_2)$ correctly transforms under framing transformations and generates the annulus instanton numbers correctly in arbitrary framing $f$.

\subsubsection{The planar free energy $F^{(0)}$}\label{sec:F0}
As discussed at the end of Section~\ref{sec:KnotFreeEnergy}, although we do not have the expression for the annulus kernel $\widehat{B}_\mathcal{K}(p_1,p_2)$ for non-torus knot augmentation curves, we conjecture that the variational formula still holds and venture to compute the planar free energy from non-torus knot augmentation curves. Recall that this is possible because the planar free energy $F^{(0)}$ does not depend on the annulus kernel according to (\ref{equ:d3F0NoB}). We perform this computation for the figure eight knot, the knot $5_2$, the knot $6_1$, and the knot $6_2$, whose augmentation curves in framing zero can be found on Ng's website\cite{NgWeb}. For readers' convenience, we also attach the augmentation polynomials of these curves in the coordinates $\alpha$ and $\beta$ in Appendix~\ref{sec:AugPoly}.

\begin{figure}
\centering
\subfloat[Framing 0]{\includegraphics[width=0.23\textwidth]{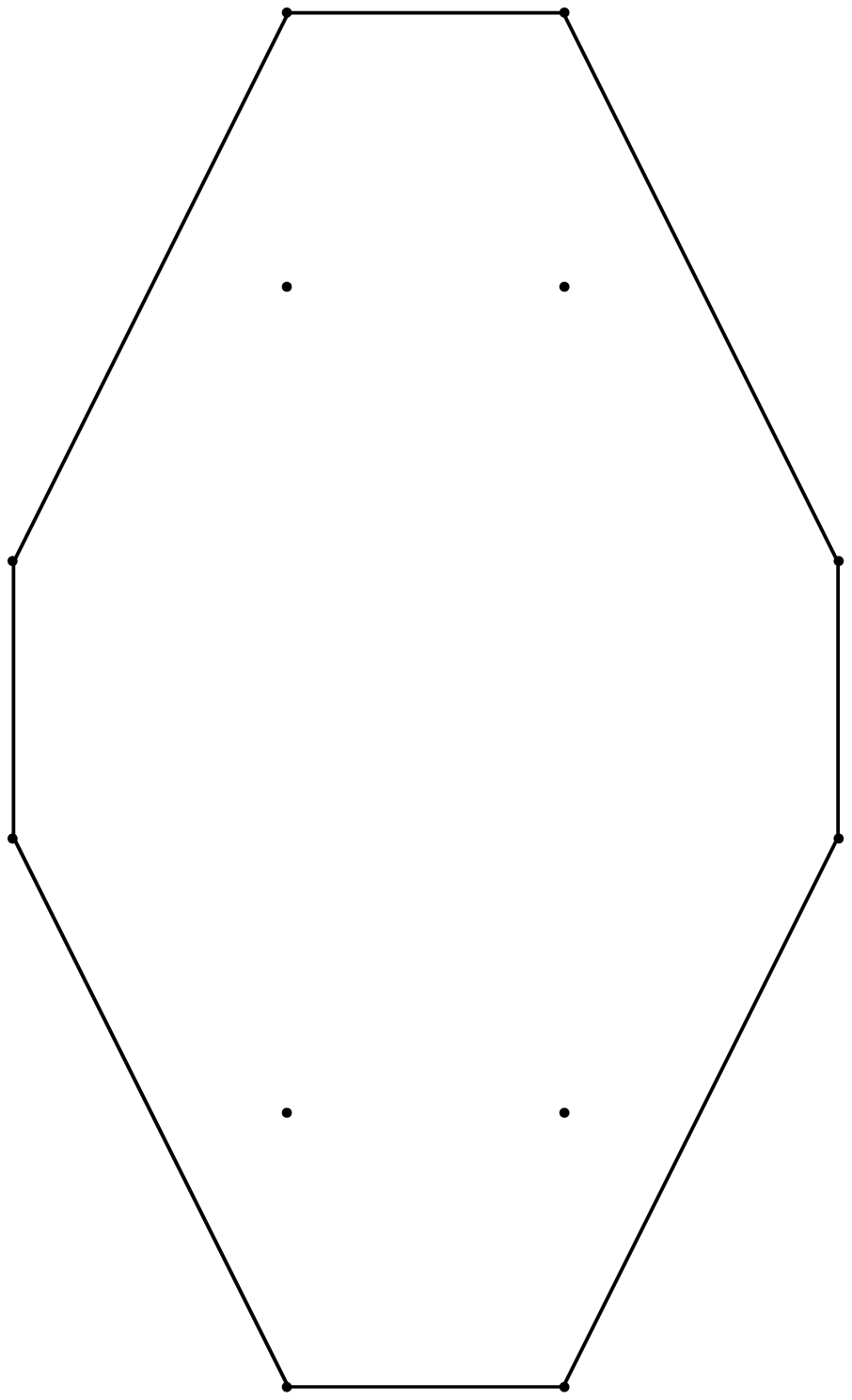}\label{fig:FigureEightNewtonPolytope1}}\hskip10ex
\subfloat[Framing 1]{\includegraphics[width=0.20\textwidth]{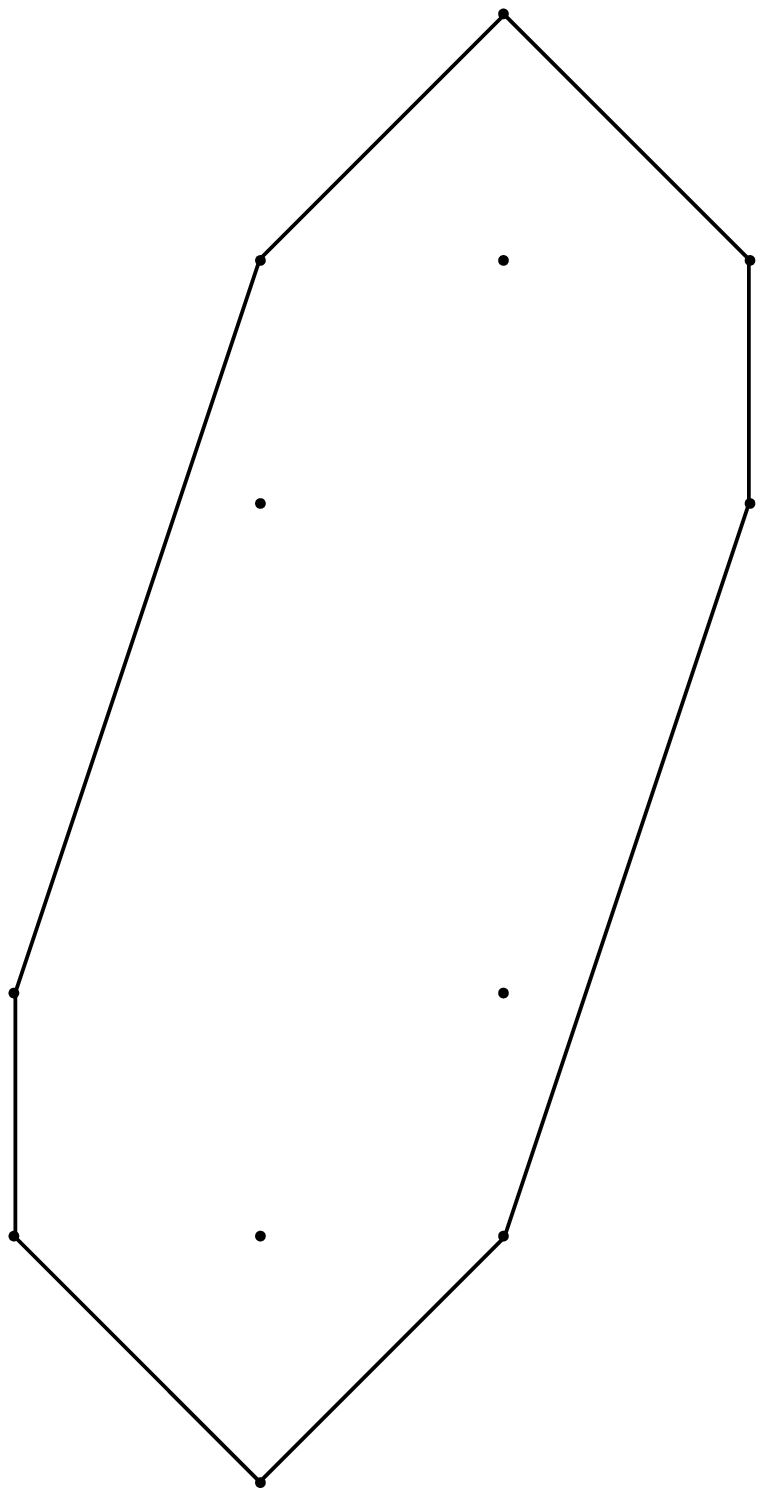}\label{fig:FigureEightNewtonPolytope2}}
\caption{Newton polytopes of the augmentation curve for the figure eight knot with different framings. Horizontal segments appear in framing 0 but not in framing 1.}\label{fig:FigureEightNewtonPolytope}
\end{figure}

Here we presume that the instanton generated part --- that is to say ignoring the classical piece --- is again framing independent, and therefore we only calculate in a particular framing in the following. In many cases, however, zero is not a good framing, as it may turn out that some of the segments of the boundary of the Newton polytope may be horizontal (vertical rays) (see Figure~\ref{fig:FigureEightNewtonPolytope} for the illustration in the case of figure eight knot), meaning that the number of ramification points with respect to $\alpha$ is reduced. See the discussion at the end of Appendix~\ref{sec:SymVariance} on this issue. As pointed out in ref.~\cite{Bouchard:2011ya}, this may cause the topological recursion to fail. So we need to choose a proper framing. In the case of the figure eight knot, we choose framing $f=1$ as a suitable framing, which amounts to carrying out the framing transformation $(\alpha, \beta) \mapsto (\alpha \beta,\beta)$ on the augmentation curve~\eqref{eq:Feight}. Using (\ref{equ:d3F0NoB}) the (derivative of) planer free energy is computed to be
\begin{equation}
	\frac{\partial^3}{\partial t^3} F^{(0)} =(2\pi i)^3 \frac{3}{N_{4_1}}	
	\left(\frac{Q}{1-Q} + \frac{4}{9} \right) \ .
\end{equation}
Again the quantum piece conforms to the planar free energy $F^{(0)}$ of the resolved confold! In fact we propose that the normalization factor $N_{4_1}$ should be chosen such that $\frac{\partial^3}{\partial t^3} F^{(0)}$ (ignoring the classical remnant) coincides with the result of resolved conifold. Then the normalization factor thus found can be used in the future topological recursion computation for the correlation differentials $\omega^{(g)}_n(p_1,\cdots,p_n)$ as well as higher genus free energies once the proper annulus kernel is at hand. Therefore for the figure eight knot, we propose
\begin{equation}
	N_{4_1} = 3\ .
\end{equation}

\begin{figure}[t]
\centering
\subfloat[Framing 0]{\includegraphics[width=0.20\textwidth]{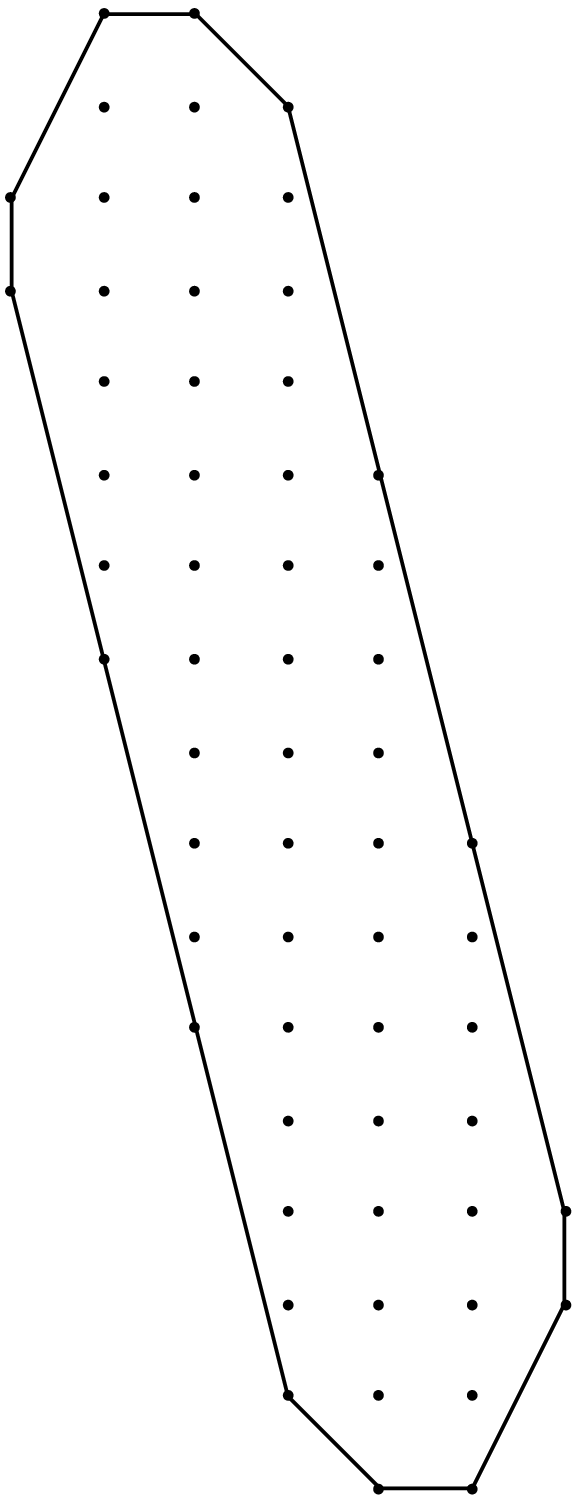}}\hskip6ex
\subfloat[Framing 1]{\includegraphics[width=0.23\textwidth]{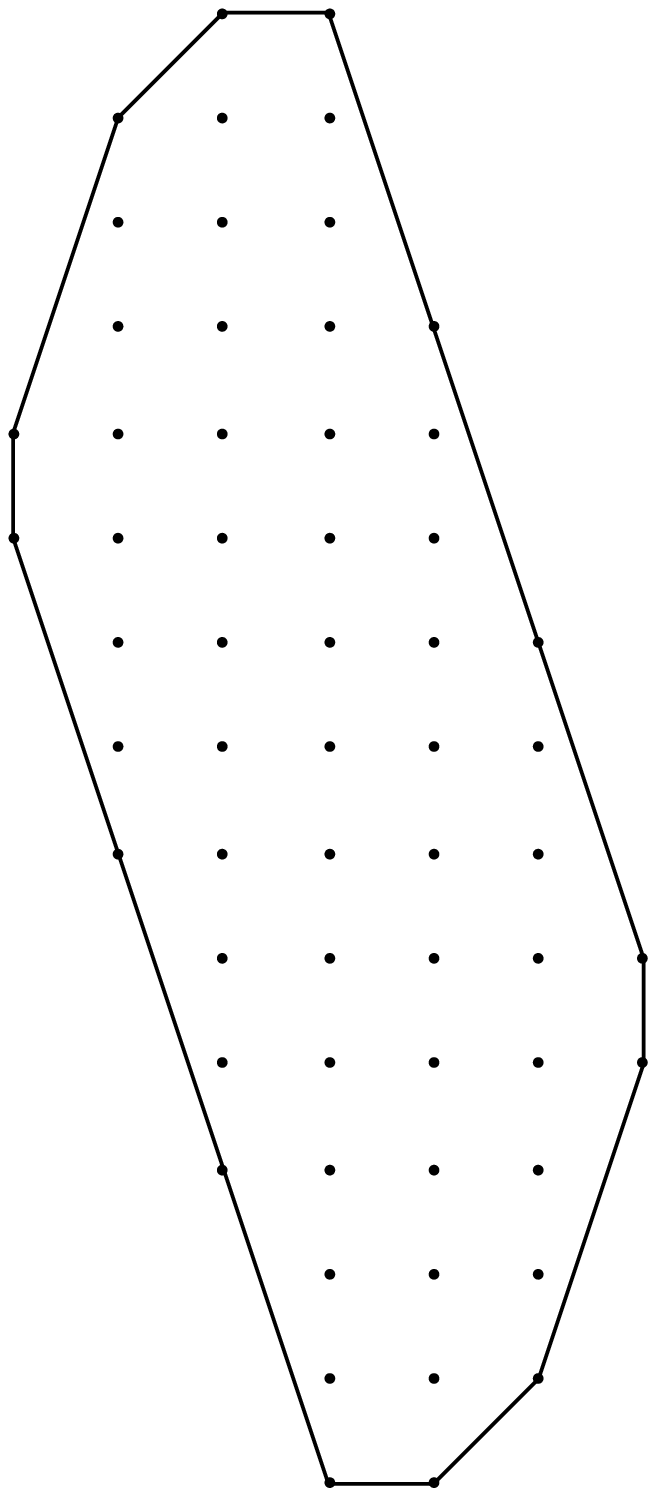}}\hskip6ex
\subfloat[Framing 2]{\includegraphics[width=0.245\textwidth]{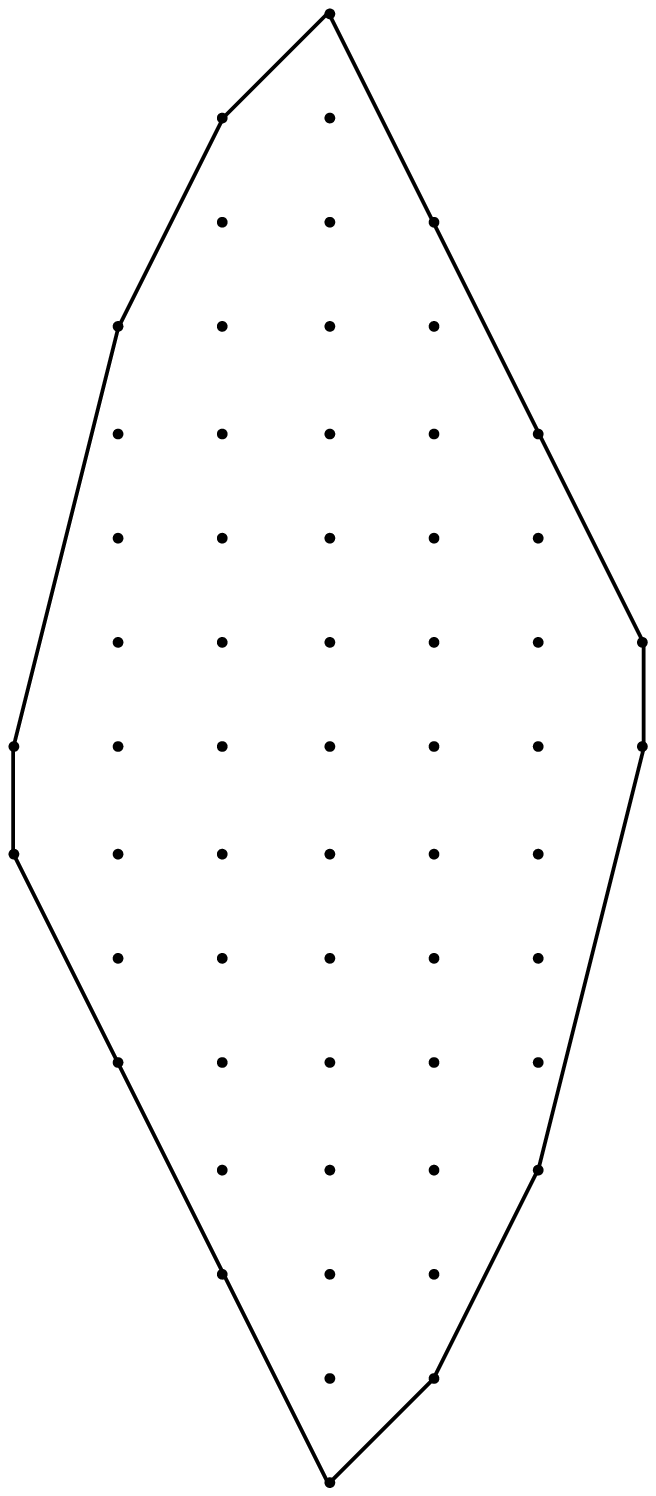}}
\caption{Newton polytopes of augmentation curves of the knot $6_2$ in the framing 0,1,2, respectively. Only the last Newton polytope has no horizontal boundary, hence no reduced number of ramification points.}\label{fig:K62NP}
\end{figure}

In the case of knot $5_2$, the computation is done in framing $f=1$. The planar free energy becomes
\begin{equation}
	\frac{\partial^3}{\partial t^3} F^{(0)} \,=\,(2\pi i)^3\frac{3}{N_{5_2}} \left(\frac{Q}{1-Q}  -\frac{1}{6} \right) \ ,
\end{equation}
which is again in agreement with the Yukawa coupling of the resolved conifold for the normalization
\begin{equation}
	N_{5_2} = 3\ .
\end{equation}
Furthermore, for the knot $6_1$ again in framing $f=1$ we get
\begin{equation}
	\frac{\partial^3}{\partial t^3} F^{(0)}\,=\,(2\pi i)^3 \frac{3}{N_{6_1}}  \left(\frac{Q}{1-Q}  +\frac{1}{9}\right)\ ,
\end{equation}
and postulate the normalization factor
\begin{equation}
	N_{6_1} \,=\, 3\ .
\end{equation}
Finally, for the knot $6_2$, we need to do the computation at least in framing $f=2$ so as to prevent the reduction of the number of ramification points (c.f., with the Newton polytopes of different framings in Figure~\ref{fig:K62NP}). We find 
\begin{equation}
	\frac{\partial^3}{\partial t^3} F^{(0)} =(2\pi i)^3 \frac{5}{N_{6_2}}\left(\frac{Q}{1-Q}  +\frac{9}{20}\right)\ ,
\end{equation}
and determine
\begin{equation}
	N_{6_2} = 5\ .
\end{equation}

It is gratifying to see that in all these examples of non-torus knots, the quantum part of the planar free energy $F^{(0)}$ of the resolved conifold is calculated correctly.

%%%%%%%%%%%%%%%%%%%%%%%%%%%%%%
\section{Topological recursion and knot theory} \label{sec:knotinv}
In this section we would like to point out a few consequences for knot theory that arise from the topological recursion point of view. While our observations are rather basic consequences from the topological recursion, they seem to imply strong statements in the context of knot theory.

In the following, our basic assumption is that the proposed topological recursion for the brane $\mathcal{L}_\mathcal{K}$ on the resolved conifold --- which as explained in Section~\ref{sec:toprec} further modifies the topological recursion of the remodelled B-model \cite{Bouchard:2007ys, Marino:2006hs} --- applies to any knot $\mathcal{K}$ (and not just to the class of torus knots $\mathcal{K}_{r,s}$ as shown in this work). Clearly, verifying the validity of the topological recursion for a non-torus knot (such as the figure eight knot) would be strong evidence in favor of this assumption. Unfortunately, we currently do not know how to construct the calibrated annulus kernel for non-torus knots to run such a check. However, at least  we can calculate from the augmentation curve of non-torus knots (without the knowledge of the calibrated annulus kernel) the planar free energy~$F^{(0)}$ of the resolved conifold as shown in Section~\ref{sec:F0}. This is a fairly strong check on the variational principle, which is an important part of the topological recursion program. More generally in topological string theories with a matrix model formulation, the matrix model Ward identities can be solved in the sense that all amplitudes can be recursively calculated from the disk and annulus amplitudes~\cite{Ambjorn:1992gw,Akemann:1996zr,Eynard:2007kz}. Hence, in such topological string theories the disk and annulus instantons should be the building blocks to construct --- via a suitable surgery operation --- all open and closed higher genus worldsheet instantons~\cite{Marino:2006hs,Eynard:2009qr}. But we cannot exclude the possibility that certain knots require yet another recursion scheme, for which our presented implications do not necessarily hold.

\subsection{Planar annulus amplitudes} \label{sec:planar}
First --- without making use of our assumption --- we would like to show that any planar annulus amplitude associated to the brane $\mathcal{L}_\mathcal{K}$ in the resolved conifold can be obtained from the HOMFLY polynomials $\mathcal{H}_\mu$ of the knot $\mathcal{K}$, colored with Young tableaus $\mu$ with at most two rows. 

The simplest example starts with considering HOMFLY invariants colored only with two boxes. With these invariants, we can extract the one-point function at winding two and the two-point function with total winding two
\begin{equation}\label{HMtwo}
\begin{aligned}
&{\cal H}_{\tableau{2}}-{\cal H}_{\tableau{1 1}}=\frac{1}{g_s}F_{(0,1)}^{(g=0)}+g_s\,F_{(0,1)}^{(g=1)}+\ldots\ ,\\
&{\cal H}_{\tableau{2}}+{\cal H}_{\tableau{1 1}}-{\cal H}_{\tableau{1}}^2=F_{(2)}^{(g=0)}+g_s^2\,F_{(2)}^{(g=1)}+\ldots\ ,
\end{aligned}
\end{equation}
where $F_{{\vec k}}^{(g)}$ is the free energy associated with winding vector $\vec{k}$ at genus $g$. Since the $g_s$ expansion of the free energy is graded by the Euler characteristic of amplitudes, we can eliminate some of the colored HOMFLY invariants at lowest order. For instance, if we are only interested in disk amplitude of winding two, i.e. $F_{(0,1)}^{(g=0)}$, then the right hand side of the second equation in (\ref{HMtwo}) is irrelevant and for this order ${\cal O}(g_s^{-1})$ we can set it to zero. This implies
\begin{equation}\label{HMtwoantisymm}
\Big[{\cal H}_{\tableau{1 1}}\Big]_{{\cal O}(g_s^{-1})}\,=\,\Big[-{\cal H}_{\tableau{2}}+{\cal H}_{\tableau{1}}^2\Big]_{{\cal O}(g_s^{-1})}\ .
\end{equation}
Here and in the following the symbol `$[\,\cdot\,]_{\mathcal{O}(g_s^\ell)}$' indicates a truncation in $g_s$ up to the order $g_s^{\ell}$. Using \eqref{HMtwoantisymm}, we can now eliminate ${\cal H}_{\tableau{1 1}}$ from the disk and we would therefore find
\begin{equation}\label{HMdisc}
F_{(0,1)}^{(g=0)}=g_s\Big[2{\cal H}_{\tableau{2}}-{\cal H}_{\tableau{1}}^2\Big]_{{\cal O}(g_s^{-1})}\ .
\end{equation}
This shows that disk amplitude at winding two can be obtained only by the knowledge of HOMFLY invariants in symmetric representations. Now, let us proceed with the next level of winding. With winding three, we can have one-point function at winding three, two-point function with total winding three, and the lowest winding three-point function. We have 
\begin{equation}\label{HMthree}
\begin{aligned}
&{\cal H}_{\tableau{3}}-{\cal H}_{\tableau{2 1}}+{\cal H}_{\tableau{1 1 1}}=\frac{1}{g_s}F_{(0,0,1)}^{(g=0)}+g_s\, F_{(0,0,1)}^{(g=1)}+\ldots\ ,\\
&{\cal H}_{\tableau{3}}-{\cal H}_{\tableau{1 1 1}}-{\cal H}_{\tableau{1}}({\cal H}_{\tableau{2}}-{\cal H}_{\tableau{1 1}})=F_{(1,1)}^{(g=0)}+g_s^2\,F_{(1,1)}^{(g=1)}+\ldots\ ,\\
&{\cal H}_{\tableau{3}}+2{\cal H}_{\tableau{2 1}}+{\cal H}_{\tableau{1 1 1}}-3{\cal H}_{\tableau{1}}({\cal H}_{\tableau{2}}+{\cal H}_{\tableau{1 1}})+2{\cal H}_{\tableau{1}}^3=g_s F_{(3)}^{(g=0)}+g_s^3\,F_{(3)}^{(g=1)}+\ldots\ .
\end{aligned}
\end{equation}
Similar to the previous case, if we are just interested in disk amplitude the right hand sides of the second and third equations of (\ref{HMthree}) are irrelevant and can be set to zero. In doing so, we can eliminate two of the colored HOMFLY invariants in favor of the third one. Let us solve ${\cal H}_{\tableau{2 1}}$ and ${\cal H}_{\tableau{1 1 1}}$ in terms of ${\cal H}_{\tableau{3}}$ and the lower colored invariants at order ${\cal O}(g_s^{-1})$. Solving the equations together with (\ref{HMtwoantisymm}), we arrive at
\begin{equation}\label{HMsolve}
\begin{aligned}
&\Big[{\cal H}_{\tableau{2 1}}\Big]_{{\cal O}(g_s^{-1})}\,=\,
  \Big[-{\cal H}_{\tableau{3}}+{\cal H}_{\tableau{1}}{\cal H}_{\tableau{2}}\Big]_{{\cal O}(g_s^{-1})}\ ,\\
&\Big[{\cal H}_{\tableau{1 1 1}}\Big]_{{\cal O}(g_s^{-1})}\,=\,\Big[{\cal H}_{\tableau{3}}-{\cal H}_{\tableau{1}}(2{\cal H}_{\tableau{2}}-{\cal H}_{\tableau{1}}^2)\Big]_{{\cal O}(g_s^{-1})}\ .
\end{aligned}
\end{equation}
Using (\ref{HMsolve}), we can state the disk amplitude at winding three in terms of only symmetric representations in the following way
\begin{equation}\label{discthree}
F_{(0,0,1)}^{(g=0)}=g_s\Big[3{\cal H}_{\tableau{3}}-3{\cal H}_{\tableau{1}}{\cal H}_{\tableau{2}}+{\cal H}_{\tableau{1}}^{3}\Big]_{{\cal O}(g_s^{-1})}\ .
\end{equation}
The annulus amplitude in (\ref{HMthree}) however starts with ${{\cal O}(g_s^{0})}$. Therefore, at this order we can only set the right hand side of the last equation of (\ref{HMthree}) to zero. This allows us to eliminate only one colored HOMFLY invariant in favor of others at this stage. We should note that at this order, (\ref{HMtwoantisymm}) will not be valid anymore. Setting the last equation of (\ref{HMthree}) to zero, we find
\begin{equation}\label{HMMM}
\Big[{\cal H}_{\tableau{1 1 1}}\Big]_{{\cal O}(g_s^{0})}\,=\,\Big[-{\cal H}_{\tableau{3}}-2{\cal H}_{\tableau{2 1}}+3{\cal H}_{\tableau{1}}({\cal H}_{\tableau{2}}+{\cal H}_{\tableau{1 1}})-2{\cal H}_{\tableau{1}}^3\Big]_{{\cal O}(g_s^{0})}\ .
\end{equation}
Now plugging (\ref{HMMM}) into (\ref{HMthree}), we can express the annulus amplitude in terms of representations with at most two rows
\begin{equation}\label{AnnHM}
F_{(1,1)}^{(g=0)}=\Big[2{\cal H}_{\tableau{3}}+2{\cal H}_{\tableau{2 1}}-4{\cal H}_{\tableau{1}}{\cal H}_{\tableau{2}}-2{\cal H}_{\tableau{1}}{\cal H}_{\tableau{1 1}}+2{\cal H}_{\tableau{1}}^3\Big]_{{\cal O}(g_s^{0})}\ .
\end{equation}

The described pattern is the base case for the proof of a recursive algorithm that allows us to express any planar annulus amplitude with arbitrary windings just in terms of HOMFLY polynomials colored with Young tableaux with at most two rows. To verify the inductive step of the proposed algorithm, we assume that the assertion is true for all planar annulus amplitudes $F_{\vec k,|\vec k|=2}^{(g=0)}$ for winding vectors $\vec k$ up to the total winding number $\sum_j j k_j <N$. This implies that up to the order $\mathcal{O}(g_s^0)$ all representations with less than $N$ boxes and with more than two rows can already be expressed in terms of HOMFLY polynomials with at most two rows, i.e.,
\begin{equation} \label{eq:indhypo}
  \Big[  \mathcal{H}_{\mu} \Big]_{{\cal O}(g_s^{0})}\,=\, \Big[f_\mu(\left\{ \mathcal{H}_\nu \middle| c_\nu \le 2, n_\nu \le n_\mu \right\})  \Big]_{{\cal O}(g_s^{0})} \quad \text{for} \quad c_\mu>2, n_\mu <N\ .
\end{equation}  
Here $f_\mu$ is a polynomial in the specified set of HOMFLYs and $n_\mu, n_\nu$ and $c_\mu, c_\nu$ denote the numbers of boxes and the numbers of rows of the representations $\mu$ and $\nu$, respectively. 

The (sum of) free energies $F_{\vec k}^{(g)}$ of total winding $N$ can in general be written as
\begin{equation}
    \sum_g g_s^{2g+|\vec k| -2} F_{\vec k}^{(g)} \,=\, \sum_{\mu, n_\mu = N} c_{\vec k,\mu} \mathcal{H}_\mu + h_{\vec k} ( \left\{ \mathcal{H}_\nu \middle|  n_\nu < N \right\}) \ \quad \text{for} \quad \sum_j j k_j = N \ ,
\end{equation} 
in terms of the polynomials $h_{\vec k}$. Thus, they depend linearly on the HOMFLYs with the maximum number of $N$ boxes (but depend non-linearly on those with less then $N$ boxes). Truncating to the order of interest, namely $\mathcal{O}(g_s^0)$ yields then the following relations
\begin{align}
   F_{\vec k}^{(g=0)} \,&=\, g_s^{|\vec k|-2}\big[ \sum_{\mu, n_\mu = N} c_{\vec k,\mu} \mathcal{H}_\mu + \tilde h_{\vec k} ( \left\{ \mathcal{H}_\nu \middle|  c_\nu\le 2, n_\nu < N \right\}) \Big]_{{\cal O}(g_s^{0})} 
   \ \text{for} \ |\vec k|\le 2 \ , \label{eq:Ftwoholes} \\
   0 \,&=\,\big[ \sum_{\mu, n_\mu = N} c_{\vec k,\mu} \mathcal{H}_\mu + \tilde h_{\vec k} ( \left\{ \mathcal{H}_\nu \middle|  c_\nu\le 2, n_\nu < N \right\}) \Big]_{{\cal O}(g_s^{0})}
    \quad \text{for} \ |\vec k|>2 \ , \label{eq:Fbigger}
\end{align}
where the polynomials $\tilde h_{\vec k}$ arise from inserting the induction hypothesis~\eqref{eq:indhypo} into the polynomials $h_{\vec k}$, and we use that $F_{\vec k}^{(g)}$ starts off at order $g_s^{2g+|\vec{k}|-2}$. 

Since we can assign to each winding vector $\vec k$ (of total winding $N=\sum_j  j k_j$) a Young tableau with $|\vec k|$ rows and $N$ boxes, we get from the identities~\eqref{eq:Fbigger} as many relations as there are Young tableaux with more than two rows. Solving for the Young tableaux with more than two rows (which appear linearly in \eqref{eq:Fbigger}), we can carry out the induction step, by noting that \eqref{eq:indhypo} gets extended by the HOMFLY polynomials with $n_\mu=N$. Thus we now have
$$
  \Big[  \mathcal{H}_{\mu}\Big]_{{\cal O}(g_s^{0})} \,=\, \Big[f_\mu(\left\{ \mathcal{H}_\nu \middle| c_\nu \le 2, n_\nu \le n_\mu \right\})  \Big]_{{\cal O}(g_s^{0})} \quad \text{for} \quad c_\mu>2, n_\mu \le N\ .
$$  
Inserting the last relations into \eqref{eq:Ftwoholes} (for $|\vec k|=2$) shows the assertion up to the order $\sum_j j k_j = N$, and it completes the verification of the described recursion algorithm by induction.

To summarize we have confirmed that the planar annulus amplitude can be calculated from the set of HOMFLY polynomials colored with representations with at most two rows. Note that as illustrated at the beginning of this subsection the fact that the planar disk amplitude~$F_{0,{\vec k},|\vec k|=1}(Q)$ just arises from symmetric representations (i.e., Young tableaux with just one row), as for instance used in ref.~\cite{Aganagic:2012jb,Jockers:2012pz}, can be argued for analogously. Furthermore, similar identities can be derived for the planar free energies with more than two holes.

\subsection{Knot invariants, mutants and the topological recursion}
Let us now contemplate, under the assumption that the proposed topological recursion is valid for any knot $\mathcal{K}$, what our results imply for knot theory. 

First, we observe that the definition of a topological recursion of a knot $\mathcal{K}$ --- denoted by $\operatorname{TopRec}(\mathcal{K})$ --- in principal assigns to a knot $\mathcal{K}$ the set of all correlation differentials $\hat\omega^{(g)}_{\vec k}$. This is equivalent to the knowledge of the set of all HOMFLY polynomials $\left\{ \mathcal{H}_\mu^\mathcal{K}(Q,q) \right\}$ colored with any representation of $SU(n)$ (for arbitrarily high $n$), i.e., 
\begin{equation} \label{eq:conj1}
  \operatorname{TopRec}(\mathcal{K}) \, \simeq \, \left\{ \mathcal{H}_\mu^\mathcal{K}(Q,q) \right\} \ .
\end{equation}  
If we now assume that our proposal of the topological recursion is universal, that is to say that for any knot $\mathcal{K}$ the topological recursion $\operatorname{TopRec}(\mathcal{K})$ is given by the modified topological recursion of the remodelled B-model based on the augmentation polynomial~$F_\mathcal{K}(\alpha,\beta;Q)$ and a calibrated annulus kernel $\widehat{B}_\mathcal{K}(p_1,p_2;Q)$, then we arrive at the equivalence
\begin{equation} \label{eq:conj2}
   \left\{ F_\mathcal{K}(\alpha,\beta;Q), \widehat{B}_\mathcal{K}(p_1,p_2;Q) \right\} \,\simeq\,  \left\{ \mathcal{H}_\mu^\mathcal{K}(Q,q) \right\} \ .
\end{equation}
Thus the knowledge of the augmentation polynomial together with the calibrated annulus kernel is as good as the knowledge of all colored HOMFLY polynomials! Furthermore, using the result of the last subsection that one can calculate the disk generating function $F_{0,{\vec k},|\vec k|=1}(Q)$ and the annulus generating function $F_{0,{\vec k},|\vec k|=2}(Q)$ just from the HOMFLY polynomials colored with representations of at most two rows, we obtain the further equivalence
\begin{equation} \label{eq:conj3}
  \left\{ \mathcal{H}_\mu^\mathcal{K}(Q,q) \middle| c_\mu \le 2 \right\} \,\simeq\,  \left\{ \mathcal{H}_\mu^\mathcal{K}(Q,q) \right\} \ .
\end{equation}
Here we have used that the disk generating function $F_{0,{\vec k},|\vec k|=1}(Q)$ calculates the augmentation polynomial \cite{Aganagic:2012jb,Jockers:2012pz,Aganagic:2013jpa} and that the calibrated annulus kernel $\widehat{B}_\mathcal{K}(p_1,p_2;Q)$ is essentially a closed expression for the annulus generating function $F_{0,{\vec k},|\vec k|=2}(Q)$. We are not aware if such or similar statements have appeared or have even been proven in the literature before.

The above listed conjectures have some remarkable consequences. First of all, we note that if two distinct knots $\mathcal{K}_1$ and $\mathcal{K}_2$ can be distinguished by any colored HOMFLY polynomial at all, then the conjecture~\eqref{eq:conj3} implies that those two knots must already be distinguishable by a colored HOMFLY polynomial $\mathcal{H}_\mu(Q,q)$ for some Young tableau $\mu$ with at most two rows $c_\mu\le 2$. 

Let us examine what the above assertion implies for mutant knots.\footnote{We would like to thank Marcos Mari\~no for drawing our attention to mutant knots in this context.} Two mutant knots are related by a flip or a $180^\circ$ rotation of a tangle component, which is called a mutation operation in knot theory.\footnote{For an elementary review, see for instance ref.~\cite{MR2079925}.} Mutants are generally interesting because they cannot be distinguished by many knot invariants including the Alexander polynomial, the Jones polynomial, the Kauffman polynomial and the HOMFLY polynomial (all taken in the fundamental representation). In fact, a theorem by Morton and Cromwell \cite{MR1395780} implies that HOMFLY polynomials colored with symmetric representations are not sufficient to distinguish a pair of mutants.\footnote{Theorem 5 in ref.\cite{MR1395780} proves that a HOMFLY polynomial colored by a representation $R$ cannot distinguish mutant pairs if the decomposition of $R\otimes R$ has no repeated summands. This is true for any symmetric representation.} Since we can construct the augmentation polynomial from all HOMFLY polynomials in symmetric representations, we claim that a pair of mutant knots must have the same augmentation variety. If, however, the mutant pair can be distinguished by a colored HOMFLY polynomial at all, our conjectures further assert that such mutants --- while having the same augmentation polynomial --- must have distinct calibrated annulus kernels, which in turn is equivalent to the statement that such mutants are distinguishable by a colored HOMFLY polynomial with two rows.

The simplest and most studied mutant pair is the Conway's knot and the Kinoshita--Terasaka knot. Both knots are non-torus knots and have 11 crossings. The Conway and Kinoshata--Terasake mutant pair is distinguishable by their HOMFLY polynomials colored with $\mu=\tableau{2 1}$ as shown in ref.~\cite{MR1395780}. Note this is a representation of two rows! Thus, even though we expect that they have the same augmentation variety, we predict that their calibrated annulus kernel are distinct. This also shows that for a general knot the calibrated annulus kernel cannot be constructed from the knowledge of the augmentation variety alone. 

%%%%%%%%%%%%%%%%%%%%%%%%%%%%%%
\section{Conclusions and outlook} \label{sec:conc}
%%%%%%%%%%%%%%%%%%%%%%%%%%%%%%
Knot invariants are encoded in Wilson loop expectation values of Chern--Simons theory on $S^3$ \cite{Witten:1988hf}. A remarkable chain of dualities relates them to open topological B-model amplitudes on $T^* S^3$ with additional branes ending on the knots, going through a holographic duality to an A-model formulation on the resolved conifold \cite{Witten:1992fb,Ooguri:1999bv}, which in turn is mirror dual to a B-model. The latter enjoys a matrix model description with a topological recursion for the amplitudes \cite{Dijkgraaf:2002fc,Marino:2006hs,Bouchard:2007ys}, which yields a powerful program for extracting knot invariants. Moreover, the BPS expansion of the open topological string reveals strong integrality properties of the colored HOMFLY polynomials \cite{Ooguri:1999bv} and suggests refinements of the latter.        

The topological recursion is based on a spectral curve together with a meromorphic one form \cite{Eynard:2007kz,Marino:2006hs,Bouchard:2007ys}. This set of data directly fixes the topological disk amplitudes. In the context of conifold geometry, it is argued in refs.~\cite{Aganagic:2012jb,Jockers:2012pz,Aganagic:2013jpa} that this information is in one-to-one correspondence to the augmentation variety of the differential graded algebra of the knot contact homology \cite{MR2116316,MR2175153,MR2376818,MR2807087,MR3070519}. As a further input, the topological recursion requires the kernel bi-differential that produces the annulus numbers. An important second claim is that these initial data unambiguously fix all open and closed amplitudes by the recursion.

One conclusion of this paper is that the kernel bi-differential is not canonically associated to the spectral curve with the meromorphic one form. This is different then in the remodelling approach to the open topological A-model on local toric Calabi--Yau spaces with Harvey--Lawson branes \cite{Marino:2006hs,Bouchard:2007ys}, where the closed string spectral curve happens to be identical to moduli space of the branes and the kernel bi-differential is canonically given by the Bergman kernel of the spectral curve. In the context of knot theory a canonical association is not to be expected because --- as we discussed in Section~\ref{sec:knotinv} --- a pair of mutant knots cannot be distinguished by the first set of data, but may differ by their annulus amplitudes (as it is the case for the Conway and Kinoshita--Terasaka mutant pair). This also implies that in general an unambiguous quantization prescription cannot be given from the spectral curve and the meromorphic one form alone, but instead requires the knowledge of some additional data.     

The main claim of this paper is that this quantization ambiguity is entirely fixed by the right choice of the annulus kernel. Indeed for the torus knots we complement the augmentation variety and the meromorphic one form by a kernel bi-differential that we call calibrated annulus kernel and give convincing evidence that the recursion produces correctly all open- and closed-string amplitudes for torus knots. In particular, we check the knot invariants arising from the three-point function at genus zero, as well as the one-point function at genus one. In addition, we calculate from the recursion the closed-string free energies at genus zero and genus one. The latter calculations are a further check that the analytic structure of the annulus kernel, which is summarized in Section~\ref{sec:PoleStructure}, is correct. We further argue that the discussed calibration on the annulus kernel is necessary to produce the correct framing dependence for torus knots invariants and show that it is required to render the proposed topological recursion consistent.

The calculation of streched annulus invariants --- performed in Section~\ref{sec:stretched} --- are checked by an explicit localization calculation (in Appendix~\ref{app:Localization}) that extends the approach of ref.~\cite{Diaconescu:2011xr} to open strings with an orientifold involution. These stretched annulus amplitudes calculate Chern--Simons Wilson loop observables in the composite representations of $U(N)$, which are evaluated in Appendix~\ref{app:composite} using the Rosso--Jones formula and are compared with the A- and B-model results. Comparison with the latter confirms again the necessity of the calibration of the annulus kernel in a simple setting. Note that the stretched annulus invariants probe the analytic structure of the calibrated annulus kernel in a non-trivial way as they are evaluated in a different phase of its parameter space after the involution \eqref{eq:invol}. We present consistency checks confirming that the recursion applies for the stretched geometries.   
 
The evidence that a spectral curve related in the standard way to the disk invariants and the calibrated annulus kernel yield via the topological recursion the Wilson loop expectation values of Chern--Simons theory on $S^3$ with more complicated knots raises a natural question. Does a matrix model description for the more general branes associated to these knots exist on the local conifold? The work \cite{Garoufalidis:2006ew} hints that Chern--Simons invariants on $S^3$ might have fairly generally a matrix model description. 

Different then for the much simpler Harvey--Lawson branes, the equivalence  of the open/closed topological string amplitudes and of the amplitudes obtained by recursion is not proven to  all orders in $g_s$ for our new proposed kernel --- not even for the torus knots. Besides such a proof the main practical challenge is to find a  conceptual way to provide for the calibrated physical annulus kernel based on more minimal information regarding the knot. As there seems to be a link between the braid word of torus knots and the construction of augmentation varieties via fractional unknots, one could speculate that the braid word of a knot could give rise to the calibrated annulus kernel by fixing its non-trivial pole structure and the residues of the kernel. This pole structure could be related to physical short distance behavior, as argued at the end of Section~\ref{sec:kernel}. It would be interesting to translate the choice of the annulus kernel and these observations about the pole structure more conceptually into a quantizing prescription of the augmentation variety. 
 
Having the annulus kernel for non-torus knots at hand, one can then compute colored HOMFLY invariants associated with these knots by our proposed recursion algorithm. This would then provide a powerful tool for computing new colored invariants, which are typically difficult to compute in the context of knot theory.

As pointed out in (\ref{eq:invol}) for the trefoil knot and in ref.~\cite{Ng:2012qq} for any given knot, the associated augmentation polynomial enjoys an involution symmetry. Employing this symmetry, as in the case of unknot, it would be interesting to develop the required techniques in the framework of B-model to compute the Kauffman invariants of the unoriented knots. A crucial step will be the construction of the correct annulus bi-differential for this case.

In this paper, we have only considered oriented knots on $S^3$. A natural question is whether our formalism extends to calculate knot invariants in other three-manifolds, in particular for those geometries, which can be embedded into geometries of known topological string constructions. These involve in general a four cycle leading to non-trivial automorphic functions governing the closed string moduli. It would be interesting to understand the relation between the closed- and open-string moduli spaces in such settings. A simple class of such examples are furnished by knots in Lens spaces. As the first step in this direction, one may start to construct the correct physical annulus kernel for torus knots in Lens spaces. In the spirit of ref.~\cite{Brini:2011wi}, torus knots in Lens spaces have recently been studied in the B-model in ref.~\cite{Stevan:2013tha}. 

We hope to address some of these questions in the future.

%%%%%%%%%%%%%%%%%%
\bigskip
\subsection*{Acknowledgments}
%%%%%%%%%%%%%%%%%%
%%
We would like to thank
Mina Aganagic,
Janko B\"ohm,
Vincent Bouchard,
Sergei Gukov,
Marcos Mari\~no,
Lenhard Ng
and
Cumrun Vafa
for useful discussions and correspondence.
J.G. is supported by the BCGS program,
H.J. and A.K. are supported by the DFG grant KL 2271/1-1,
and M.S. is supported by NSF FRG grant DMS 1159049 and NSF PHY 1053842.

%%%%%%%%%%%%%%%%%%%
%\bigskip
\newpage
\appendix
%%%%%%%%%%%%%%%%%%%

%%%%%%%%%%%%%%%%%%%%%%%%%%%%%%%%%%%%%
\section{Physical annulus kernel of the trefoil knot} \label{app:B23}
%%%%%%%%%%%%%%%%%%%%%%%%%%%%%%%%%%%%%
Here we record the physical annulus kernel of the trefoil torus knot, which is given by
\begin{equation}
  B_{2,3}(\alpha_1,\beta_1,\alpha_2,\beta_2;Q)d\alpha_1d\alpha_2 = 
  \frac{U_{2,3}(\alpha_1,\beta_1,\alpha_2,\beta_2;Q)+U_{2,3}(\alpha_2,\beta_2,\alpha_1,\beta_1;Q)}
    {V_{2,3}(\alpha_1,\beta_1,\alpha_2,\beta_2;Q)+V_{2,3}(\alpha_2,\beta_2,\alpha_1,\beta_1;Q)}d\alpha_1d\alpha_2 \ ,
\end{equation}
while the calibrated annulus kernel of the trefoil knot reads
\begin{equation}
  \widehat{B}_{2,3}(\alpha_1,\beta_1,\alpha_2,\beta_2;Q)d\alpha_1d\alpha_2 =   \left[B_{2,3}(\alpha_1,\beta_1,\alpha_2,\beta_2;Q)-\frac{1}{(\alpha_1-    \alpha_2)^2}\right] d\alpha_1d\alpha_2 \ ,
\end{equation}
in terms of the rather lengthy polynomials
{\scriptsize{
$$
\begin{aligned}
U&{}_{2,3}(\alpha_1,\beta_1,\alpha_2,\beta_2;Q)=(1-Q) \beta_2^3  \\
\times&\Big[Q^2 \alpha _2 \left(24 Q \alpha _1 \beta _2^8+(11-24 Q) \alpha _1 \beta _2^7-6 Q^4 \beta _2^6-2 (Q-4) Q^3 \beta _2^5+4 Q^2 (6 Q-5) \beta _2^4
   +2 Q \left(4 Q^2-9 Q+5\right) \beta _2^3\right.\\
&\left.+4 (1-3 Q) Q \beta _2^2-6 (Q-1) Q \beta_2+4 Q\right) \beta _1^{10}+\alpha _2 \left(-6 Q^4 \alpha _1 \beta _2^{10}+6 (Q-1) Q^3 \alpha _1 \beta _2^9+2 Q^2 (4 Q-19) \alpha _1 \beta _2^8\right.\\
&\left.+Q \left(-2 Q^2+61 Q-65\right) \alpha _1 \beta _2^7-2 \left(16 Q^2-41Q+25\right) \alpha _1 \beta _2^6+9 Q^4 \beta _2^5+3 Q^2 \left(7 Q^2-18 Q+5\right) \beta _2^3\right.\\
&\left.+6 (Q-1)^2 Q^2 \beta _2^2+9 Q^2 \beta _2-6 (Q-1) Q^2\right) \beta _1^9+\alpha _2 \left(-2 (Q-4) Q^3 \alpha _1 \beta _2^{10}+Q^2
   \left(2 Q^2-10 Q+17\right) \alpha _1 \beta _2^9\right.\\
&\left.-75 (Q-1) Q \alpha _1 \beta _2^8+\left(24 Q^5+\left(2 Q^3+57 Q^2-240 Q+275\right) \alpha _1\right) \beta _2^7+\left(2 (7 Q-19) Q^4+\left(-3 Q^2+95 Q-140\right) \alpha_1\right) \beta _2^6\right.\\
&\left.+\left(2 \left(Q^2-41 Q+40\right) Q^3+\left(2 Q^2+29 Q-74\right) \alpha _1\right) \beta _2^5-4 Q^2 \left(12 Q^2-15 Q+5\right) \beta _2^4+2 Q \left(-4 Q^3+37 Q^2-38 Q+5\right) \beta _2^3\right.\\
&\left.+4 (1-3 Q)^2 Q
   \beta _2^2+6 (Q-1)^2 Q \beta _2+4 (1-3 Q) Q\right) \beta _1^8+\left(\alpha _2 \left(4 Q^2 (6 Q-5) \alpha _1 \beta _2^{10}-4 Q \left(6 Q^2-11 Q+5\right) \alpha _1 \beta _2^9\right.\right.\\
&\left.\left.-8 Q (3 Q-2) \alpha _1 \beta _2^8-26 Q^4 \beta
   _2^7+\left((59 Q-71) Q^3+(46 Q-48) \alpha _1\right) \beta _2^6+\left(Q^2 \left(11 Q^2-97 Q+132\right)-6 (3 Q-4) \alpha _1\right) \beta _2^5\right.\right.\\
&\left.\left.-8 Q \left(14 Q^2-29 Q+15\right) \beta _2^4+\left(-80 Q^3+197 Q^2-132 Q+25\right)
   \beta _2^3+\left(-8 Q^3+74 Q^2-76 Q+10\right) \beta _2^2\right.\right.\\
&\left.\left.+3 \left(7 Q^2-18 Q+5\right) \beta _2+2 \left(4 Q^2-9 Q+5\right)\right)-13 Q^6 \beta _2^4\right) \beta _1^7-\alpha _2 \left(-2 Q \left(4 Q^2-9 Q+5\right) \alpha _1
   \beta _2^{10}\right.\\
&\left.+\left(8 Q^3-47 Q^2+82 Q-25\right) \alpha _1 \beta _2^9+3 \left(2 Q^4+\left(7 Q^2-43 Q+40\right) \alpha _1\right) \beta _2^8+\left(2 (19 Q-22) Q^3+\left(8 Q^2+67 Q-97\right) \alpha _1\right) \beta _2^7\right.\\
&\left.+4
   (Q-1) \left(3 (Q-4) Q^2-4 \alpha _1\right) \beta _2^6-2 \left(2 Q \left(27 Q^2-52 Q+25\right)+(20-3 Q) \alpha _1\right) \beta _2^5-4 (2-3 Q)^2 Q \beta _2^4\right.\\
&\left.+\left(8 \left(14 Q^2-29 Q+15\right)+13 \alpha _1\right) \beta
   _2^3+4 \left(12 Q^2-15 Q+5\right) \beta _2^2-24 Q+20\right) \beta _1^6+\left(Q^2 \left(9 Q^4 \beta _2^4+2 (20-3 Q) Q^2 \beta _2^2\right.\right.\\
&\left.\left.-6 Q (3 Q-4) \beta _2+2 Q^2+29 Q-74\right) \beta _2^2+\alpha _2 \left(4 (1-3 Q) Q \alpha _1
   \beta _2^{10}+\left(9 Q^4+2 \left(9 Q^2-14 Q+5\right) \alpha _1\right) \beta _2^9\right.\right.\\
&\left.\left.+\left(\left(-6 Q^2+42 Q-20\right) \alpha _1-9 (Q-1) Q^3\right) \beta _2^8+\left((187-105 Q) Q^2+(38-14 Q) \alpha _1\right)
   \beta_2^7-\left(3 Q \left(16 Q^2-57 Q+55\right)\right.\right.\right.\\
&\left.\left.\left.+4 (Q+7) \alpha _1\right) \beta _2^6+\left(329 Q^2-690 Q+275\right) \beta _2^5+4 \left(27 Q^2-52 Q+25\right) \beta _2^4+\left(11 Q^2-97 Q+132\right) \beta _2^3\right.\right.\\
&\left.\left.+2 \left(Q^2-41
   Q+40\right) \beta _2^2+9 \beta _2-2 Q+8\right)\right) \beta _1^5+\left(\beta _2 \left(4 Q^6 \beta _2^6-4 Q^4 (Q+7) \beta _2^4+16 (Q-1) Q^3 \beta _2^3+2 Q^2 (23 Q-24) \beta _2^2\right.\right.\\
&\left.\left.+Q \left(-3 Q^2+95 Q-140\right) \beta _2-32
   Q^2+82 Q-50\right)+\alpha _2 \left(2 Q \left(2 Q^3-3 (Q-1) \alpha _1\right) \beta _2^{10}+\left(3 \left(2 Q^2-4 Q+5\right) \alpha _1\right.\right.\right.\\
&\left.\left.\left.-4 (Q-1) Q^3\right) \beta _2^9+2 Q^2 (3 Q-19) \beta _2^8+2 \left(5 Q \left(Q^2+3
   Q-4\right)+3 (Q-4) \alpha _1\right) \beta _2^7+4 \left(Q^3-19 Q^2+41 Q-25\right) \beta _2^6\right.\right.\\
&\left.\left.-3 \left(16 Q^2-57 Q-3 \alpha _1+55\right) \beta _2^5-12 \left(Q^2-5 Q+4\right) \beta _2^4+(59 Q-71) \beta _2^3+2 (7 Q-19) \beta
   _2^2-6\right)\right) \beta _1^4\\
&+\left(2 Q \left(3 Q^2+2 \alpha _1\right) \alpha _2 \beta _2^{10}-2 \left(9 Q^2+5 (Q-1) \alpha _1\right) \alpha _2 \beta _2^9+\left(Q \left(-6 Q^2+43 Q-25\right)+(6 Q-20) \alpha _1\right)
   \alpha _2 \beta _2^8\right.\\
&\left.+\left(6 Q^5+\left(49 Q^2-290 Q+6 \alpha _1+275\right) \alpha _2\right) \beta _2^7+2 \left(3 (Q-4) Q^4+\left(5 \left(Q^2+3 Q-4\right)+2 \alpha _1\right) \alpha _2\right) \beta _2^6\right.\\
&\left.+\left(2 (19-7 Q)
   Q^3+(187-105 Q) \alpha _2\right) \beta _2^5+\left(\left(-8 Q^2-67 Q+97\right) Q^2+(44-38 Q) \alpha _2\right) \beta _2^4-26 \alpha _2 \beta _2^3\right.\\
&\left.+\left(2 Q^3+57 Q^2-240 Q+24 \alpha _2+275\right) \beta _2^2+\left(-2 Q^2+61
   Q-65\right) \beta _2-24 Q+11\right) \beta _1^3+\left(2 Q^2 (3 Q-10) \alpha _2 \beta _2^{10}\right.\\
&\left.-2 Q \left(3 Q^2-13 Q+10\right) \alpha _2 \beta _2^9+16 Q \alpha _2 \beta _2^8+\left(2 (3 Q-10) Q^4+\left(-6 Q^2+43 Q-25\right)
   \alpha _2\right) \beta _2^7+2 (3 Q-19) \alpha _2 \beta _2^6\right.\\
&\left.+\left(-2 \left(3 Q^2-21 Q+10\right) Q^2-9 (Q-1) \alpha _2\right) \beta _2^5-3 \left(Q \left(7 Q^2-43 Q+40\right)+2 \alpha _2\right) \beta _2^4-8 Q (3 Q-2) \beta
   _2^3-75 (Q-1) \beta _2^2\right.\\
&\left.+(8 Q-38) \beta _2+24\right) \beta _1^2+\beta _2 \left(-10 (Q-1) Q \alpha _2 \beta _2^9+\left(16 Q^2-32 Q+25\right) \alpha _2 \beta _2^8-2 \left(3 Q^2-13 Q+10\right) \alpha _2 \beta _2^7\right.\\
&\left.-2 \left(5
   (Q-1) Q^3+9 \alpha _2\right) \beta _2^6+\left(3 Q^2 \left(2 Q^2-4 Q+5\right)-4 (Q-1) \alpha _2\right) \beta _2^5+\left(2 Q \left(9 Q^2-14 Q+5\right)+9 \alpha _2\right) \beta _2^4\right.\\
&\left.+\left(-8 Q^3+47 Q^2-82 Q+25\right) \beta
   _2^3-4 \left(6 Q^2-11 Q+5\right) \beta _2^2+\left(2 Q^2-10 Q+17\right) \beta _2+6 (Q-1)\right) \beta _1\\
&+2 \beta _2 \left(2 Q \alpha _2 \beta _2^9-5 (Q-1) \alpha _2 \beta _2^8+(3 Q-10) \alpha _2 \beta _2^7+\left(2 Q^3+3
   \alpha _2\right) \beta _2^6+\left(2 \alpha _2-3 (Q-1) Q^2\right) \beta _2^5\right.\\
&\left.+2 (1-3 Q) Q \beta _2^4+\left(4 Q^2-9 Q+5\right) \beta _2^3+2 (6 Q-5) \beta _2^2-(Q-4) \beta _2-3\right)  \Big] \ ,
\end{aligned}
$$}}
{\scriptsize{
$$
\begin{aligned}
V&{}_{2,3}(\alpha_1,\beta_1,\alpha_2,\beta_2;Q)=\\
&24 (Q-4) \alpha _1^2 \alpha _2^3 \beta _2^9 \beta _1^9-6 Q \left(4 Q^2-23 Q+25\right) \alpha _1 \alpha _2^3 \beta _2^9 \beta _1^9-2 \left(61 Q^3-447 Q^2+975 Q-625\right) \alpha _1^2 \alpha _2^2 \beta _2^9 \beta _1^9\\
&+36 Q^5
   \alpha _1 \alpha _2^2 \beta _2^9 \beta _1^9+18 (Q-1) Q^4 \alpha _1 \alpha _2^2 \beta _2^8 \beta _1^9-90 (Q-1) Q \alpha _1 \alpha _2^3 \beta _2^7 \beta _1^9+36 Q^3 \left(Q^2-6 Q+5\right) \alpha _1 \alpha _2^2 \beta _2^7
   \beta _1^9\\
&-6 Q^2 \left(16 Q^2-35 Q+25\right) \alpha _1 \alpha _2^2 \beta _2^6 \beta _1^9-18 Q^2 \left(Q^2+4 Q-5\right) \alpha _1 \alpha _2^2 \beta _2^5 \beta _1^9+9 Q \left(3 Q^2-28 Q+25\right) \alpha _1 \alpha _2^2 \beta
   _2^4 \beta _1^9\\
&+18 (5-3 Q) Q^2 \alpha _1 \alpha _2^2 \beta _2^3 \beta _1^9+(132-24 Q) \alpha _1^2 \alpha _2^3 \beta _2^9 \beta _1^8+6 Q \left(4 Q^2-21 Q+35\right) \alpha _1 \alpha _2^3 \beta _2^9 \beta _1^8+\left(82
   Q^3-1185 Q^2\right.\\
&\left.+3120 Q-2125\right) \alpha _1^2 \alpha _2^2 \beta _2^9 \beta _1^8+12 Q^4 (14 Q-23) \alpha _1 \alpha _2^2 \beta _2^9 \beta _1^8+4 \left(10 Q^3+3 Q^2-63 Q+50\right) \alpha _1^2 \alpha _2^2 \beta _2^8 \beta
   _1^8\\
&-6 Q^3 \left(37 Q^2-92 Q+55\right) \alpha _1 \alpha _2^2 \beta _2^8 \beta _1^8+54 (Q-1) Q \alpha _1 \alpha _2^3 \beta _2^7 \beta _1^8-6 Q^2 \left(6 Q^3-14 Q^2-17 Q+25\right) \alpha _1 \alpha _2^2 \beta _2^7 \beta
   _1^8\\
&+6 Q \left(16 Q^3-98 Q^2+225 Q-125\right) \alpha _1 \alpha _2^2 \beta _2^6 \beta _1^8+6 Q \left(9 Q^3-32 Q^2+73 Q-50\right) \alpha _1 \alpha _2^2 \beta _2^5 \beta _1^8-9 Q \left(17 Q^2-52 Q\right.\\
&\left.+35\right) \alpha _1 \alpha
   _2^2 \beta _2^4 \beta _1^8-6 Q \left(Q^2-33 Q+50\right) \alpha _1 \alpha _2^2 \beta _2^3 \beta _1^8-36 \alpha _1^2 \alpha _2^3 \beta _2^9 \beta _1^7+20 \left(Q^2-5 Q-5\right) \alpha _1 \alpha _2^3 \beta _2^9 \beta _1^7\\
&-3\left(8 Q^3-163 Q^2+518 Q-375\right) \alpha _1^2 \alpha _2^2 \beta _2^9 \beta _1^7+2 Q^3 \left(47 Q^2-487 Q+530\right) \alpha _1 \alpha _2^2 \beta _2^9 \beta _1^7+6 \left(4 Q^3-43 Q^2+104 Q\right.\\
&\left.-65\right) \alpha _1^2 \alpha
   _2^2 \beta _2^8 \beta _1^7+Q^2 \left(-94 Q^3+1167 Q^2-2448 Q+1375\right) \alpha _1 \alpha _2^2 \beta _2^8 \beta _1^7+12 \left(Q^2+4 Q-5\right) \alpha _1 \alpha _2^3 \beta _2^7 \beta _1^7\\
&+12 \left(4 Q^2-5 Q+1\right) \alpha
   _1^2 \alpha _2^2 \beta _2^7 \beta _1^7+Q \left(-373 Q^3+2013 Q^2-3015 Q+1375\right) \alpha _1 \alpha _2^2 \beta _2^7 \beta _1^7+2 \left(72 Q^3+163 Q^2\right.\\
&\left.-950 Q+625\right) \alpha _1 \alpha _2^2 \beta _2^6 \beta _1^7-2 \left(6
   Q^4+10 Q^3+9 Q^2+225 Q-250\right) \alpha _1 \alpha _2^2 \beta _2^5 \beta _1^7+6 \left(7 Q^3+23 Q^2-55 Q\right.\\
&\left.+25\right) \alpha _1 \alpha _2^2 \beta _2^4 \beta _1^7+20 \left(Q^3+8 Q^2-25 Q+25\right) \alpha _1 \alpha _2^2 \beta
   _2^3 \beta _1^7-12 Q^3 \left(2 Q^2-7 Q+5\right) \alpha _2^3 \beta _2^9 \beta _1^6+4 \left(6 Q^3-26 Q^2\right.\\
&\left.-8 Q+55\right) \alpha _1 \alpha _2^3 \beta _2^9 \beta _1^6+24 \left(5 Q^2-18 Q+10\right) \alpha _1^2 \alpha _2^2 \beta
   _2^9 \beta _1^6+2 Q^2 \left(-61 Q^3+256 Q^2-338 Q+125\right) \alpha _1 \alpha _2^2 \beta _2^9 \beta _1^6\\
&+36 Q^7 \alpha _1 \alpha _2 \beta _2^9 \beta _1^6+54 (Q-1) Q^6 \alpha _2^2 \beta _2^8 \beta _1^6-12 \left(10 Q^2-57
   Q+38\right) \alpha _1^2 \alpha _2^2 \beta _2^8 \beta _1^6+Q \left(-40 Q^4+438 Q^3-2451 Q^2\right.\\
&\left.+5320 Q-3375\right) \alpha _1 \alpha _2^2 \beta _2^8 \beta _1^6+12 Q^6 (14 Q-23) \alpha _1 \alpha _2 \beta _2^8 \beta _1^6-36 (Q-1)
   Q^3 \alpha _2^3 \beta _2^7 \beta _1^6-36 (3 Q-2) \alpha _1^2 \alpha _2^2 \beta _2^7 \beta _1^6\\
&+18 Q^5 \left(2 Q^2-7 Q+5\right) \alpha _2^2 \beta _2^7 \beta _1^6-\left(24 Q^5-204 Q^4+329 Q^3+915 Q^2-1725 Q+625\right)
   \alpha _1 \alpha _2^2 \beta _2^7 \beta _1^6\\
&+2 Q^5 \left(47 Q^2-433 Q+476\right) \alpha _1 \alpha _2 \beta _2^7 \beta _1^6-72 (Q-1) \alpha _1^2 \alpha _2^2 \beta _2^6 \beta _1^6-12 Q^4 \left(17 Q^2-37 Q+20\right) \alpha
   _2^2 \beta _2^6 \beta _1^6\\
&+2 \left(68 Q^4+67 Q^3-1261 Q^2+2055 Q-875\right) \alpha _1 \alpha _2^2 \beta _2^6 \beta _1^6-2 Q^4 \left(245 Q^2-745 Q+554\right) \alpha _1 \alpha _2 \beta _2^6 \beta _1^6\\
&-36 Q^4 \left(2 Q^2-3
   Q+1\right) \alpha _2^2 \beta _2^5 \beta _1^6+2 \left(24 Q^4-50 Q^3-90 Q^2+441 Q-325\right) \alpha _1 \alpha _2^2 \beta _2^5 \beta _1^6+18 Q^3 \left(3 Q^2-8 Q\right.\\
&\left.+5\right) \alpha _2^2 \beta _2^4 \beta _1^6-24 (Q-1)^2 (2 Q-5)
   \alpha _1 \alpha _2^2 \beta _2^4 \beta _1^6+36 (Q-1) Q^4 \alpha _2^2 \beta _2^3 \beta _1^6-4 \left(6 Q^3+19 Q^2-98 Q+100\right) \alpha _1 \alpha _2^2 \beta _2^3 \beta _1^6\\
&-90 (Q-1) Q^5 \alpha _2 \beta _2^2 \beta _1^6+36
   (Q-1) Q^4 \alpha _2 \beta _1^6-18 Q^4 \left(2 Q^2-7 Q+5\right) \alpha _2 \beta _2 \beta _1^6+12 \left(Q^2+10 Q-11\right) \alpha _1 \alpha _2^3 \beta _2^9 \beta _1^5\\
&+12 \left(4 Q^2-17 Q+13\right) \alpha _1^2 \alpha _2^2
   \beta _2^9 \beta _1^5+Q \left(-323 Q^3+1593 Q^2-2145 Q+875\right) \alpha _1 \alpha _2^2 \beta _2^9 \beta _1^5-24 \left(2 Q^2-13 Q\right.\\
&\left.+11\right) \alpha _1^2 \alpha _2^2 \beta _2^8 \beta _1^5+\left(206 Q^4-2094 Q^3+6888
   Q^2-8750 Q+3750\right) \alpha _1 \alpha _2^2 \beta _2^8 \beta _1^5-36 (Q-1) \alpha _1 \alpha _2^3 \beta _2^7 \beta _1^5\\
&-108 (Q-1) \alpha _1^2 \alpha _2^2 \beta _2^7 \beta _1^5-12 \left(2 Q^4-58 Q^3+241 Q^2-310
   Q+125\right) \alpha _1 \alpha _2^2 \beta _2^7 \beta _1^5+243 Q^4 \left(Q^2-4 Q+3\right) \alpha _1 \alpha _2 \beta _2^7 \beta _1^5\\
&+288 (Q-1) Q \alpha _1 \alpha _2^2 \beta _2^6 \beta _1^5-Q^3 \left(188 Q^3-891 Q^2+588
   Q+115\right) \alpha _1 \alpha _2 \beta _2^6 \beta _1^5+8 \left(13 Q^3-9 Q^2+6 Q-10\right) \alpha _1 \alpha _2^2 \beta _2^5 \beta _1^5\\
&-2 Q^2 \left(319 Q^3-1755 Q^2+2661 Q-1225\right) \alpha _1 \alpha _2 \beta _2^5 \beta
   _1^5-18 \left(7 Q^2-16 Q+9\right) \alpha _1 \alpha _2^2 \beta _2^4 \beta _1^5-24 \left(4 Q^2+Q-5\right) \alpha _1 \alpha _2^2 \beta _2^3 \beta _1^5\\
&+4 Q \left(14 Q^3-21 Q^2-18 Q+25\right) \alpha _2^3 \beta _2^9 \beta
   _1^4-12 \left(4 Q^2-17 Q+13\right) \alpha _1 \alpha _2^3 \beta _2^9 \beta _1^4-108 (Q-1) Q^5 \alpha _2^2 \beta _2^9 \beta _1^4\\
&-72 (Q-1) \alpha _1^2 \alpha _2^2 \beta _2^9 \beta _1^4+2 \left(130 Q^4-749 Q^3+1944 Q^2-2575
   Q+1250\right) \alpha _1 \alpha _2^2 \beta _2^9 \beta _1^4-126 (Q-1) Q^5 \alpha _1 \alpha _2 \beta _2^9 \beta _1^4\\
&+72 (Q-1) \alpha _1^2 \alpha _2^2 \beta _2^8 \beta _1^4-18 Q^4 \left(5 Q^2-16 Q+11\right) \alpha _2^2 \beta
   _2^8 \beta _1^4+\left(64 Q^4-349 Q^3+315 Q^2+845 Q-875\right) \alpha _1 \alpha _2^2 \beta _2^8 \beta _1^4\\
&-6 Q^4 \left(41 Q^2-127 Q+86\right) \alpha _1 \alpha _2 \beta _2^8 \beta _1^4+12 Q \left(Q^2+4 Q-5\right) \alpha
   _2^3 \beta _2^7 \beta _1^4-6 Q^3 \left(10 Q^3-87 Q^2+132 Q-55\right) \alpha _2^2 \beta _2^7 \beta _1^4\\
&+6 \left(8 Q^4-73 Q^3+259 Q^2-319 Q+125\right) \alpha _1 \alpha _2^2 \beta _2^7 \beta _1^4-2 Q^3 \left(100 Q^3-861
   Q^2+1566 Q-805\right) \alpha _1 \alpha _2 \beta _2^7 \beta _1^4\\
&+4 Q^2 \left(77 Q^3-147 Q^2+45 Q+25\right) \alpha _2^2 \beta _2^6 \beta _1^4+\left(-208 Q^3+708 Q^2-540 Q+40\right) \alpha _1 \alpha _2^2 \beta _2^6 \beta
   _1^4+6 Q^2 \left(145 Q^3-439 Q^2\right.\\
&\left.+419 Q-125\right) \alpha _1 \alpha _2 \beta _2^6 \beta _1^4+12 Q^2 \left(10 Q^3-31 Q^2+26 Q-5\right) \alpha _2^2 \beta _2^5 \beta _1^4-6 \left(16 Q^3-87 Q^2+132 Q-61\right) \alpha _1 \alpha
   _2^2 \beta _2^5 \beta _1^4\\
&+Q \left(400 Q^4-2773 Q^3+6543 Q^2-6295 Q+2125\right) \alpha _1 \alpha _2 \beta _2^5 \beta _1^4-6 Q \left(15 Q^3-17 Q^2-23 Q+25\right) \alpha _2^2 \beta _2^4 \beta _1^4\\
&+48 \left(Q^2-5 Q+4\right)
   \alpha _1 \alpha _2^2 \beta _2^4 \beta _1^4+72 (Q-1) Q^5 \alpha _2 \beta _2^4 \beta _1^4-2 (Q-1)^2 \left(53 Q^2+350 Q-625\right) \alpha _1 \alpha _2 \beta _2^4 \beta _1^4\\
&-60 (Q-1)^2 Q^2 \alpha _2^2 \beta _2^3 \beta
   _1^4+12 \left(4 Q^2-17 Q+13\right) \alpha _1 \alpha _2^2 \beta _2^3 \beta _1^4+108 (Q-1) Q^4 \alpha _2 \beta _2^3 \beta _1^4+30 Q^3 \left(5 Q^2-16 Q+11\right) \alpha _2 \beta _2^2 \beta _1^4\\
&-60 (Q-1)^2 Q^2 \alpha _2 \beta
   _1^4+30 (Q-1)^2 Q^2 (2 Q-5) \alpha _2 \beta _2 \beta _1^4+6 Q \left(14 Q^2-43 Q+35\right) \alpha _2^3 \beta _2^9 \beta _1^3-24 (4 Q-7) \alpha _1 \alpha _2^3 \beta _2^9 \beta _1^3\\
&-36 Q^4 \alpha _2^2 \beta _2^9 \beta
   _1^3-24 (Q-4) \alpha _1^2 \alpha _2^2 \beta _2^9 \beta _1^3+\left(700 Q^3-4164 Q^2+7320 Q-4000\right) \alpha _1 \alpha _2^2 \beta _2^9 \beta _1^3-36 Q^4 (3 Q-2) \alpha _1 \alpha _2 \beta _2^9 \beta _1^3\\
&+12 (2 Q-11) \alpha
   _1^2 \alpha _2^2 \beta _2^8 \beta _1^3-27 Q^3 \left(7 Q^2-22 Q+15\right) \alpha _2^2 \beta _2^8 \beta _1^3+\left(8 Q^3+819 Q^2-2244 Q+1525\right) \alpha _1 \alpha _2^2 \beta _2^8 \beta _1^3\\
&-6 Q^3 \left(93 Q^2-266
   Q+155\right) \alpha _1 \alpha _2 \beta _2^8 \beta _1^3+54 (Q-1) Q \alpha _2^3 \beta _2^7 \beta _1^3+36 \alpha _1^2 \alpha _2^2 \beta _2^7 \beta _1^3-9 Q^2 \left(14 Q^3-51 Q^2+62 Q-25\right) \alpha _2^2 \beta _2^7 \beta
   _1^3
\end{aligned}
$$
$$
\begin{aligned}
&+\left(116 Q^3-597 Q^2+1260 Q-815\right) \alpha _1 \alpha _2^2 \beta _2^7 \beta _1^3+2 Q^2 \left(-348 Q^3+2122 Q^2-3239 Q+1375\right) \alpha _1 \alpha _2 \beta _2^7 \beta _1^3\\
&+6 Q \left(117 Q^3-386 Q^2+400
   Q-125\right) \alpha _2^2 \beta _2^6 \beta _1^3-12 \left(31 Q^2-60 Q+23\right) \alpha _1 \alpha _2^2 \beta _2^6 \beta _1^3-36 Q^6 \alpha _2 \beta _2^6 \beta _1^3\\
&+2 Q \left(47 Q^4+758 Q^3-4147 Q^2+5985 Q-2625\right) \alpha
   _1 \alpha _2 \beta _2^6 \beta _1^3+12 Q \left(25 Q^3-69 Q^2+69 Q-25\right) \alpha _2^2 \beta _2^5 \beta _1^3\\
&-12 \left(23 Q^2-49 Q+26\right) \alpha _1 \alpha _2^2 \beta _2^5 \beta _1^3+72 (Q-1) Q^5 \alpha _2 \beta _2^5
   \beta _1^3+\left(1859 Q^4-10557 Q^3+19323 Q^2-14375 Q\right.\\
&\left.+3750\right) \alpha _1 \alpha _2 \beta _2^5 \beta _1^3-9 Q \left(9 Q^2-44 Q+35\right) \alpha _2^2 \beta _2^4 \beta _1^3+72 (Q-1) \alpha _1 \alpha _2^2 \beta _2^4 \beta
   _1^3+126 (Q-1) Q^4 \alpha _2 \beta _2^4 \beta _1^3\\
&-2 \left(100 Q^4-592 Q^3+1827 Q^2-2710 Q+1375\right) \alpha _1 \alpha _2 \beta _2^4 \beta _1^3-6 Q \left(29 Q^2-73 Q+50\right) \alpha _2^2 \beta _2^3 \beta _1^3+24 (7
   Q-10) \alpha _1 \alpha _2^2 \beta _2^3 \beta _1^3\\
&+12 Q^3 \left(4 Q^2-11 Q+10\right) \alpha _2 \beta _2^3 \beta _1^3-4 \left(200 Q^3-1041 Q^2+1623 Q-800\right) \alpha _1 \alpha _2 \beta _2^3 \beta _1^3+9 Q^2 \left(17
   Q^2-42 Q+25\right) \alpha _2 \beta _2^2 \beta _1^3\\
&-6 Q \left(29 Q^2-73 Q+50\right) \alpha _2 \beta _1^3+6 Q \left(29 Q^3-129 Q^2+225 Q-125\right) \alpha _2 \beta _2 \beta _1^3-2 \left(20 Q^3+24 Q^2-165 Q+175\right) \alpha
   _2^3 \beta _2^9 \beta _1^2\\
&+12 (2 Q-11) \alpha _1 \alpha _2^3 \beta _2^9 \beta _1^2+6 Q^3 \left(33 Q^2-100 Q+85\right) \alpha _2^2 \beta _2^9 \beta _1^2-3 \left(46 Q^3-385 Q^2+928 Q-625\right) \alpha _1 \alpha _2^2 \beta
   _2^9 \beta _1^2\\
&+18 Q^3 \left(3 Q^2-13 Q+10\right) \alpha _1 \alpha _2 \beta _2^9 \beta _1^2+12 Q^3 \left(2 Q^2-13 Q+11\right) \alpha _2^2 \beta _2^8 \beta _1^2-12 (Q-1)^2 (2 Q-5) \alpha _1 \alpha _2^2 \beta _2^8 \beta
   _1^2\\
&+6 Q^2 \left(15 Q^3-37 Q^2+47 Q-25\right) \alpha _1 \alpha _2 \beta _2^8 \beta _1^2-90 (Q-1) \alpha _2^3 \beta _2^7 \beta _1^2
   +2 Q \left(20 Q^4-2 Q^3-423 Q^2+655 Q-250\right) \alpha _2^2 \beta _2^7 \beta _1^2\\
&-12
   \left(2 Q^3-17 Q^2+52 Q-37\right) \alpha _1 \alpha _2^2 \beta _2^7 \beta _1^2-162 (Q-1) Q^6 \alpha _2 \beta _2^7 \beta _1^2+Q \left(114 Q^4-1015 Q^3+2541 Q^2-3015 Q\right.\\
&\left.+1375\right) \alpha _1 \alpha _2 \beta _2^7 \beta _1^2-2
   \left(28 Q^4+124 Q^3-798 Q^2+1325 Q-625\right) \alpha _2^2 \beta _2^6 \beta _1^2+12 \left(6 Q^2-37 Q+22\right) \alpha _1 \alpha _2^2 \beta _2^6 \beta _1^2\\
&-6 Q^5 (7 Q-25) \alpha _2 \beta _2^6 \beta _1^2+\left(-222 Q^4+1015
   Q^3-2535 Q^2+2475 Q-625\right) \alpha _1 \alpha _2 \beta _2^6 \beta _1^2-10 (Q-1)^2 \left(8 Q^2+15 Q\right.\\
&\left.-50\right) \alpha _2^2 \beta _2^5 \beta _1^2+24 \left(2 Q^2-13 Q+11\right) \alpha _1 \alpha _2^2 \beta _2^5 \beta
   _1^2+324 (Q-1)^2 Q^4 \alpha _2 \beta _2^5 \beta _1^2-2 \left(114 Q^4-946 Q^3+2817 Q^2-3360 Q\right.\\
&\left.+1375\right) \alpha _1 \alpha _2 \beta _2^5 \beta _1^2+\left(-48 Q^3+423 Q^2-900 Q+525\right) \alpha _2^2 \beta _2^4 \beta
   _1^2+18 Q^3 \left(26 Q^2-71 Q+45\right) \alpha _2 \beta _2^4 \beta _1^2-12 \left(10 Q^3\right.\\
&\left.-97 Q^2+212 Q-125\right) \alpha _1 \alpha _2 \beta _2^4 \beta _1^2+2 \left(20 Q^3+9 Q^2-225 Q+250\right) \alpha _2^2 \beta _2^3 \beta
   _1^2+(132-24 Q) \alpha _1 \alpha _2^2 \beta _2^3 \beta _1^2-6 Q^2 \left(27 Q^3\right.\\
&\left.-117 Q^2+233 Q-125\right) \alpha _2 \beta _2^3 \beta _1^2+3 \left(38 Q^3-313 Q^2+694 Q-455\right) \alpha _1 \alpha _2 \beta _2^3 \beta _1^2-2 Q
   \left(41 Q^3+114 Q^2-405 Q\right.\\
&\left.+250\right) \alpha _2 \beta _2^2 \beta _1^2+12 \left(2 Q^2-13 Q+11\right) \alpha _1 \alpha _2 \beta _2^2 \beta _1^2+2 \left(20 Q^3+9 Q^2-225 Q+250\right) \alpha _2 \beta _1^2+\left(-40 Q^4-161
   Q^3\right.\\
&\left.+1476 Q^2-2525 Q+1250\right) \alpha _2 \beta _2 \beta _1^2+\left(-80 Q^2+40 Q+220\right) \alpha _2^3 \beta _2^9 \beta _1+36 \alpha _1 \alpha _2^3 \beta _2^9 \beta _1+2 Q^2 \left(88 Q^2-53 Q\right.\\
&\left.-125\right) \alpha _2^2 \beta
   _2^9 \beta _1-3 \left(61 Q^2-194 Q+145\right) \alpha _1 \alpha _2^2 \beta _2^9 \beta _1+45 Q^2 \left(3 Q^2-8 Q+5\right) \alpha _1 \alpha _2 \beta _2^9 \beta _1+Q \left(184 Q^3-1629 Q^2\right.\\
&\left.+3570 Q-2125\right) \alpha _2^2 \beta
   _2^8 \beta _1-12 \left(5 Q^2-28 Q+23\right) \alpha _1 \alpha _2^2 \beta _2^8 \beta _1+3 Q \left(47 Q^3-272 Q^2+475 Q-250\right) \alpha _1 \alpha _2 \beta _2^8 \beta _1\\
 &+5 \left(12 Q^4-43 Q^3-39 Q^2+195 Q-125\right) \alpha
   _2^2 \beta _2^7 \beta _1-36 (Q-1)^2 \alpha _1 \alpha _2^2 \beta _2^7 \beta _1+\left(151 Q^4-1007 Q^3+2631 Q^2-3025 Q\right.\\
&\left.+1250\right) \alpha _1 \alpha _2 \beta _2^7 \beta _1-10 \left(22 Q^3+71 Q^2-286 Q+175\right) \alpha _2^2
   \beta _2^6 \beta _1+36 (3 Q-2) \alpha _1 \alpha _2^2 \beta _2^6 \beta _1-2 Q^4 \left(47 Q^2-325 Q\right.\\
&\left.+368\right) \alpha _2 \beta _2^6 \beta _1+\left(24 Q^4-437 Q^3+1467 Q^2-1215 Q+125\right) \alpha _1 \alpha _2 \beta _2^6
   \beta _1+\left(-76 Q^3+24 Q^2+702 Q-650\right) \alpha _2^2 \beta _2^5 \beta _1\\
&+72 (Q-1) \alpha _1 \alpha _2^2 \beta _2^5 \beta _1-15 Q^3 \left(29 Q^2-100 Q+71\right) \alpha _2 \beta _2^5 \beta _1-6 \left(41 Q^3-219
   Q^2+303 Q-125\right) \alpha _1 \alpha _2 \beta _2^5 \beta _1\\
&-24 \left(Q^2+4 Q-5\right) \alpha _2^2 \beta _2^4 \beta _1+2 Q^2 \left(100 Q^3-717 Q^2+1242 Q-625\right) \alpha _2 \beta _2^4 \beta _1-24 \left(2 Q^3-11 Q^2+19
   Q-10\right) \alpha _1 \alpha _2 \beta _2^4 \beta _1\\
&+4 \left(2 Q^2+53 Q-100\right) \alpha _2^2 \beta _2^3 \beta _1-36 \alpha _1 \alpha _2^2 \beta _2^3 \beta _1+2 Q \left(436 Q^3-2581 Q^2+4235 Q-2000\right) \alpha _2 \beta
   _2^3 \beta _1+3 \left(Q^2-32 Q\right.\\
&\left.+43\right) \alpha _1 \alpha _2 \beta _2^3 \beta _1+162 (Q-1) Q^5 \beta _2^2 \beta _1+\left(-114 Q^4+829 Q^3-1815 Q^2+1725 Q-625\right) \alpha _2 \beta _2^2 \beta _1+24 \left(Q^2-5 Q\right.\\
&\left.+4\right)
   \alpha _1 \alpha _2 \beta _2^2 \beta _1+4 \left(2 Q^2+53 Q-100\right) \alpha _2 \beta _1-72 (Q-1) Q^4 \beta _2 \beta _1+\left(-179 Q^3+519 Q^2-465 Q+125\right) \alpha _2 \beta _2 \beta _1\\
&+36 (Q-1) \alpha _1 \alpha _2
   \beta _2 \beta _1-4 \left(4 Q^2-50 Q+73\right) \alpha _2^3 \beta _2^9+2 Q \left(46 Q^3-493 Q^2+965 Q-500\right) \alpha _2^2 \beta _2^9-24 (Q-4) \alpha _1 \alpha _2^2 \beta _2^9\\
&-36 Q^6 \alpha _2 \beta _2^9+18 (5-3 Q) Q^2
   \alpha _1 \alpha _2 \beta _2^9+2 \left(8 Q^4-168 Q^3+1464 Q^2-3125 Q+1875\right) \alpha _2^2 \beta _2^8+12 (2 Q-11) \alpha _1 \alpha _2^2 \beta _2^8\\
&+12 (14-5 Q) Q^5 \alpha _2 \beta _2^8-6 Q \left(Q^2-33 Q+50\right) \alpha
   _1 \alpha _2 \beta _2^8+\left(16 Q^4-228 Q^3+981 Q^2-1180 Q+375\right) \alpha _2^2 \beta _2^7\\
&+36 \alpha _1 \alpha _2^2 \beta _2^7-2 Q^4 \left(38 Q^2-361 Q+413\right) \alpha _2 \beta _2^7+20 \left(Q^3+8 Q^2-25 Q+25\right)
   \alpha _1 \alpha _2 \beta _2^7+\left(-48 Q^3+548 Q^2\right.\\
&\left.-1108 Q+500\right) \alpha _2^2 \beta _2^6+2 Q^3 \left(145 Q^2-692 Q+655\right) \alpha _2 \beta _2^6-4 \left(6 Q^3+19 Q^2-98 Q+100\right) \alpha _1 \alpha _2 \beta
   _2^6+\left(-32 Q^3\right.\\
&\left.+348 Q^2-1086 Q+770\right) \alpha _2^2 \beta _2^5+Q^2 \left(152 Q^3-873 Q^2+696 Q+25\right) \alpha _2 \beta _2^5-24 \left(4 Q^2+Q-5\right) \alpha _1 \alpha _2 \beta _2^5-76 Q^4\\
&+72 (Q-1) \alpha _2^2 \beta
   _2^4+2 Q \left(-92 Q^3+327 Q^2-360 Q+125\right) \alpha _2 \beta _2^4+12 \left(4 Q^2-17 Q+13\right) \alpha _1 \alpha _2 \beta _2^4+518 Q^3+36 Q^5 \beta _2^3\\
&+4 \left(4 Q^2-32 Q+55\right) \alpha _2^2 \beta _2^3+\left(-76
   Q^4-298 Q^3+4088 Q^2-7500 Q+3750\right) \alpha _2 \beta _2^3+24 (7 Q-10) \alpha _1 \alpha _2 \beta _2^3-550 Q^2\\
&-6 Q^4 (17 Q+1) \beta _2^2+\left(98 Q^3-651 Q^2+1320 Q-875\right) \alpha _2 \beta _2^2+(132-24 Q) \alpha _1
   \alpha _2 \beta _2^2+4 \left(4 Q^2-32 Q+55\right) \alpha _2\\
&+2 Q^3 \left(38 Q^2-253 Q+305\right) \beta _2+\left(-16 Q^3+315 Q^2-888 Q+625\right) \alpha _2 \beta _2-36 \alpha _1 \alpha _2 \beta _2 \ .
\end{aligned}
$$}}
In terms of these polynomials, the numerator of the kernel~$B_{2,3}(\alpha_1,\beta_1,\alpha_2,\beta_2;Q)$ is of degree one in $\alpha_\mu$ and degree $13$ in $\beta_\mu$ for $\mu=1,2$, while the denominator of $B_{2,3}(\alpha_1,\beta_1,\alpha_2,\beta_2;Q)$ is of degree three in $\alpha_\mu$ and degree nine in $\beta_\mu$ for $\mu=1,2$.

%%%%%%%%%%%%%%%%%%%%%%%%%%%%%%%%%%%%%
\section{Annulus instantons from localization} \label{app:Localization}
%%%%%%%%%%%%%%%%%%%%%%%%%%%%%%%%%%%%%
\def\la{\lambda}
\def\IE{\mathbb{E}}
The aim of this section is to calculate on the resolved conifold some leading order annulus instanton numbers for the branes associated to the torus knots $\mathcal{K}_{r,s}$ directly by means of localization on the relevant space of stable maps with two boundaries on the branes $\mathcal{L}_{r,s}$ along the lines of ref.~\cite{Diaconescu:2011xr}. We extend the results of ref.~\cite{Diaconescu:2011xr} by simultaneously considering the brane~$\mathcal{L}_{r,s}$ and the image brane~$\iota_*\mathcal{L}_{r,s}$ of the involution~\eqref{eq:invol}. That is to say we want to compute the open Gromov--Witten invariants that are schematically given by
\begin{align} 
   A_{g,d,h_1,h_2}^{(\mathcal{L}_{r,s},\mathcal{L}_{r,s})} \,&=\, \int_{\overline{\mathcal{M}}_{g,d,h_1,h_2}(X;\mathcal{L}_{r,s} \coprod \mathcal{L}_{r,s})^\text{vir}} \mathbf{1} \ , \label{eq:Aone} \\
   A_{g,d,h_1,h_2}^{(\mathcal{L}_{r,s},\iota_*\mathcal{L}_{r,s})} \,&=\, \int_{\overline{\mathcal{M}}_{g,d,h_1,h_2}(X;\mathcal{L}_{r,s} \cup \iota_*\mathcal{L}_{r,s})^\text{vir}} \mathbf{1} \ . \label{eq:Atwo}
\end{align}
Here $A_{g,d,h_1,h_2}^{(\mathcal{L}_{r,s},\mathcal{L}_{r,s})}$ and $A_{g,d,h_1,h_2}^{(\mathcal{L}_{r,s},\iota_*\mathcal{L}_{r,s})}$ enumerate the genus $g$ stable maps $f_g : \operatorname{Ann} \rightarrow X$ from the annulus $\operatorname{Ann}$ to the resolved conifold $X$ with $f_g[\operatorname{Ann}] = d [\mathbb{P}^1] +h_1 [S^1] + h_2 [S^1]$ in the relative homology classes $H_2(X,\mathcal{L}_{r,s} \coprod\mathcal{L}_{r,s})$ and $H_2(X,\mathcal{L}_{r,s} \cup \iota_*\mathcal{L}_{r,s})$, respectively. Here the former integral enumerates annulus instantons with both boundaries on the brane $\mathcal{L}_{r,s}$, while the latter integral counts the annulus instantons stretching between the branes $\mathcal{L}_{r,s}$ and $\iota_*\mathcal{L}_{r,s}$.

To evaluate the integrals~\eqref{eq:Aone} and \eqref{eq:Atwo}, as pioneered in ref.~\cite{Kontsevich:1994na} we take advantage of the Atiyah--Bott fixed point formula and localize on the fixed point locus with respect to the $\mathbb{C}^*$ symmetries in the moduli spaces $\overline{\mathcal{M}}_{g,d,h_1,h_2}$, which are induced from the $\mathbb{C}^*$-action of the toric description of the resolved conifold description. Then evaluating the above integrals --- which in this case are of virtual dimension zero --- amounts to summing over suitable $\mathbb{C}^*$-equivariant characters defined on the fixed point loci. In practice these fixed point loci can represented in terms of graphs, to which one then assigns the corresponding equivariant classes. For details on this somewhat technical construction we refer the reader to refs.~\cite{Kontsevich:1994na,MR1666787,Katz:2001vm,Li:2001sg,Graber:2001dw,Diaconescu:2011xr}. 

To evaluate the integrals~\eqref{eq:Aone} and \eqref{eq:Atwo} explicitly, we first need to assemble the relevant equivariant classes associated to the fixed point loci depicted by graphs. For the given geometry~$X$ with the torus knot branes $\mathcal{L}_{r,s}$ a detailed derivation of the various contributions has been given in the existing literature, and therefore we simply collect the necessary ingredients here.

%%%%%%%%%%%%%%%%%%%%%%
\begin{figure}
\begin{center}
\subfloat[~\hskip18ex~]{\includegraphics[width=0.29\textwidth]{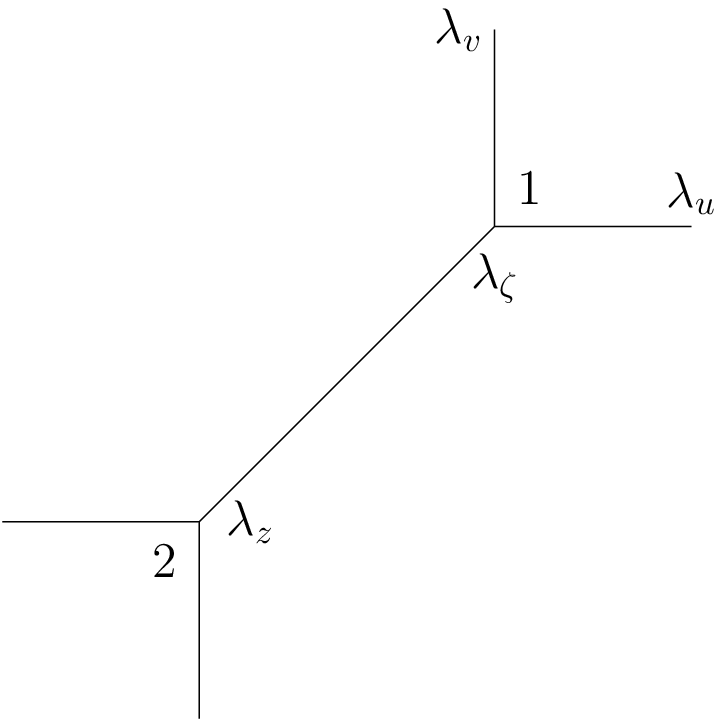}\label{fig:loc1}}\hskip3ex 
\subfloat[~\hskip18ex~]{\includegraphics[width=0.29\textwidth]{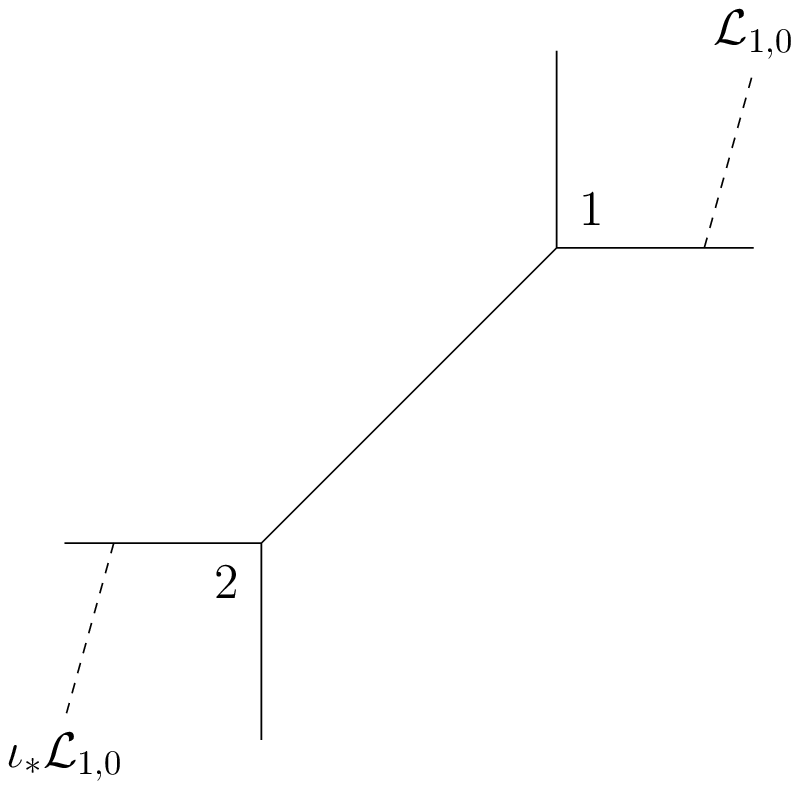}\label{fig:loc2}}\hskip3ex
\subfloat[~\hskip18ex~]{\includegraphics[width=0.29\textwidth]{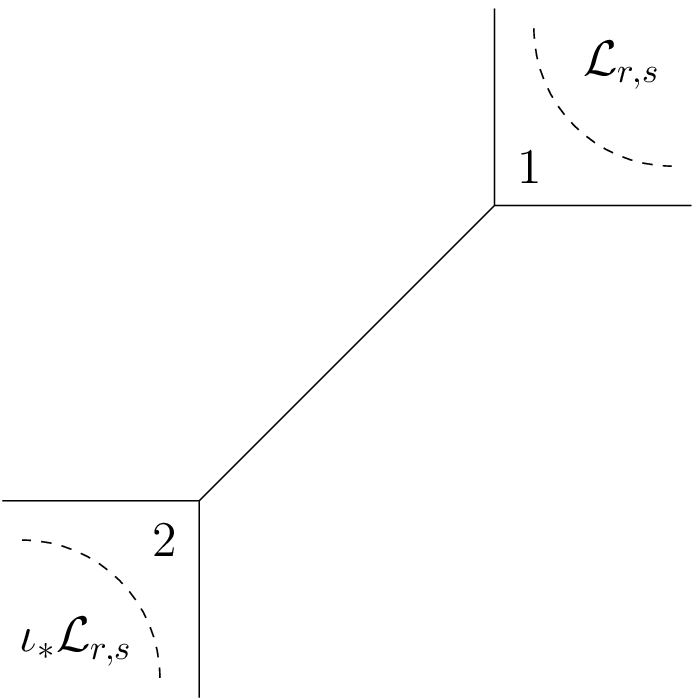}\label{fig:loc3}}
\end{center}
\caption{The figure (a) shows the weights assigned to the toric skeleton of the resolved conifold. In figure (b) we exhibit the location of the toric brane $\mathcal{L}_{1,0}$, while figure (c) schematically depicts the torus knot branes $\mathcal{L}_{r,s}$.} 
\end{figure}
%%%%%%%%%%%%%%%%%%%%%%

%%%%%%%%%%%%%%
\subsection{Weights of the equivariant classes}
%%%%%%%%%%%%%%
The fixed point loci of the resolved conifold with respect to the $\mathbb{C}^*$ symmetries are given by its toric skeleton, to which we assign the following vertices and weights as in Figure~\ref{fig:loc1}.
This means that we assign to the vertices the weights $\lambda_1$ and $\lambda_2$, which arise from the $\mathbb{C}^*$-action on the homogeneous coordinates $[x_1:x_2]$ of the $\mathbb{P}^1$-cycle of the resolved conifold. Then the other weights are the weights of the local coordinates $(\zeta=\frac{x_2}{x_1},u,v)$, where $u$ and $v$ are the fiber coordinates in the local patch $\zeta$ with assigned weights $\lambda_u$ and $\lambda_v$, respectively. Note that we have the relations $\lambda_\zeta = \lambda_2 - \lambda_1$ and $\la_z = - \la_\zeta= \la_1 - \la_2$ due to the coordinate transformation $z=\zeta^{-1}=\frac{x_1}{x_2}$ to the other local patch of $\mathbb{P}^1$. 

The relevant graphs associated to closed string Gromov--Witten invariants can be assembled from graphs with vertices and edges. The equivariant class $\frac{i^*\phi}{e(N^{\rm vir})}$ assigned to such a closed-string graph of the resolved conifold geometry is given in terms of the above weights as \cite{Kontsevich:1994na,Klemm:1999gm}
\begin{equation}
\begin{aligned}
  \frac{i^*\phi}{e(N^{\rm vir})} \,=\,&
  \prod_e \frac{(-1)^{d_e}}{(d_e!)^2\,\la_\zeta^{2d_e}} 
  \prod_\nu \prod_{j\ne i(\nu)} ( \la_{i(\nu)}-\la_j )^{{\rm val}(\nu)-1} \\[1ex]
  &\cdot\prod_\nu 
  \begin{cases}
    \left( \sum_F w_F^{-1}\right)^{{\rm val}(\nu)-3} \cdot \prod_{F\ni \nu} w_F^{-1} & g(\nu)=0 \\[2ex]
    \prod_{j\ne i(\nu)} P_{g(\nu)}(\la_{i(\nu)}-\la_{j(\nu)},\IE^*)\cdot \prod_{F\ni\nu}\frac1{w_F-\psi_{\nu,F}} & g(\nu)>0 
  \end{cases} \\[1ex]
  % i^* part
  &\cdot \prod_\nu \left((-\la_u+\la_1-\la_\nu)(-\la_v+\la_1-\la_\nu)\right)^{{\rm val}(\nu)-1} \\
  &\cdot \prod_\nu P_{g(\nu)}(-\la_u+\la_1-\la_\nu,\IE^*)P_{g(\nu)}(-\la_v+\la_1-\la_\nu,\IE^*) \\
  &\cdot \prod_e \prod_{\buildrel{a,b<0}\over{a+b=-d_e}} \left(\la_u+\la_1+\frac{a \la_1+b\la_2}{d_e} \right)\left(\la_v+\la_1+\frac{a \la_1+b\la_2}{d_e} \right)
   \ . 
\end{aligned}
\end{equation}   
Here the products are taken over the edges $e$ and the vertices $\nu$ of the graph. $d_e$ denotes the degree of an edge, while $\operatorname{val}(\nu)$ describes the valence of a vertex, and $w_F=\frac{\la_i(F)-\la_j(F)}{d_F}$ is the flag of degree $d_F$ with respect to the vertices $i(F)$ and $j(F)$ of the flag $F$\cite{Kontsevich:1994na}. Furthermore, $g(\nu)$ is the genus of a component mapped by $f_g$ to the vertex $\nu$, and $P_g(\la,\IE^*)$ denotes the Chern class of the rank $g$ dual Hodge bundle~$\mathbb{E}^*$ \cite{Witten:1990hr,Kontsevich:1992ti,MR1714822}
\begin{equation}
  P_g(\la,\IE^*)\,=\, \sum_{k=0}^g \la^k c_{g-k}(\IE^*) \,=\, (-1)^g \sum_{k=0}^g (-\la)^k c_{g-k}(\IE) \ .
\end{equation}
Here we use that the Hodge bundle $\IE$ obeys $c_k(\IE^*)=(-1)^k c_k(\IE)$.

The open-string graphs --- which are of interest to us so as to the evalutation of the integrals~\eqref{eq:Aone} and \eqref{eq:Atwo} --- are further decorated by adding legs to the closed-string graphs to capture the boundary components of the Gromov--Witten invariants. First we briefly recall the result for the unknot brane $\mathcal{L}_{1,0}$ \cite{Katz:2001vm,Li:2001sg,Graber:2001dw}. Let us consider a disk contribution along the edge associated to the local coordinate $u$ attached to the brane $\mathcal{L}_{1,0}$ as shown in Figure~\ref{fig:loc2}. To be specific a component of winding $h$ is mapped to the resolved conifold according to
\begin{equation}
  f: D_h \rightarrow U \subset X,\  t \mapsto (\zeta,u,v)=(0,t^h,0) \ . 
\end{equation}
Here $U$ is the local coordinate patch of $X$. The weight contribution of such a disk component $D_h$ is graphically described by a leg $\ell(h)$ attached to a closed-string graph. The equivariant class of such a leg reads
\begin{equation} \label{eq:leg}
   \ell(h) \,=\,- \frac{\prod_{k=1}^{h-1}\left(k\,\la_u + h \la_z \right)}{ {(h-1)!\, \la_u^h}}  \ ,
\end{equation} 
where the flag of the disk $D_h$ is given by $w_F(D_h)\,=\, - \frac{\la_u}h$. Note that the presence of the brane, which realizes the boundary condition of the open Gromov--Witten invariant breaks part of the $\mathbb{C}^*$ symmetries of the resolved conifold \cite{Katz:2001vm,Li:2001sg,Graber:2001dw}.  As a consequence, in the presence of a brane also the symmetries on the corresponding moduli space of stable maps are reduced, which gives rise to relations among the weights. For the unknot toric brane depicted in Figure~\ref{fig:loc2} the corresponding relations read \cite{Katz:2001vm,Li:2001sg,Graber:2001dw}
\begin{equation} \label{eq:const1}
  -\la_z\,=\,\la_\zeta\,=\,a\,\la_u \ , \qquad \la_v \,=\, \frac{1-a}{a} \la_z = (a-1) \la_u  \ ,
\end{equation}  
where the integer $a$ relates to the framing in eq.~\eqref{eq:unknot} of the Lagrangian brane of the unknot according to $f=1-a$. 

Let us now turn to the leg contribution of the image unknot brane $\iota_*\mathcal{L}_{1,0}$. Note that the weights entering the image component $\iota_*D_h$ are give by $\la_{\iota_*z}=-\la_z$, $\lambda_{\iota_*u} = \lambda_u + \lambda_z$, and $\lambda_{\iota_*v} = \lambda_v + \lambda_z$. Expressing the weight constraints~\eqref{eq:const1} in terms of the image weights $\la_{\iota_*z}$, $\la_{\iota_*u}$, and $\lambda_{\iota_*v}$ yields $-\la_{\iota_*z} \,=\, a\,\la_{\iota_*v}$ and $\la_{\iota_*u} \,=\, (a-1) \la_{\iota_*v}$, where the weights for the two line bundles get exchanged. In order to preserve the same $\mathbb{C}^*$~symmetries for both disk components $D_h$ and $\iota_*D_h$, we compensate this exchange in the image leg contribution accordingly. In terms of $\la_z$, $\la_u$ and $\la_v$ we therefore arrive at
\begin{equation} \label{eq:ileg}
  \iota_*\ell(h) \,=\, - \frac{\prod_{k=1}^{h-1}\left(k\,(\la_v+\la_z) - h \la_z \right)}{ {(h-1)!\, (\la_v+\la_z)^h}} \ .
\end{equation}   
Imposing the constraints \eqref{eq:const1}, the image brane $\iota_*\mathcal{L}_{1,0}$ is now in the same framing. The flag of the image disk $\iota_*D_h$ reads $w_F(\iota_*D_h)\,=\, - \frac{\la_v+\la_z}h$. Note that this structure of the weights of legs and flags ensures that diagrams are manifest invariant with respect to the involution $\iota$ after inserting the constraints \eqref{eq:const1}.\footnote{We would like to thank the referee for pointing out to us the involutive symmetry on the level of diagrams. It allowed us to fix an error in eq.~\eqref{eq:ileg} in a previous version of this manuscript.}

Our main interest here, however, is on the branes $\mathcal{L}_{r,s}$ associated to non-trivial torus knots $\mathcal{K}_{r,s}$. A localization scheme for these branes has carefully been worked out in ref.~\cite{Diaconescu:2011xr}. Schematically, the toric skeleton together with these Lagrangian branes $\mathcal{L}_{r,s}$ and $\iota_*\mathcal{L}_{r,s}$ are depicted in Figure~\ref{fig:loc3}. Embedding the disk component $D_s$ according to\footnote{For technical reasons the coprime integers $r$ and $s$ must obey $1\le s <r$ \cite{Diaconescu:2011xr}.}
\begin{equation}
   f: D_s \rightarrow U \subset X,\  t \mapsto (\zeta,u,v)=(0,t^s,t^r) \ , \quad 1\le s<r \ ,
\end{equation}   
which describes a disk component of winding one with its boundary mapped to the brane~$\mathcal{L}_{r,s}$ \cite{Diaconescu:2011xr}. The leg contribution of such a disk component turns out to be also given by the formula~\eqref{eq:leg}. However, the $\mathbb{C}^*$ symmetries broken by the presence of the branes $\mathcal{L}_{r,s}$ and $\iota_*\mathcal{L}_{r,s}$ are distinct. This is reflected on the modified constraints on the weights \cite{Diaconescu:2011xr}
\begin{equation} \label{eq:rsrels}
   -\la_z\,=\,\la_\zeta\,=\,\frac{a}{s}\,\la_u \ , \qquad \la_v\,=\,-\frac{r}{a}\,\la_z\,=\,\frac{r}{s} \la_u \ , \qquad a=r+s \ . 
\end{equation}
Here we do not have a choice for the framing $a$. In order to still be able to apply the localization technique with respect to an unbroken symmetry, we are required to fix the integer $a$ as stated. Geometrically, this amounts to choosing the canonical framing $r\cdot s$ for the torus knot branes $\mathcal{L}_{r,s}$. Note that higher order windings $h$ are simply calculated by considering the leg contributions $\ell(s\cdot h)$ (with the same relations~\eqref{eq:rsrels} on the weights).
As for the unknot, we can realize a disk component of winding $h$ on the image brane $\iota_*\mathcal{L}_{r,s}$ by considering the leg contribution $\iota_*\ell(s\cdot h)$ together with the same constraint \eqref{eq:rsrels}, which again ensure that diagrams are manifest symmetric with respect to the involutive symmetry $\iota$. 
  
%%%%%%%%%%%%%%
\subsection{Annulus instanton numbers for some torus knots}
%%%%%%%%%%%%%%
In order to check some results in the main text, we evaluate the following annulus numbers explicitly (for $1\le s<r$)
\begin{equation}
\begin{aligned}
   A_{0,0,h_1,h_2}^{(\mathcal{L}_{r,s},\mathcal{L}_{r,s})} \,&=\,  \left.\left[\quad  \xleftarrow{\ \ \ell(sh_1)\ \ }\!\!\!\!\bullet\!\!\!\!\xrightarrow{\ \ \ell(sh_2)\ \ } \quad \right]\right|_{s \la_z = -(r+s)\la_u}\ , \\[1ex]
   A_{0,1,h_1,h_2}^{(\mathcal{L}_{r,s},\mathcal{L}_{r,s})} \,&=\,   \left.\left[\quad
      \raisebox{-2ex}{$\xleftarrow{\ \ \ell(sh_1)\ \ }
      \!\!\!\!\bullet
      \!\!\rule[0.5ex]{0.5pt}{6ex}\!\!\!\;\mbox{\raisebox{6ex}{$\bullet$}}\!\mbox{\raisebox{3ex}{${}^{d=1}$}}\!\!\!\!\!\!\!\!\!\!
      \xrightarrow{\ \ \ell(sh_2)\ \ }$}
      \quad  \right] \right|_{s \la_z = -(r+s)\la_u}\ , \\[1ex]
    A_{0,1,h_1,h_2}^{(\mathcal{L}_{r,s},\iota_*\mathcal{L}_{r,s})} \,&=\, \left.\left[\quad 
       \xleftarrow{\ \iota_*\ell(sh_1)\ \ }\!\!\!\bullet\!\!{\buildrel{d=1}\over{\rule[0.5ex]{8ex}{.5pt}}}\!\!\bullet\!\!\!\xrightarrow{\ \ \ell(sh_2)\ \ } \quad \right]\right|_{s \la_z = -(r+s)\la_u} \ ,
\end{aligned}
\end{equation}
where $h_1$ and $h_2$ denote the winding numbers at the two boundary components. The first two amplitude enumerate annulus instantons with boundaries only on the branes $\mathcal{L}_{r,s}$ of degree zero and one, while the last amplitude calculate stretched annulus numbers for degree one between the branes $\mathcal{L}_{r,s}$ and $\iota_*\mathcal{L}_{r,s}$. 

Note also that the annulus numbers $A_{0,0,h_1,h_2}^{(\mathcal{L}_{r,s},\mathcal{L}_{r,s})}$ could be rational, if $h_1$ and $h_2$ have a non-trivial common multiple. In this case the instanton numbers are determined by taking into account the multi-covering contributions, which then enumerates the integral annulus instanton numbers. The other two amplitudes cannot contain any multi-covering contributions as they are all of degree one. As a result they are expected to directly yield the integral annulus numbers.

To easily compare with the numbers obtained from the calibrated annulus kernel, we actually present the numbers multiplied by $h_1\cdot h_2$, i.e., $\widetilde{A}_{g,d,h_1,h_2} = h_1h_2 A_{g,d,h_1,h_2}$.\footnote{The calibrated annulus kernel integrated to the annulus amplitude $A_2^{(0)}$ --- as for instance in eq.~\eqref{equ:AnnulusAmplitude} --- gives rise to the annulus instanton numbers~$A_{g,d,h_1,h_2}$. The expansion numbers from the calibrated annulus kernel are annulus instanton numbers multiplied by $h_1\cdot h_2$ and hence should be compared to $\widetilde{A}_{g,d,h_1,h_2}$.} For the branes $\mathcal{L}_{3,2}$, $\mathcal{L}_{5,2}$ and $\mathcal{L}_{4,3}$ (in the framing $6$, $10$, and $12$), respectively, we find then from the above graphs the explicit numbers
%%%%%%%%%%%%%%%%%%%%%%%%%%%%%%%%
\def\XLstrut{\vphantom{\rule[-1.5ex]{1pt}{4.5ex}}}
\begin{equation} \label{eq:locK23}
\vbox{\offinterlineskip
\halign{\strut\vrule~\hfil$#$~\vrule&~\hfil$#$~&~\hfil$#$~&~\hfil$#$~&~\hfil$#$~\vrule\cr
\noalign{\hrule}
\XLstrut\widetilde{A}_{0,0,h_1,h_2}^{(\mathcal{L}_{3,2},\mathcal{L}_{3,2})}&h_1\!=\!1&2&3&4\cr
\noalign{\hrule}
h_2\!=\!1 &60&1\,680& 45\,045&1\,209\,312\cr
2 &1\,680&52\,920&1\,513\,512&42\,325\,920\cr
3 &45\,045&1\,513\,512& 45\,090\,045&1\,296\,987\,120\cr
\noalign{\hrule}
\noalign{\phantom{X}}
\noalign{\hrule}
\XLstrut\widetilde{A}_{0,1,h_1,h_2}^{(\mathcal{L}_{3,2},\mathcal{L}_{3,2})}&h_1\!=\!1&2&3&4\cr
\noalign{\hrule}
h_2\!=\!1 &144&6\,048& 216\,216&7\,255\,872\cr
2 &6\,048 & 254\,016&9\,081\,072&304\,746\,624\cr
3 & 216\,216&9\,081\,072  &324\,648\,324&10\,894\,691\,808\cr
\noalign{\hrule}
\noalign{\phantom{X}}
\noalign{\hrule}
\XLstrut\widetilde{A}_{0,1,h_1,h_2}^{(\mathcal{L}_{3,2},\iota_*\mathcal{L}_{3,2})}&h_1\!=\!1&2&3&4\cr
\noalign{\hrule}
h_2\!=\!1 & -9 & -168 & -3\,861&-95\,472 \cr
2 & -168 & -3\,136 & -72\,072 &-1\,782\,144\cr
3 & -3\,861& -72\,072  & -1\,656\,369&-40\,957\,488\cr
\noalign{\hrule}
}}
\end{equation}
%%%%
\begin{equation}
\vbox{\offinterlineskip
\halign{\strut\vrule~\hfil$#$~\vrule&~\hfil$#$~&~\hfil$#$~&~\hfil$#$~&~\hfil$#$~\vrule\cr
\noalign{\hrule}
\XLstrut\widetilde{A}_{0,0,h_1,h_2}^{(\mathcal{L}_{5,2},\mathcal{L}_{5,2})}&h_1\!=\!1&2&3&4\cr
\noalign{\hrule}
h_2\!=\!1 &315&20\,020& 1\,220\,940&74\,594\,520\cr
2 &20\,020&1\,431\,430& 93\,117\,024&5\,926\,120\,200\cr
3 &1\,220\,940& 93\,117\,024& 6\,309\,817\,920&413\,040\,513\,600\cr
\noalign{\hrule}
\noalign{\phantom{X}}
\noalign{\hrule}
\XLstrut\widetilde{A}_{0,1,h_1,h_2}^{(\mathcal{L}_{5,2},\mathcal{L}_{5,2})}&h_1\!=\!1&2&3&4\cr
\noalign{\hrule}
h_2\!=\!1 &900&85\,800& 6\,976\,800&532\,818\,000\cr
2 &85\,800 & 8\,179\,600&  665\,121\,600&50\,795\,316\,000\cr
3 & 6\,976\,800 &665\,121\,600&54\,084\,153\,600&4\,130\,405\,136\,000\cr
\noalign{\hrule}
\noalign{\phantom{X}}
\noalign{\hrule}
\XLstrut\widetilde{A}_{0,1,h_1,h_2}^{(\mathcal{L}_{5,2},\iota_*\mathcal{L}_{5,2})}&h_1\!=\!1&2&3&4\cr
\noalign{\hrule}
h_2\!=\!1 & -25 & -1\,100 & -58\,140 &-3\,289\,000 \cr
2 & -1\,100 & -48\,400 & -2\,558\,160 &  -144\,716\,000\cr
3 & -58\,140& -2\,558\,160   & -135\,210\,384& -7\,648\,898\,400\cr
\noalign{\hrule}
}}
\end{equation}
%%%%
\begin{equation}
\vbox{\offinterlineskip
\halign{\strut\vrule~\hfil$#$~\vrule&~\hfil$#$~&~\hfil$#$~&~\hfil$#$~&~\hfil$#$~\vrule\cr
\noalign{\hrule}
\XLstrut\widetilde{A}_{0,0,h_1,h_2}^{(\mathcal{L}_{4,3},\mathcal{L}_{4,3})}&h_1\!=\!1&2&3&4\cr
\noalign{\hrule}
h_2\!=\!1 &1\,050&120\,120& 13\,226\,850&1\,460\,244\,240\cr
2 &120\,120&15\,459\,444&1\,815\,781\,968&208\,814\,926\,320\cr
3 &13\,226\,850&1\,815\,781\,968&222\,158\,172\,600&26\,278\,138\,130\,400\cr
\noalign{\hrule}
\noalign{\phantom{X}}
\noalign{\hrule}
\XLstrut\widetilde{A}_{0,1,h_1,h_2}^{(\mathcal{L}_{4,3},\mathcal{L}_{4,3})}&h_1\!=\!1&2&3&4\cr
\noalign{\hrule}
h_2\!=\!1 &3\,600&617\,760& 90\,698\,400&12\,516\,379\,200\cr
2 & 617\,760&106\,007\,616 & 15\,563\,845\,440&2\,147\,810\,670\,720 \cr
3 & 90\,698\,400 & 15\,563\,845\,440 & 2\,285\,055\,489\,600&315\,337\,657\,564\,800\cr
\noalign{\hrule}
\noalign{\phantom{X}}
\noalign{\hrule}
\XLstrut\widetilde{A}_{0,1,h_1,h_2}^{(\mathcal{L}_{4,3},\iota_*\mathcal{L}_{4,3})}&h_1\!=\!1&2&3&4\cr
\noalign{\hrule}
h_2\!=\!1 & -100 & -7\,920 & -755\,820 & -77\,261\,600\cr
2 & -7\,920  & -627\,264 & -59\,860\,944 & -6\,119\,118\,720\cr
3 & -755\,820& -59\,860\,944  & -5\,712\,638\,724&-583\,958\,625\,120\cr
\noalign{\hrule}
}}
\end{equation}
Note that the symmetry $\widetilde{A}_{0,1,h_1,h_2}^{(\mathcal{L}_{r,s},\iota_*\mathcal{L}_{r,s})}=\widetilde{A}_{0,1,h_2,h_1}^{(\mathcal{L}_{r,s},\iota_*\mathcal{L}_{r,s})}$ for the stretched annulus numbers in the tables demonstrates the involutive symmetry $\iota$.

%%%%%%%%%%%%%%%%%%%%%%%%%%%%%%%%%%%%%
\section{Composite representations and stretched annuli} \label{app:composite}

In Section~\ref{sec:stretched}, we found the generating function of stretched annulus amplitudes (\ref{eq:B23stretch}) from the B-model perspective. In this section, we calculate the stretched annulus amplitudes from the perspective of Chern--Simons theory. 

The most general irreducible representation of $U(N)$ is characterized by a pair of Young tableaux. The composite representation $[\mu,\nu]$ is then defined as (for a review on the subject see for instance \cite{Marino:2009mw})
\begin{equation}\label{eq:defcomp}
[\mu,\nu]\equiv\sum_{\rho,\alpha,\beta}(-1)^{|\rho|}\,N^{\mu}_{\rho\alpha}\,
N^{\nu}_{\rho^{t}\beta}\,(\alpha\otimes\bar{\beta})\ ,
\end{equation}
where $\bar{\beta}$ is the conjugate representation associated with representation $\beta$, $\rho^{t}$ denotes the transposed Young tableau associated with $\rho$, and $N^{\mu}_{\rho\alpha}$ are the standard Littlewood-Richardson coefficients. Notice that the sums in \eqref{eq:defcomp} include the empty partition as well. The composite representation $[\mu,\nu]$, in addition to the tensor product of the two representations $\mu$ and $\nu$, contains `corrections' from the lower representations. \par
Defining the composite representation in \eqref{eq:defcomp}, one can naturally compute the associated quantum dimension
\begin{equation}\label{qdimComp}
{\rm dim}_{q}[\mu,\nu]=\sum_{\rho,\alpha,\beta}(-1)^{|\rho|}\,N^{\mu}_{\rho\alpha}
\,N^{\nu}_{\rho^{t}\beta}\,{\rm dim}_{q}\alpha\,{\rm dim}_{q}\beta\ .
\end{equation}
Using the topological vertex, we can present a closed formula for all quantum dimensions of composite representations. We notice that the quantum dimension of $[\mu,\nu]$ corresponds to the normalized open-string amplitude in which one inserts a Lagrangian brane in an outer leg of the resolved conifold in representation $\mu$, and another Lagrangian brane in the other outer phase on the symmetric point-reflected leg in representation $\nu$. Computing the normalized open-string amplitude, we arrive at
\begin{equation}\label{qdimCompVert}
\begin{aligned}
{\rm dim}_{q}[\mu,\nu]&=Q^{-\frac12(|\mu|+|\nu|)}s_{\mu}(q^{-\rho})s_{\nu}(q^{-\rho})\,
\frac{\prod_{i=1}^{c_{\mu^{t}}}\prod_{j=1}^{c_{\nu^{t}}}(1-Q\,q^{i+j-\mu^{t}_{i}-
\nu^{t}_{j}-1}) }{ \prod_{i=1}^{c_{\mu^{t}}}\prod_{j=1}^{c_{\nu^{t}}}(1-Q\,q^{i+j-1})}\\
&\times \Big(\prod_{i=1}^{c_{\mu^{t}}}\prod_{j=c_{\nu^{t}}-\mu^{t}_{i}+1}^{c_{\nu^{t}}}
(1-Q\,q^{i+j-1})\Big)\,\Big(\prod_{i=1}^{c_{\nu^{t}}}\prod_{j=c_{\mu^{t}}-\nu^{t}_{i}+1}
^{c_{\mu^{t}}}(1-Q\,q^{i+j-1})\Big)\ ,
\end{aligned}
\end{equation}
where $c_{\mu}$ is the number of rows of the Young tableau associated with representation $\mu$, and $\mu_{i}$ is the $i-$th component of the partition $\mu=(\mu_{1},\mu_{2},\ldots,\mu_{c_{\mu}})$. Having the Schur-Weyl duality in mind, here we are using Greek letters to indicate both irreducible representations and the Young Tableau (partitions) associated with them. Although there are several apparent factors in the denominator of \eqref{qdimCompVert}, the final result is a polynomial in terms of $Q$, just as ordinary quantum dimensions.

Now, we would like to compute the HOMFLY polynomials of torus knots colored with composite representations. Recall that due to the Rosso--Jones formula \cite{MR1209320}, the HOMFLY invariant of an $(r,s)$ torus knot (with $rs$ units of framing) colored with an irreducible representation $\mu$ of $SU(N)$ can be computed via quantum dimensions as
\begin{equation}\label{Homf}
{\cal H}^{(r,s)}_{\mu}(Q,q)=\sum_{|\nu|=s|\mu|}c_{\mu,s}^{\nu}\,Q^{-\frac{r}{2s}|\mu|}\,q^{-\frac{r}{2s}\kappa_{\mu}}\,{\rm dim}_{q}\nu\ ,
\end{equation}
where $c_{\mu,s}^{\nu}$ is the coefficient of Adams operation and it can be easily computed by using the Frobenius formula
\begin{equation}\label{AdamCo}
c_{\mu,n}^{\nu}=\sum_{\ell(\vec{k})=|\mu|}\frac1{z_{\vec{k}}}\,\chi_{\mu}(C_{\vec{k}})\,\chi_{\nu}(C_{\vec{k}^{(n)}})\ .
\end{equation}
It turns out that the Rosso--Jones formula can be generalized to the case of composite representations as well. The HOMFLY polynomial of an $(r,s)$ torus knot in the composite representation $[\mu,\nu]$ -- with $rs$ units of framing -- is then expressed in terms of quantum dimensions in composite representations
\begin{equation}\label{HomComp}
{\cal H}_{[\mu,\nu]}^{(r,s)}(Q,q)=\sum_{\alpha,\beta}c_{[\mu,\nu],s}^{[\alpha,\beta]}\,
Q^{-\frac{r}{2s}(|\alpha|+|\beta|)}\,q^{-\frac{r}{2s}(\kappa_{\alpha}+\kappa_{\beta})}\,{\rm dim}_{q}[\alpha,\beta]\ .
\end{equation}
where the sums are over all partitions including the empty partition. In \eqref{HomComp},  $c_{[\mu,\nu],s}^{[\alpha,\beta]}$ are the coefficients of Adams operation for composite representations. Using the formulae in \cite{Kioke}, these coefficients can be explicitly calculated. They are given by
\begin{equation}\label{AdamComp}
c_{[\mu,\nu],n}^{[\alpha,\beta]}=\sum_{\tau,\lambda,\zeta,\xi,\rho,\sigma}
(-1)^{|\tau|}\,N^{\mu}_{\tau\zeta}\,N^{\nu}_{\tau^{t}\xi}\,c^{\rho}_{\zeta,n}\,
c^{\sigma}_{\xi,n}\,N^{\rho}_{\lambda\alpha}\,N^{\sigma}_{\lambda\beta}\ ,
\end{equation}
where again all sums are performed over all partitions, including the empty partition. Notice that $c_{[\mu,\nu],s}^{[\alpha,\beta]}$ can only be nonzero if $|\alpha|=js$ and $|\beta|=ks$ where $j\in\{0,1,\cdots,|\mu|\}$ and $k\in\{0,1,\cdots,|\nu|\}$. From \eqref{AdamComp}, it is easy to see that $c_{[\mu,\varnothing],s}^{[\alpha,\beta]}=c_{\mu,s}^{\alpha}\delta_{\beta,\varnothing}$ and $c_{[\varnothing,\nu],s}^{[\alpha,\beta]}=c_{\nu,s}^{\beta}\delta_{\alpha,\varnothing}$, and the Rosso-Jones formula \eqref{HomComp} for composite representations reduces to the ordinary Rosso-Jones formula \eqref{Homf}. Similar to~\eqref{Homf}, the Rosso-Jones formula \eqref{HomComp} is invariant under the exchange of $r$ and $s$.

Using \eqref{HomComp}, we can now work out all HOMFLY invariants of torus knots colored with composite representations. For instance for the trefoil knot, we expose the associated HOMFLY invariants colored by composite representations with at most three boxes. For composite representation $[\tableau{1},\tableau{1}]$, one finds
\begin{equation}\label{TreComp}
\begin{aligned}
{\cal H}_{[\tableau{1},\tableau{1}]}^{(3,2)}&=Q^{-3}q^{-3}{\rm dim}_{q}[\tableau{2},\tableau{2}]-Q^{-3}{\rm dim}_{q}[\tableau{2},\tableau{1 1}]-Q^{-3}{\rm dim}_{q}[\tableau{1 1},\tableau{2}]\\
&+Q^{-3}q^{3}{\rm dim}_{q}[\tableau{1 1},\tableau{1 1}]+{\rm dim}_{q}[\varnothing,\varnothing]\\
&=1+Q^{-3}\Big(q^{-3}{\rm dim}_{q}[\tableau{2},\tableau{2}]-2{\rm dim}_{q}[\tableau{2},\tableau{1 1}]+q^{3}{\rm dim}_{q}[\tableau{1 1},\tableau{1 1}]\Big)\ .
\end{aligned}
\end{equation}
For the composite representation $[\tableau{2},\tableau{1}]$, we find
\begin{equation}\label{TreCompsym}
\begin{aligned}
{\cal H}_{[\tableau{2},\tableau{1}]}^{(3,2)}&=Q^{-\frac32}q^{-\frac{21}{2}}\Big({\rm dim}_{q}[\tableau{4},\tableau{2}]-q^{3}{\rm dim}_{q}[\tableau{4},\tableau{1 1}]-q^{6}{\rm dim}_{q}[\tableau{3 1},\tableau{2}]+q^{9}{\rm dim}_{q}[\tableau{3 1},\tableau{1 1}]
\\
&+q^{9}{\rm dim}_{q}[\tableau{2 2},\tableau{2}]-q^{12}{\rm dim}_{q}[\tableau{2 2},\tableau{1 1}]+Q^{3}q^{9}{\rm dim}_{q}[\tableau{2},\varnothing]-Q^{3}q^{12}{\rm dim}_{q}[\tableau{1 1},\varnothing] \Big)\ ,
\end{aligned}
\end{equation}
while for representation $[\tableau{1 1},\tableau{1}]$, one has
\begin{equation}\label{TreCompansymm}
\begin{aligned}
{\cal H}_{[\tableau{1 1},\tableau{1}]}^{(3,2)}&=Q^{-\frac32}q^{-\frac32}\Big({\rm dim}_{q}[\tableau{2 2},\tableau{2}]-q^{3}{\rm dim}_{q}[\tableau{2 2},\tableau{1 1}]-q^{3}{\rm dim}_{q}[\tableau{2 1 1},\tableau{2}]+q^{6}{\rm dim}_{q}[\tableau{2 1 1},\tableau{1 1}]\\
&+q^{9}{\rm dim}_{q}[\tableau{1 1 1 1},\tableau{2}]-q^{12}{\rm dim}_{q}[\tableau{1 1 1 1},\tableau{1 1}]+Q^{3}{\rm dim}_{q}[\tableau{2},\varnothing]-Q^{3}q^{3}{\rm dim}_{q}[\tableau{1 1},\varnothing] \Big)\ .
\end{aligned}
\end{equation}
In order to compute higher winding stretched annulus amplitudes, we need to compute composite HOMFLY invariants in higher representations. For instance, for total winding four, we need to compute the composite HOMFLY invariants of the trefoil up to total four boxes. As before, these invariants are computed using (\ref{HomComp}). The composite invariant in totally symmetric representation with three boxes and the fundamental representation is given by 
\begin{equation}\label{TreComp31}
\begin{aligned}
{\cal H}_{[\tableau{3},\tableau{1}]}^{(3,2)}&=Q^{-3}q^{-24}\Big({\rm dim}_{q}[\tableau{6},\tableau{2}]-q^{3}{\rm dim}_{q}[\tableau{6},\tableau{1 1}]-q^{9}{\rm dim}_{q}[\tableau{5 1},\tableau{2}]\\
&+q^{12}{\rm dim}_{q}[\tableau{5 1},\tableau{1 1}]+q^{15}{\rm dim}_{q}[\tableau{4 2},\tableau{2}]-q^{18}{\rm dim}_{q}[\tableau{4 2},\tableau{1 1}]-q^{18}{\rm dim}_{q}[\tableau{3 3},\tableau{2}]\\
&+q^{21}{\rm dim}_{q}[\tableau{3 3},\tableau{1 1}]+Q^{3}q^{15}{\rm dim}_{q}[\tableau{4},\varnothing]-Q^{3}q^{21}{\rm dim}_{q}[\tableau{3 1},\varnothing]+Q^3 q^{24}{\rm dim}_{q}[\tableau{2 2},\varnothing]\Big)\ , \\
\end{aligned}
\end{equation}
while for the composite representation $[\tableau{2 1},\tableau{1}]$, we find the following invariant for the trefoil
\begin{equation}\label{TreComp21_1}
\begin{aligned}
{\cal H}_{[\tableau{2 1},\tableau{1}]}^{(3,2)}&=Q^{-3}q^{-9}\Big({\rm dim}_{q}[\tableau{4 2},\tableau{2}]-q^{3}{\rm dim}_{q}[\tableau{4 2},\tableau{1 1}]-q^3 {\rm dim}_{q}[\tableau{4 1 1},\tableau{2}]\\
&+q^6 {\rm dim}_{q}[\tableau{4 1 1},\tableau{1 1}]-q^3 {\rm dim}_{q}[\tableau{3 3},\tableau{2}]+q^6 {\rm dim}_{q}[\tableau{3 3},\tableau{1 1}]+q^{12}{\rm dim}_{q}[\tableau{3 1 1 1},\tableau{2}]\\
&-q^{15}{\rm dim}_{q}[\tableau{3 1 1 1},\tableau{1 1}]+q^{12}{\rm dim}_{q}[\tableau{2 2 2},\tableau{2}]-q^{15}{\rm dim}_{q}[\tableau{2 2 2},\tableau{1 1}]-q^{15}{\rm dim}_{q}[\tableau{2 2 1 1},\tableau{2}]\\
&+q^{18}{\rm dim}_{q}[\tableau{2 2 1 1},\tableau{1 1}]+Q^{3}{\rm dim}_{q}[\tableau{4},\varnothing]-Q^{3}q^{6}{\rm dim}_{q}[\tableau{3 1},\varnothing]\\
&+2Q^{3}q^{9}{\rm dim}_{q}[\tableau{2 2},\varnothing]-Q^{3}q^{12}{\rm dim}_{q}[\tableau{2 1 1},\varnothing]+Q^{3}q^{18}{\rm dim}_{q}[\tableau{1 1 1 1},\varnothing]\ . \\
\end{aligned}
\end{equation}
For the composite representation with totally antisymmetric three boxes and the fundamental representation, the composite invariant of the trefoil id given by
\begin{equation}\label{TreComp1111}
\begin{aligned}
{\cal H}_{[\tableau{1 1 1},\tableau{1}]}^{(3,2)}&=Q^{-3}\Big(q^3 {\rm dim}_{q}[\tableau{2 2 2},\tableau{2}]-q^6 {\rm dim}_{q}[\tableau{2 2 2},\tableau{1 1}]-q^6 {\rm dim}_{q}[\tableau{2 2 1 1},\tableau{2}]+q^9 {\rm dim}_{q}[\tableau{2 2 1 1},\tableau{1 1}]\\
&+q^{12} {\rm dim}_{q}[\tableau{2 1 1 1 1},\tableau{2}]-q^{15} {\rm dim}_{q}[\tableau{2 1 1 1 1},\tableau{1 1}]-q^{21} {\rm dim}_{q}[\tableau{1 1 1 1 1 1},\tableau{2}]+q^{24} {\rm dim}_{q}[\tableau{1 1 1 1 1 1},\tableau{1 1}]\\
&+Q^{3}{\rm dim}_{q}[\tableau{2 2},\varnothing]-Q^3 q^3 {\rm dim}_{q}[\tableau{2 1 1},\varnothing]+Q^3 q^9 {\rm dim}_{q}[\tableau{1 1 1 1},\varnothing]\Big)\ . \\
\end{aligned}
\end{equation}
For the composite representation $[\tableau{2},\tableau{2}]$, the HOMFLY invariant is given by
\begin{equation}\label{TreComp22}
\begin{aligned}
{\cal H}_{[\tableau{2},\tableau{2}]}^{(3,2)}&=Q^{-6}q^{-18}\Big({\rm dim}_{q}[\tableau{4},\tableau{4}]-2q^6 {\rm dim}_{q}[\tableau{4},\tableau{3 1}]+2q^9 {\rm dim}_{q}[\tableau{4},\tableau{2 2}]\\
&+q^{12}{\rm dim}_{q}[\tableau{3 1},\tableau{3 1}]-2q^{15}{\rm dim}_{q}[\tableau{3 1},\tableau{2 2}]+q^{18}{\rm dim}_{q}[\tableau{2 2},\tableau{2 2}]\\
&+Q^3 q^{15} {\rm dim}_{q}[\tableau{2},\tableau{2}]-2Q^3 q^{18}{\rm dim}_{q}[\tableau{2},\tableau{1 1}]+Q^3 q^{21}{\rm dim}_{q}[\tableau{1 1},\tableau{1 1}]+Q^6 q^{18}\Big)\ , \\
\end{aligned}
\end{equation}
while the composite representation in the totally symmetric and totally antisymmetric representations with two boxes is found 
\begin{equation}\label{TreComp2_11}
\begin{aligned}
{\cal H}_{[\tableau{2},\tableau{1 1}]}^{(3,2)}&=Q^{-6}q^{-9}\Big({\rm dim}_{q}[\tableau{4},\tableau{2 2}]-q^3 {\rm dim}_{q}[\tableau{4},\tableau{2 1 1}]+q^9 {\rm dim}_{q}[\tableau{4},\tableau{1 1 1 1}]-q^6 {\rm dim}_{q}[\tableau{3 1},\tableau{2 2}]\\
&+q^9 {\rm dim}_{q}[\tableau{3 1},\tableau{2 1 1}]-q^{15}{\rm dim}_{q}[\tableau{3 1},\tableau{1 1 1 1}]+q^9 {\rm dim}_{q}[\tableau{2 2},\tableau{2 2}]-q^{12}{\rm dim}_{q}[\tableau{2 2},\tableau{2 1 1}]\\
&+q^{18} {\rm dim}_{q}[\tableau{2 2},\tableau{1 1 1 1}]+Q^3 q^6 {\rm dim}_{q}[\tableau{2},\tableau{2}]-2Q^3 q^9 {\rm dim}_{q}[\tableau{2},\tableau{1 1}]+Q^3 q^{12} {\rm dim}_{q}[\tableau{1 1},\tableau{1 1}]\Big)\ . 
\end{aligned}
\end{equation}
The last composite invariant that we need to compute is the HOMFLY invariant of trefoil in the composite representation whose both components are totally antisymmetric representation with two boxes
\begin{equation}\label{TreComptwo1111}
\begin{aligned}
{\cal H}_{[\tableau{1 1},\tableau{1 1}]}^{(3,2)}&=Q^{-6}q^{-3}\Big(q^3 {\rm dim}_{q}[\tableau{2 2},\tableau{2 2}]-2q^6 {\rm dim}_{q}[\tableau{2 2},\tableau{2 1 1}]+2q^{12}{\rm dim}_{q}[\tableau{2 2},\tableau{1 1 1 1}]+q^9 {\rm dim}_{q}[\tableau{2 1 1},\tableau{2 1 1}]\\
&-2q^{15}{\rm dim}_{q}[\tableau{2 1 1},\tableau{1 1 1 1}]+q^{21}{\rm dim}_{q}[\tableau{1 1 1 1},\tableau{1 1 1 1}]+Q^3 {\rm dim}_{q}[\tableau{2},\tableau{2}]-2Q^3 q^3 {\rm dim}_{q}[\tableau{2},\tableau{1 1}]\\
&+Q^3 q^6 {\rm dim}_{q}[\tableau{1 1},\tableau{1 1}]+Q^6 q^3\Big)\ . \\
\end{aligned}
\end{equation}

The composite HOMFLY invariants enable us to compute all kinds of open-string stretched amplitudes. With the amplitudes computed above, we will be able to compute stretched amplitude with total winding of four. Since we are interested in  providing another independent check for the validity of the physical annulus kernel introduced in Section \ref{sec:curves}, we will first extract the stretched annulus amplitudes. Following the standard procedure for converting the A-model amplitudes to connected B-model amplitudes,  stretched annuli of trefoil  
\begin{equation}\label{StrAnn}
\begin{aligned}
A_{(1,1)}^{(f)}(Q)&=Q(-9 + 16\, Q - 9\, Q^2 + Q^4)\ ,\\
A_{(2,1)}^{(f)}(Q)&=2Q(Q-1)(-24 + 18 f + 78 Q - 41 f Q - 84 Q^2 + 34 f Q^2 + 24 Q^3\\
&- 9 f Q^3 + 12 Q^4 - 2 f Q^4 - 6 Q^5 + f Q^5)\ ,\\
A_{(2,2)}^{(f)}(Q)&=2Q(-128 + 855 Q - 2376 Q^2 + 3296 Q^3 - 2088 Q^4 - 9 Q^5 + 808 Q^6 \\
&- 432 Q^7 + 72 Q^8 + Q^9-24f(Q-1)^5 (Q-2)(Q^2+Q-4)\\
&+2f^2 (Q-2)^2 (Q-1)^2 ( Q^4- 9 Q^2+16 Q -9 ))\ ,\\
A_{(3,1)}^{(f)}(Q)&=\frac{3}{2}Q(Q-1)(6(Q-1)^3(-27 + 88 Q - 52 Q^2 - 16 Q^3 + 13 Q^4)\\
&-f(246 - 1242 Q+ 2399 Q^2 - 2209 Q^3+ 859 Q^4 + 59 Q^5 - 145 Q^6 \\
&+ 31 Q^7)+3f^2 (Q-1)(Q-2)^2  (-9 + 16 Q - 9 Q^2 + Q^4))\ .
\end{aligned}
\end{equation}
where $f$ is the framing of the two Lagrangians (we take them to be the same), and the subscript $(m,n)$ indicates the windings at the two boundaries of the corresponding stretched annulus amplitude.

We can also use the obtained result in this section for composite HOMFLY invariant in composite representation to extract the three-point function for few low windings and compare its planar part against the result obtained from the recursion in \eqref{equ:Stretched3pt}. For the first nontrivial winding the stretched three-point function at genus zero is found 
\begin{equation}\label{eq:Str3pta}
\begin{aligned}
A_{(1,1,\underline{1})}^{(f)}(Q)&=-Q(Q-1)(36 (Q-1)^4 (Q+2)-12 f(Q-1)^3 \\
&(Q^2+Q-4)+f^2( Q-2) (-9 + 16 Q - 9 Q^2 + Q^4))\ ,
\end{aligned}
\end{equation}
where the subscript $(m,n,\underline{k})$ indicates the windings of the three holes: the first two are the windings of the two holes on the brane ${\cal L}_{\cal K}$ and the third (and underlined) element indicates the winding of the hole on $\iota_* {\cal L}_{\cal K}$. For total winding of four, we have two ways two distribute the windings on the brane and the image brane. In the first case two holes of different windings are on the one side and one hole with winding one is on the other brane. The amplitude is then given by
\begin{equation}\label{eq:Str3ptb}
\begin{aligned}
A_{(1,2,\underline{1})}^{(f)}(Q)&=\frac{1}{3}Q(Q-1)(108 (Q-1)^4 (15 - 41 Q - Q^2 + 15 Q^3)\\
&-18f ( Q-1)^3 (-147 + 392 Q - 210 Q^2 - 72 Q^3 + 53 Q^4)\\
&+6f^2 (246 - 1242 Q + 2399 Q^2 - 2209 Q^3 + 859 Q^4 + 59 Q^5\\
& - 145 Q^6 + 31 Q^7)-12f^3 (Q-2)^2 (Q-1) (-9 + 16 Q\\
& - 9 Q^2 + Q^4))\ .
\end{aligned}
\end{equation}
In the second case, the two holes on the one side both have winding one, and the hole on the other brane has winding two. The corresponding stretched amplitude is found
\begin{equation}\label{eq:Str3ptc}
\begin{aligned}
A_{(1,1,\underline{2})}^{(f)}(Q)&=2Q(Q-1)^2(12 ( Q-1)^3 (16 - 46 Q - 12 Q^2 + 23 Q^3 + Q^4)\\
&-2f (Q-1)^2 (-136 + 356 Q - 154 Q^2 - 107 Q^3 + 58 Q^4 + Q^5)\\
&+18 f^2(Q-2) (Q-1)^3 (-4 + Q + Q^2)\\
&-f^3(Q-2)^2 (-9 + 16 Q - 9 Q^2 + Q^4))\ .
\end{aligned}
\end{equation}
%%%%%%%%%%%%%%%%%%%%%%%%%%%%%%

%%%%%%%%%%%%%%%%%%%
\section{$\alpha$--$\beta$ (a)symmery of $F^{(1)}$} \label{sec:SymVariance}

An intriguing feature of topological recursion is the symplectic invariance of the free energies $F^{(g)}$ \cite{Eynard:2013csa,Eynard:2007nq,Eynard:2007kz}. In particular this implies the invariance of $F^{(g)}$ under the exchange of the two meromorphic functions $\alpha$ and $\beta$. However all the past proofs were concerned with the original topological recursion formalism, and there are actually some straightforward counterexamples in the remodelled scenarios. Consider the spectral curve (\ref{eq:unknot}) of the mirror manifold of the resolved conifold with framing $f$. It is a Riemann sphere, on which $\beta$ can serve as a global coordinate. The projection to the $\beta$-plane has no ramification points, hence all the stable correlation differentials computed by (\ref{equ:TopRecursion}) vanish. As a consequence, the free energies defined by 
(\ref{equ:FreeEnergyVariation}) and (\ref{equ:F0Variation}) are zero as well. If one computes for instance %$F^{(0)}$ and 
$F^{(1)}$ using the projection to the $\alpha$-plane, the result is non-trivial, namely
\begin{equation}
	\frac{\partial F^{(1)}	}{\partial t} = 2\pi i\left( \frac{Q}{12(1-Q)} - \left( f+\frac{1}{f}-1	\right)\frac{1}{24} \right) \ .
\end{equation}
where $t$ is the complexified K\"{a}hler modulus of the resolved conifold.  Note the framing dependence is confined to the classical remnants in the second term, while the quantum piece in the first term is in accordance with the A-model computation.

So what's the cause of the breakdown of the invariance of $F^{(1)}$ under the exchange of $\alpha$ and $\beta$? We try to follow the proof of the invariance of $F^{(1)}$ in the original  topological recursion formalism \emph{without remodelling} in \cite{Eynard:2007kz}. The argument there is as follows. Let $\hat{F}^{(1)}$ and $\hat{\omega}^{(1)}_1(p)$  be the genus 1 free energy and genus 1 one-point correlator computed using $\beta$-plane projection. They also satisfy the variational formula
\begin{equation}
	\delta_\Omega \hat{F}^{(1)} = - \int_{\partial \Omega} \hat{\omega}^{(1)}_1(p)\Lambda(p) \ .
\end{equation}
Here a minus sign arises because the canonical 1-form $\hat{\Phi} =  \alpha \, d\beta$ with $\beta$-plane projection differs from the usual one in the original topological recursion, which is $\Phi = \beta \,d\alpha$, by a minus up to a total differential. Then one can represent the variation of the difference of the two free energies by
\begin{equation}\label{equ:F1difference}
  \delta_\Omega (F^{(1)} - \hat{F}^{(1)}) \,=\, 
   \int_{\partial \Omega} \left( \omega^{(1)}_1(p) + \hat{\omega}^{(1)}_1(p) \right) \Lambda(p)	 \ .	
\end{equation}
The goal is to prove this vanishes identically. It can be shown that both $\omega^{(1)}_1(p)$ and $\hat{\omega}^{(1)}_1(p)$ can be cast as the sum of residues of the same bilinear differential $f(p,q)$
\begin{align}
	\omega^{(1)}_1(p) =& -\sum_i \mathop{\Res}_{q \rightarrow a_i} f(p,q) \ , \\
	\hat{\omega}^{(1)}_1(p) =& -\sum_j \mathop{\Res}_{q \rightarrow b_j} f(p,q) \ ,
\end{align}
where $a_i$ are zeros of $d\alpha/\alpha$ and $b_j$ zeroes of $d\beta/\beta$. The bilinear differential $f(p,q)$ is
\begin{equation}
	f(p,q) = \mathop{\Res}_{r \rightarrow q} 	\frac{  \int_{\xi=r}^q B(p,\xi) }{ 2(\beta(q)-\beta(r)) (\alpha(q)-\alpha(r)) } B(q,r)
\end{equation}
Here comes the crucial step. One cuts open the spectral curve so that no ramification points are on the boundary of the fundamental domain $\mathcal{D}$. Then when $\omega^{(1)}_1(p)+\hat{\omega}^{(1)}_1(p)$ is evaluated one unwinds the contours and computes instead the residue of $f(p,q)$ at $q\rightarrow p$, which is the only other pole of $f(p,q)$ for $q$. Note the integration of $f(p,q)$ over $q$ along the boundary $\partial \mathcal{D}$ of the fundamental domain does not contribute because the segments of $\partial \mathcal{D}$ are pairwise identified with opposite orientations. Then
\begin{align}
	\omega^{(1)}_1(p)+\hat{\omega}^{(1)}_1(p)	&= \mathop{\Res}_{q \rightarrow p} \mathop{\Res}_{r \rightarrow q} 
	\frac{  \int_{\xi=r}^qB(p,\xi) }{ 2(\beta(q)-\beta(r)) (\alpha(q)-\alpha(r)) } B(q,r) \nonumber \\
	=& -\mathop{\Res}_{q\rightarrow p} \frac{B(q,p)}{( \beta(q)-\beta(p) )( \alpha(q) - \alpha(p) )}	\ .
\end{align}
The last line is argued in \cite{Eynard:2007kz} to vanish inside the integral in (\ref{equ:F1difference}). 

Let's see what happens to this argument in the remodelled formalism. We again cut open the Riemann surface so that the ramification points as well as the punctures are not on the boundary of the fundamental domain $\mathcal{D}$. In the spirit of the remodelling, one would expect $\omega^{(1)}_1(p)+\hat{\omega}^{(1)}_1(p)$ to be the sum of the residues of the following bilinear form $\tilde{f}(p,q)$ at all the ramification points $a_i, b_j$ of the spectral curve $\mathcal{C}$
\begin{equation}
	\tilde{f}(p, q) = \mathop{\Res}_{r \rightarrow q} 
	\frac{  \int_{\xi=r}^qB(p,\xi) }{ 2(\log \beta(q)- \log \beta(r)) ( \log \alpha(q)-\log \alpha(r)) } B(q,r) \ .
\end{equation}
However now there is a catch in the unwinding-contour-in-the-fundamental-domain argument. $q$ in $\tilde{f}(p, q)$ has poles not only at the ramification points and the point $p$ but possibly also at the punctures. Evaluate the residue inside $\tilde{f}(p, q)$ one gets the explicit expression
\begin{equation}\label{equ:hatfq}
	\tilde{f}(p, q) = \frac{B(p,q) \beta(q) \alpha'(q)}{24 \alpha(q) \beta'(q)} + \frac{B(p,q) \alpha(q) \beta'(q)}{24 \beta(q) \alpha'(q)}  + \ldots \ .		
\end{equation}
Terms in $\cdots$ only have poles at the ramification points or $p$, and the derivative $'$ is with respect to an arbitrary local affine coordinate. Equation (\ref{equ:hatfq}) shows that $\tilde{f}(p,q)$ has a pole at a puncture if a puncture (zero or pole) of $\alpha$ but not a puncture of $\beta$ or the other way around. We call it an \emph{asymmetric puncture}. When this is the case, after unwinding the contours around ramification points $\omega^{(1)}_1(p)+\hat{\omega}^{(1)}_1(p)$ has extra piece coming from the residue of $\tilde{f}(p,q)$ at the asymmetric puncture, which invalidates the argument for the invariance of $F^{(1)}$ under the exchange of the projection planes.

\begin{figure}
\centering
\subfloat[Framing 0]{\includegraphics[width=0.32\textwidth]{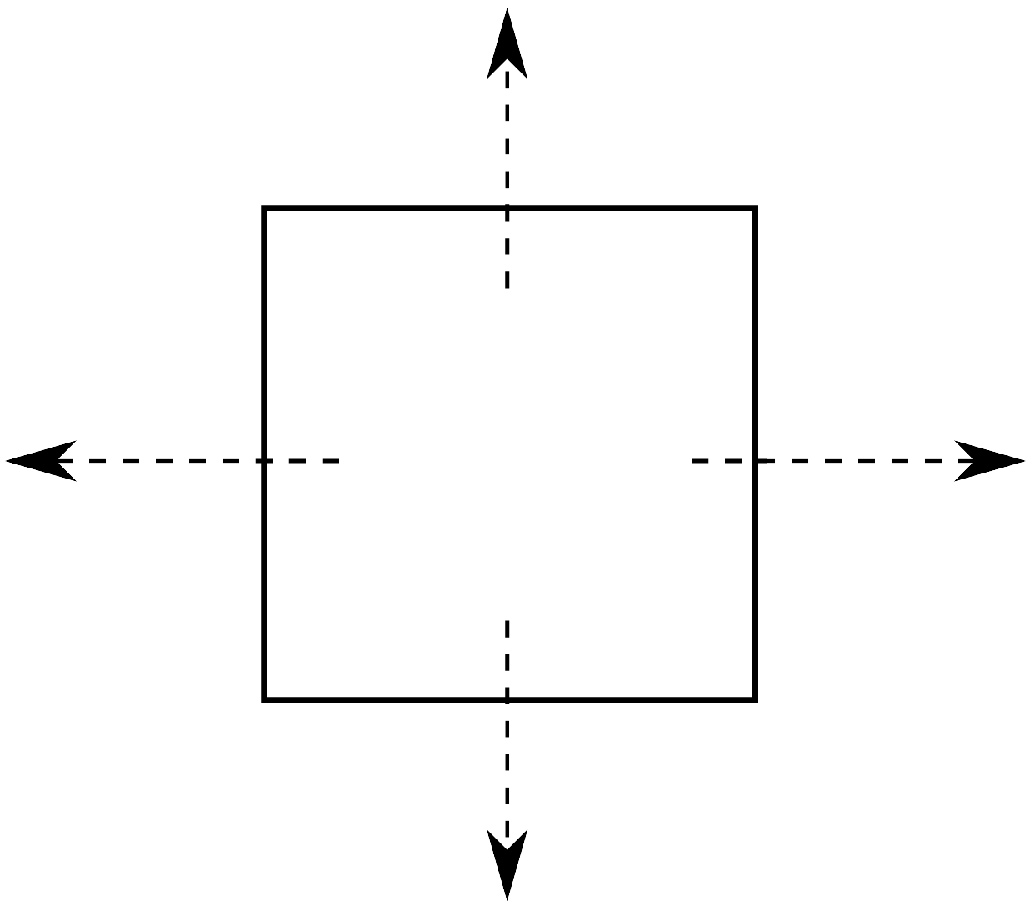}}
\subfloat[Framing 1]{\includegraphics[width=0.25\textwidth]{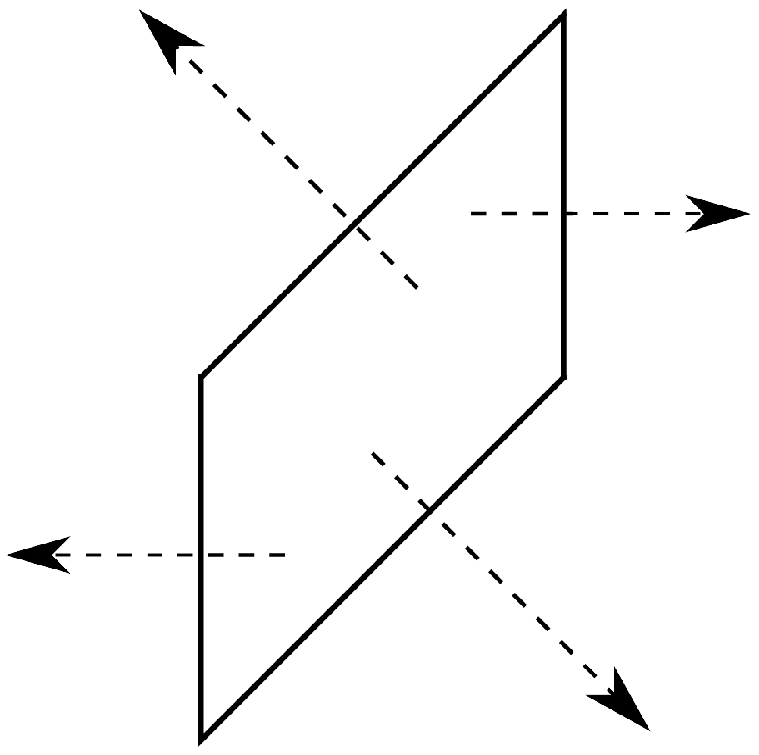}}
\subfloat[Framing 3]{\includegraphics[width=0.17\textwidth]{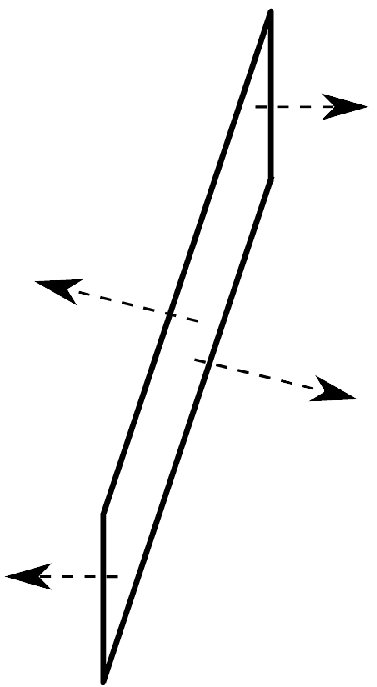}}
\caption{Newton polytopes of the spectral curves of the unknot with different choices of framing}\label{fig:UnknotNewtonPolytope}
\end{figure}

The appearance of asymmetric punctures can be readily seen by the Newton polytope. Recall the definition of Newton polytope. Let the monomials in the spectral curve $H(\alpha,\beta)$ be $\alpha^i \beta^j$, then the Newton polytope is the convex hull of the set of points $(i,j)$ in a 2d integral lattice, $i$ increasing in the horizontal direction and $j$ increasing in the vertical direction. Since we only care about the punctures of the curve, we do not triangulate the Newton polytope. Instead we just draw a ray orthogonal to each segment of the boundary of the Newton polytope pointing outwards. Each ray corresponds to a puncture of the spectral curve. Furthermore, these rays can be identified in the following way:
\begin{itemize}
\item Pointing to the left: zeros of $\alpha$;
\item Pointing to the right: poles of $\alpha$;
\item Pointing down: zeros of $\beta$;
\item Pointing up: poles of $\beta$.
\end{itemize}
If a ray is vertical, it is only a puncture of $\beta$ but not a puncture of $\alpha$, hence an asymmetric puncture. Similarly if a ray is horizontal, it corresponds to a puncture in $\alpha$ only and also an asymmetric puncture. In the case of the spectral curve of the resolved conifold with an arbitrary framing, there are always two rays horizontal (Figure~\ref{fig:UnknotNewtonPolytope}), so $F^{(1)}$'s computed with $\alpha$-plane projection and $\beta$-plane projection are never the same. 

Finally we want to show another viewpoint towards the asymmetric punctures, which is related to the observation made by Bouchard and Su\l kowski in ref.~\cite{Bouchard:2011ya}. Consider the meromorphic 1-forms $\omega_\alpha = d\alpha/\alpha$  and $\omega_\beta = d\beta/\beta$. The punctures of the 1-form $\omega_\alpha$ are the zeros and poles of the meromorphic function $\alpha$. It is known that the number of zeroes of $\omega_\alpha$ minus the number of poles of $\omega_\alpha$ is $2g-2$. In the generic case when no asymmetric puncture appears, viz. no ray in the Newton polytope is either horizontal or vertical, the poles of $\omega_\alpha$ include all the punctures of the spectral curve, in which case the number of ramification points with respect to $\alpha$, i.e. the number of zeroes of $\omega_\alpha$, gets maximized. The same thing can be said of the form $\omega_\beta$. If there is an asymmetric puncture, say, a ray is horizontal, the number of poles of $\omega_\beta$ as well as the number of ramification points with respect to $\beta$ is reduced, which can be interpreted as some ramification points being sent to the asymmetric punctures, giving rise to the residues of $\tilde{f}(p,q)$ at the latter positions. This is similar to the scenarios discussed in ref.~\cite{Bouchard:2011ya}. It is revealed there that the framing independence of $F^{(g)}, g>2$, which is interpreted as a type of symplectic transformation as well, is destroyed at some framings where the number of ramification points is reduced. 
To conclude, in the remodelled scenario, $F^{(1)}$ can be different under the exchange of $\alpha$ and $\beta$ if asymmetric punctures arise. However, we know that in computing $\omega^{(0)}_{1}$ the right `open-string' coordinate is the one which gives us the disks in considered region of the moduli space. Therefore, the $F^{(1)}$ which is computed by projecting onto the right disk coordinate should be the `legitimate' one.

%%%%%%%%%%%%%%%%%%%
\section{Propositions and Proofs}\label{sec:Props}

Here we collect some propositions used in the main text and their proofs.

\subsection{Position of poles of calibrated annulus kernel}\label{sec:PolePositions}

In Section~\ref{sec:PoleStructure} we use the statement that the zero loci of $\Theta(q_i)-\Theta(\bar{q_i})$ and the pole loci of $\widehat{B}_{r,s}(q_i,q_j)$ intersect only at the tuples $(a_{i_1},\cdots,a_{i_h})$, where $a_{i_k}$ are ramification points of the augmentation curve. This is due to the following lemma.

\begin{lem}\label{prop:PoleStruture}
Let the second point of the annulus kernel $\widehat{B}_{r,s}(q_1,q_2)$ be fixed at a ramification point $a_i$ of the augmentation curve $F_{r,s}(\alpha,\beta;Q)$. Then the kernel only has (double) poles when $q_1$ approaches a ramification point, which may be different from $a_i$.
\end{lem}

\noindent \textbf{Proof}. Recall from the discussion in Section~\ref{sec:PoleStructure} that when $q_2$ in $\widehat{B}_{r,s}(q_1,q_2)$ approaches a ramification point of the augmentation curve $F_{r,s}(\alpha,\beta;Q)$, one $\rho$-component of $\beta$, say, $\rho^{(\ell_1)}$ approaches a ramification point of the auxiliary curve $h_{r,s}(\zeta,\rho)$, while the other $\rho$ components which correspond to the same $\zeta$ value do not. Then 
\[		0=\frac{\zeta'(\rho^{(\ell_1)})}{ \zeta'(\rho^{(\ell_k)}) } = \frac{d\rho^{(\ell_k)}}{d \rho^{(\ell_1)}}	, \quad k=2,\ldots r \ ,	\]
at a ramification point, meaning that $d\rho^{(\ell_k)} = 0,k=2,\ldots,r$. Therefore
\begin{align*}
	\widehat{B}_{r,s}(q_1,a_i) &= \sum_{m,n=1}^r \frac{r^2 \rho^{(\ell_m)}(q_1)^{r-1} \rho^{(\ell_n)}(q_2)^{r-1} d\rho^{(\ell_m)}(q_1) d\rho^{(\ell_n)}(q_2) }{(\rho^{(\ell_m)}(q_1)^r-\rho^{(\ell_n)}(q_2)^r)^2} -\frac{(r-1)d\alpha(q_1) d\alpha(q_2)}{(\alpha(q_1)-\alpha(q_2))^2}\Big|_{q_2 \rightarrow a_i}\\
	&= \sum_{m=1}^r \frac{r^2 \rho^{(\ell_m)}(q_1)^{r-1} \rho^{(\ell_1)}(q_2)^{r-1} d\rho^{(\ell_m)}(q_1) d\rho^{(\ell_1)}(q_2) }{(\rho^{(\ell_m)}(q_1)^r-\rho^{(\ell_1)}(q_2)^r)^2} \Big|_{q_2 \rightarrow a_i} \ .
\end{align*}

Now for $\widehat{B}_{r,s}(q_1,a_i)$ to develop a pole, one of $\rho^{(\ell_m)}(q_1)$ has to approach $\rho^{(\ell_1)}(a_i)$. If only one $\rho^{(\ell_m)}(q_1)$ component of $\beta(q_1)$ approaches the ramification point $\rho^{(\ell_1)}(a_i)$ one has the familiar case where $\beta(q_1)$ itself is a ramification point on the augmentation curve $F_{r,s}(\alpha,\beta;Q)$. If however two $\rho^{(\ell_m)}(q_1)$ components of $\beta(q_1)$ approach the ramification point $\rho^{(\ell_1)}(a_i)$ on the auxiliary curve $h_{r,s}(\zeta,\rho)$, which is the only other possibility,  then these two $\rho^{(\ell_m)}(q_1)$ components are actually conjugate points to each other, and one can show by local computation that their contributions to the principal part of $\widehat{B}_{r,s}(q_1,a_i)$ cancel. \qed

\subsection{Rauch variational formula for calibrated annulus kernel}\label{sec:RauchCalibratedKernel}

\begin{lem}[Rauch variational formula]\label{lem:Rauch}
Given the augmentation curve of a $(r,s)$ torus knot with canonical framing $rs$, the annulus kernel $\widehat{B}_{r,s}(p_1,p_2)$ satisfies 
\begin{equation}\label{equ:RauchTorus}
	\delta_\Omega \widehat{B}_{r,s}(p_1,p_2)\Big|_{\alpha(p_1),\alpha(p_2)} = \frac{1}{N_{r,s}}\sum_i\mathop{\Res}_{q\rightarrow a_i} \frac{\Omega(q) \widehat{B}_{r,s}(p_1,q) \widehat{B}_{r,s}(p_2,q) \alpha(q)\beta(q)}{ d\alpha(q)d\beta(q)}
\end{equation}
where the normalization factor $N_{r,s}$ is
\begin{equation}
	N_{r,s} = {r+s-2 \choose r-1} \ .
\end{equation}
\end{lem}

\noindent \textbf{Proof}. Since the annulus generating kernel $\widehat{B}_{r,s}(p_1,p_2)$ is the sum of Bergman kernels, the natural starting point is the Rauch variational formula for the Bergman kernel. We use the construction of $\widehat{B}_{r,s}(p_1,p_2)$ in \eqref{eq:BphysNew} and denote $B^{(\ell_m,\ell_n)}_{r,s}(\zeta_1,\zeta_2)d\zeta_1 d\zeta_2$ also by $B^h_{r,s}(\tilde{p}_1^{(\ell_m)}, \tilde{p}_2^{(\ell_n)})$ to stress the kernel is defined on the auxiliary curve $h_{r,s}(\zeta,\rho)$. The Bergman kernel $B^{h}_{r,s}(\tilde{p}^{(\ell_m)}_1,\tilde{p}^{(\ell_n)}_2)$ is still not in the standard form. To achieve the latter, we consider instead of $h_{r,s}(\zeta,\rho)$ the genus zero Riemann surface
\begin{equation}
	H_{r,s}(X,Y) = (1-Q Y)^r - X Y^s(1-Y)^r \ ,
\end{equation}
which is related to $h_{r,s}(\zeta,\rho)$ by 
\begin{equation}
	\hat{p}(X,Y):\quad X = \zeta^r,\quad Y= \rho^r.
\end{equation}
This curve has two ramification points $\hat{a}_a,\hat{a}_b$ corresponding to the two sets of ramification points of the curve $h_{r,s}(\zeta,\rho)$, namely, $(X_a,Y_a) = (\zeta_a^r,\rho_a^r), (X_b,Y_b) = (\zeta_b^r,\rho_b^r)$, $(\zeta_a,\rho_a)$ and $(\zeta_b,\rho_b)$ being the coordinates of the representatives $\tilde{a}_a,\tilde{a}_b$ of the two sets of ramification points on $h_{r,s}(\zeta,\rho)$ respectively. Changing the representative does not change the point on $H_{r,s}(X,Y)$. (A detailed discussion on the ramification points on curve $h_{r,s}(\zeta,\rho)$ is given in Appendix~\ref{sec:RamPoints}). The standard Bergman kernel of this curve is precisely equal to $B_{r,s}^{h}(\tilde{p}^{(\ell_m)}_1,\tilde{p}^{(\ell_n)}_2)$
\begin{equation}
	B_{r,s}^{H}(\hat{p}^{(\ell_m)}_1,\hat{p}^{(\ell_n)}_2) = \frac{dY(\hat{p}^{(\ell_m)}_1) dY(\hat{p}^{(\ell_n)}_2)}{\left(Y(\hat{p}^{(\ell_m)}_1)-Y(\hat{p}^{(\ell_n)}_2)\right)^2} =  B_{r,s}^{h}(\tilde{p}^{(\ell_m)}_1,\tilde{p}^{(\ell_n)}_2) \ ,
\end{equation}
and it satisfies the Rauch variational formula
\begin{equation}
	\delta  B_{r,s}^{H}(\hat{p}^{(\ell_m)}_1,\hat{p}^{(\ell_n)}_2)\Big|_{X(\hat{p}^{(\ell_m)}_1),X(\hat{p}^{(\ell_n)}_2)} 
	= \sum_{\mu=a,b}\mathop{\Res}_{\hat{q}\rightarrow \hat{a}_\mu}\frac{\Omega^H(\hat{q}) B_{r,s}^{H}(\hat{p}^{(\ell_m)}_1,\hat{q}) B_{r,s}^{H}(\hat{p}^{(\ell_n)}_2,\hat{q}) }{ dX(\hat{q}) dY(\hat{q})	}		\label{equ:BHRauch}
\end{equation}
where the variation 1-form $\Omega^H(q)$ is 
\[		\Omega^H(\hat{q}) = \delta Y(\hat{q})|_{X(\hat{q})} d X(\hat{q}) \ .	\]
Let us compute the left hand side of (\ref{equ:RauchTorus}):
\begin{align}
	\delta B_{r,s}(p_1,p_2)\Big|_{\alpha(p_1),\alpha(p_2)} =&	\sum_{m,n=1}^r \delta B_{r,s}^{H}(\hat{p}^{(\ell_m)}_1,\hat{p}^{(\ell_n)}_2)-(r-1)\delta\left(\frac{d\alpha_1 d\alpha_2}{(\alpha_1-\alpha_2)^2}\right)\Big|_{X(\hat{p}^{(\ell_m)}_1),X(\hat{p}^{(\ell_n)}_2)} \nonumber\\
	=& \sum_{m,n=1}^r	\sum_{\mu=a,b}\mathop{\Res}_{\hat{q}\rightarrow \hat{a}_\mu} \frac{\Omega^H(\hat{q}) B_{r,s}^{H}(\hat{p}^{(\ell_m)}_1,\hat{q}) B_{r,s}^{H}(\hat{p}^{(\ell_n)}_2,\hat{q})  }{dX(\hat{q}) dY(\hat{q})}	\nonumber\\
	=& \sum_{m,n=1}^r	\sum_{\mu=a,b}\mathop{\Res}_{\tilde{q}\rightarrow \tilde{a}_\mu} \frac{ \delta \rho|_{\zeta(\tilde{q})} d\zeta(\tilde{q}) B_{r,s}^{h}(\tilde{p}^{(\ell_m)}_1,\tilde{q}) B_{r,s}^{h}(\tilde{p}^{(\ell_n)}_2,\tilde{q})  }{d\rho(\tilde{q}) d\zeta(\tilde{q})}		\label{equ:RauchTorusLHS}
\end{align}
From the first line to the second we use $\delta \alpha|_{X} = 0$, and (\ref{equ:BHRauch}). Next we compute the right hand side of (\ref{equ:RauchTorus}), the identity we want to prove. Recall the variation on the augmentation curve $F_{r,s}(\alpha,\beta;Q)$ is given by
\begin{equation}
	\Omega(q) = \delta \log(\beta) |_\alpha d\alpha/\alpha =  \frac{\delta \beta|_\alpha d\alpha}{\alpha\beta}
\end{equation}

The integrand on the right hand side of (\ref{equ:RauchTorus}) is 
\begin{equation}\label{equ:Integrand}
 \frac{\Omega(q) \widehat{B}_{r,s}(p_1,q) \widehat{B}_{r,s}(p_2,q) \alpha(q)\beta(q)}{d\alpha(q) d\beta(q)}
	\,=\, \frac{ \delta \beta(q)|_{\alpha(q)} d\alpha(q) }{d\alpha(q) d\beta(q)} \widehat{B}_{r,s}(p_1,q) \widehat{B}_{r,s}(p_2,q) \ .	
\end{equation}
We know this integrand has only simple pole at a ramification point, so that when the residue is evaluated, only the leading terms in the numerator and denominator of the integrand contribute. Furthermore, recall that $\beta = (-1)^{r+1}\rho^{(\ell_1)}\rho^{(\ell_2)}\cdots\rho^{(\ell_r)}$ and when it approaches a ramification point of the augmentation curve $F_{r,s}(\alpha,\beta;Q)$, one of the $\rho$-components, say, $\rho^{(\ell_1)}$ approaches a ramification point of the auxiliary curve $h_{r,s}(\zeta,\rho)$, while the other $\rho^{(\ell_k)}, k=2,\ldots,r$ do not, which nevertheless correspond to the same $\zeta$. Then
\[		\frac{d\rho^{(\ell_k)}}{d \rho^{(\ell_1)}}=\frac{\zeta'(\rho^{(\ell_1)})}{ \zeta'(\rho^{(\ell_k)}) } \rightarrow 0,\quad \rho^{(\ell_1)} \textrm{ approaches a ramification point}		\]
Therefore, when $\rho^{(\ell_1)}$ approaches a ramification point, the other $d\rho^{(\ell_k)},k=2,\cdots,r$ are infinitesimally small compared to $d\rho^{(\ell_1)}$, and hence can be dropped. 

Now let us consider each piece on the right hand side of eq.~\eqref{equ:Integrand}. The first piece to be considered is
\[		d\beta = (-1)^{r+1} \left(\rho^{(\ell_2)}\cdots\rho^{(\ell_r)} d\rho^{(\ell_1)}+ \cdots +\rho^{(\ell_1)}\cdots\rho^{(\ell_{r-1})}d\rho^{(\ell_r)} \right) \ .	\]
Its leading term is $(-1)^{r+1} \rho^{(\ell_2)}\cdots\rho^{(\ell_r)} d\rho^{(\ell_1)}$. The next piece to be considered is
\[		\delta\beta|_\alpha d\alpha = (-1)^{r+1} \left(\rho^{(\ell_2)}\cdots\rho^{(\ell_r)}\delta\rho^{(\ell_1)}|_\zeta+ \cdots +\rho^{(\ell_1)}\cdots\rho^{(\ell_{r-1})}\delta\rho^{(\ell_r)}|_\zeta \right) r \zeta^{r-1}d\zeta \	.		\]
Now given the auxiliary curve $h_{r,s}(\zeta,\rho; Q)$ where $t=\frac{1}{2\pi i} \log Q$ the complexified K\"{a}hler modulus is the actual parameter of the curve we are going to vary, the variation $\delta\rho|_\zeta d\zeta$ is given by
\begin{equation}	
			\delta h_{r,s} = \frac{\partial h_{r,s}}{\partial \rho} \delta \rho|_\zeta + \frac{\partial h_{r,s}}{\partial t} \delta t = 0 
			\quad \Rightarrow \quad
			\delta\rho|_\zeta d\zeta = -\frac{\partial_t h_{r,s}}{\partial_\rho h_{r,s}} \delta t d\zeta
			= \frac{\partial_t h_{r,s}}{\partial_\zeta h_{r,s}} \delta t d\rho \ .
\end{equation}
Therefore, $\delta \rho^{(\ell_k)} d\zeta \propto d\rho^{(\ell_k)}, k=1,\cdots,r$, and the leading contribution from $\delta \beta|_\alpha d\alpha$ is the first term $(-1)^{r+1}r \zeta^{r-1} \rho^{(\ell_2)}\cdots\rho^{(\ell_r)} \delta \rho^{(\ell_1)}|_\zeta d\zeta$. The third piece we consider is
\begin{align*}
	\widehat{B}_{r,s}(p_1,q)\Big|_{q\rightarrow a_i} =& \sum_{m,h=1}^r B^{h}_{r,s}(\tilde{p}^{(\ell_m)}_1, \tilde{q}^{(\ell_h)}) -(r-1)\frac{d\alpha(p_1) d\alpha(q)}{(\alpha(p_1)-\alpha(q))^2} \Big|_{\tilde{q}^{(\ell_1)}\rightarrow\tilde{a}_i, q\rightarrow a_i  }		\\
	=& \sum_{m,h=1}^r \frac{r^2  (\rho^{(\ell_m)}_1)^{r-1} (\rho^{(\ell_h)})^{r-1} d\rho^{(\ell_m)}_1 d\rho^{(\ell_h)} }{ \left( (\rho^{(\ell_m)}_1)^{r} - (\rho^{(\ell_h)})^{r}  \right)^2 }\Big|_{\tilde{q}^{(\ell_1)}\rightarrow\tilde{a}_i} \ ,
\end{align*}
where $\rho^{(\ell_m)}_1=\rho(\tilde{p}^{(\ell_m)}_1)$ and $\rho^{(\ell_h)} = \rho(\tilde{q}^{(\ell_h)})$. The leading contributions from the annulus kernel $\widehat{B}_{r,s}(p_1,q)$ are $\sum_{m=1}^r B^{h}_{r,s}(\tilde{p}^{(\ell_m)}_1,\tilde{q}^{(\ell_1)})$. Likewise, the leading contributions from $\widehat{B}_{r,s}(p_2,q)$ are $\sum_{n=1}^r B^{h}_{r,s}(\tilde{p}^{(\ell_n)}_2,\tilde{q}^{(\ell_1)})$. Putting everything together, the residue on the right hand side of (\ref{equ:RauchTorus}) is found to be
\begin{equation}\label{equ:RauchTorusRHS}
	\mathop{\Res}_{\tilde{q}^{(\ell_1)}\rightarrow \tilde{a}_i} \sum_{m,n=1}^r \frac{\delta \rho(\tilde{q}^{(\ell_1)})|_{\zeta(\tilde{q}^{(\ell_1)})}  d\zeta(\tilde{q}^{(\ell_1)}) }{d\zeta(\tilde{q}^{(\ell_1)}) d\rho(\tilde{q}^{(\ell_1)})} B^{h}_{r,s}(\tilde{p}^{(\ell_m)}_1, \tilde{q}^{(\ell_1)}) B^{h}_{r,s}(\tilde{p}^{(\ell_n)}_2, \tilde{q}^{(\ell_1)})		\ .		
\end{equation}

Finally identify $\tilde{q}^{(\ell_1)}$ in (\ref{equ:RauchTorusRHS}) with $\tilde{q}$ in (\ref{equ:RauchTorusLHS}), and recall from the proof of Proposition~\ref{prop:RamPoints} that ${r+s-2\choose r-1}$ ramification points on the augmentation curve correspond to a single set of ramification points on the auxiliary curve, one can arrive at the desired identity (\ref{equ:RauchTorus}).		\qed

\subsection{The number of ramification points of augmentation curve}\label{sec:RamPoints}

\begin{prop}\label{prop:RamPoints}
Given the augmentation curve $F_{r,s}(\alpha,\beta;Q)$ of a torus knot $(r,s)$ with canonical framing $rs$, the number of ramification points, viz. the zero of $d\alpha/\alpha$, for a generic value of $Q$ is
\begin{equation}
	2 {r+s-2\choose r-1}\ .
\end{equation}
\end{prop}

\noindent \textbf{Proof}. We recall that a ramification point on the augmentation curve $F_{r,s}(\alpha,\beta;Q)$ with canonical framing $rs$ is described by \eqref{equ:AugPolyRam} and that a ramification point of the augmentation curve corresponds to a ramification point of the auxiliary curve as well. In order to study the latter, we solve $\zeta$ as a function of $\rho$ from the auxiliary curve,
\begin{equation}
	\zeta(\rho) = \frac{\rho ^{-s} \left(Q \rho ^r-1\right)}{\rho ^r-1} \ .
\end{equation}
$d\zeta/\zeta=0$ requires
\begin{equation}
	Q s \rho ^{2 r}+\rho ^r (Q r-Q s-r-s)+s = 0\ .
\end{equation}
which for a generic value of $Q$ has $2r$ solutions. The $2r$ ramification points of the auxiliary curve $h_{r,s}(\zeta,\rho)$ can be divided into two groups $S_a, S_b$. Within each group the values of $\rho$ and hence the values of $\zeta$ as well differ only by a phase shift. Choose a representative point $(\zeta_a,\rho_a)$ in $S_a$, out of which the way to construct a ramification point on the augmentation curve is as follows. Let $\alpha$ be $(\zeta_a)^r$, and let $\rho^{(\ell_1)}, \rho^{(\hat\ell_1)}$ in (\ref{equ:AugPolyRam}) approach $\rho_a$. $\rho^{(\ell_2)},\ldots,\rho^{(\ell_r)}$ can be any $r-1$ different $\rho^{(k)}$'s chosen from the rest $(r+s-2)$ $\rho^{(k)}$ associated with $\zeta_a$. So from $(\zeta_a,\rho_a)$ alone we can find ${r+s-2}\choose{r-1}$ ramification points on the augmentation curve, denoted by $(\alpha_a,\beta_{a,i}), i=1,\ldots, {{r+s-2}\choose{r-1}}$, where all share the same branch point $\alpha_a$. For the other points on the auxiliary curve in the group $S_a$ the only difference is a phase shift $\eta$, the $r$-th root of unity, in $\zeta_a$ and a corresponding phase shift in $\rho^{(\ell_k)},k=1,\ldots,r$, which do not change either $\alpha$ or $\beta$. Therefore, no new ramification points on the augmentation curve can be found. The case is similar for the group $S_b$. With a representative $(\zeta_b,\rho_b)$ we can find another ${r+s-2}\choose{r-1}$ ramification points on the augmentation curve, denoted by $(\alpha_b,\beta_{b,i}), i=1,\ldots,{{r+s-2}\choose{r-1}}$, where all project onto the same branch point $\alpha_b$ on the $\alpha$ plane. Other members of $S_b$ bring no new ramification point. Therefore, the total number of ramification points on the augmentation curve for torus knot $(r,s)$ is $2\cdot {{r+s-2}\choose{r-1}}$.
\qed

\bigskip
Note that although this proposition is proved for the canonical framing $rs$, according to the discussion at the end of Appendix~\ref{sec:SymVariance}, the number of ramification points does not change when the framing changes, since the number of punctures of the curve stays fixed, as one can easily see from the Newton polytope, except at some critical framing where some segments of the boundary of the Newton polytope become horizontal. Therefore the number of ramification points for a generic framing is also $2\cdot {r+s-2\choose r-1}$.

\section{Augmentation curves of some non-torus knots}\label{sec:AugPoly}
Further augmentation curves are listed on Ng's website~\cite{NgWeb} in terms of the multiplicative generators $\lambda, \mu$ and $U$ introduced in ref.~\cite{MR2807087}.\footnote{Note that this choice of generators differs a bit from ref.~\cite{Ng:2012qq}; see also the discussion of conventions in the appendix of this paper.} They are related to the generators $\alpha, \beta$ and $U$ according to
\begin{equation} \label{eq:Conversion}
	U = 1/Q\ , \quad \lambda = \alpha \ , \quad \mu = - 1/\beta \ .
\end{equation}
In terms of the generators $\alpha, \beta$ and $Q$, the following augmentation polynomials (in framing zero) read for the figure-eight knot
\begin{equation} \label{eq:Feight}
\begin{aligned}
	F_{4_1}(\alpha,\beta;Q) &= \beta^2 -Q \beta^3 + \alpha(-1+2\beta-2Q^2\beta^4+Q^2\beta^5)\\
	&+ \alpha^2(1-2 Q\beta +2 Q^2\beta^4-Q^3 \beta^5) + \alpha^3(-Q^2\beta^2+Q^2\beta^3)\ ,
\end{aligned}
\end{equation}
for the $5_2$ knot
\begin{equation}
\begin{aligned}
	F_{5_2}(\alpha,\beta;Q) &=- Q \beta^7 + Q^2 \beta^8 \\
	&+ \alpha (-\beta^2 + 2 \beta^3 - \beta^4 - 2 Q \beta^5 - 3 Q \beta^6 + 3 Q^2 \beta^6 + 4 Q^2 \beta^7 - 
    2 Q^3 \beta^8) \\
	&+  \alpha^2 (1 - 3 \beta + 5 \beta^2 - 4 Q \beta^2 - 3 \beta^3 + 3 Q \beta^3 - 4 Q \beta^4 + 6 Q^2 \beta^4 - 
    3 Q \beta^5 \\
	&+ 3 Q^2 \beta^5 + 5 Q^2 \beta^6 - 4 Q^3 \beta^6 - 3 Q^3 \beta^7 + Q^4 \beta^8)	\\
    &+\alpha^3 (-2 + 4 \beta - 3 \beta^2 + 3 Q \beta^2 - 2 Q \beta^3 - Q \beta^4 + 2 Q^2 \beta^5 - Q^3 \beta^6)\\ 
	&+\alpha^4 (1 - \beta) \ ,
\end{aligned}
\end{equation}
for the $6_1$ knot
\begin{equation}
\begin{aligned}
	F_{6_1}(\alpha,\beta;Q) &= Q \beta^4 - Q^2 \beta^5  \\
 &+\alpha (-1 + 2 \beta - \beta^2 + 2 Q \beta^4 + Q^2 \beta^5 - Q^2 \beta^6 - 4 Q^3 \beta^6 + 2 Q^3 \beta^7) \\ 
& +\alpha^2 (2 - 2 \beta - 4 Q \beta + 3 Q \beta^2 + 4 Q \beta^3 - 3 Q^2 \beta^5 + Q^3 \beta^6 + 2 Q^3 \beta^7 \\
&-  6 Q^4 \beta^7 + 4 Q^4 \beta^8 - Q^4 \beta^9) \\
&+  \alpha^3 (-1 + 4 Q \beta + 2 Q \beta^2 - 6 Q^2 \beta^2 + Q^2 \beta^3 - 3 Q^2 \beta^4 + 4 Q^3 \beta^6\\
& +   3 Q^4 \beta^7 - 2 Q^4 \beta^8 - 4 Q^5 \beta^8 + 2 Q^5 \beta^9) \\ 
&+ \alpha^4 (2 Q^2 \beta^2 - Q^2 \beta^3 - 4 Q^3 \beta^3 + Q^3 \beta^4 + 2 Q^3 \beta^5 - Q^4 \beta^7 + 
    2 Q^5 \beta^8 - Q^6 \beta^9)\\
&+ \alpha^5 (-Q^4 \beta^4 + Q^4 \beta^5) \ ,
\end{aligned}
\end{equation}
and for the $6_2$ knot 
\begin{equation}
\begin{aligned}
	F_{6_2}(\alpha,\beta) &=-Q^4 \beta^{13} + Q^5 \beta^{14} + \alpha^6 (Q^2 \beta^2 - Q^2 \beta^3)\\
	&+  \alpha (3 Q^3 \beta^9 - 4 Q^3 \beta^{10} - 4 Q^4 \beta^{10} - 2 Q^4 \beta^{11} + 2 Q^5 \beta^{11} + 
    7 Q^5 \beta^{12} + 4 Q^5 \beta^{13} \\
	&- 4 Q^6 \beta^{13} - 2 Q^5 \beta^{14} - 2 Q^6 \beta^{14} + 
    Q^6 \beta^{15} + 2 Q^7 \beta^{15} - Q^7 \beta^{16}) \\ 
&+ \alpha^2 (-3 Q^2 \beta^5 + 6 Q^2 \beta^6 + 5 Q^3 \beta^6 - 6 Q^2 \beta^7 + 4 Q^3 \beta^7 - 
    3 Q^4 \beta^7 - 3 Q^3 \beta^8 \\
	&- 14 Q^4 \beta^8 + Q^5 \beta^8 - 14 Q^4 \beta^9 + 12 Q^5 \beta^9 + 
    7 Q^4 \beta^{10} + 20 Q^5 \beta^{10} - 4 Q^6 \beta^{10} \\
	&- 8 Q^4 \beta^{11} + 28 Q^5 \beta^{11} - 
    18 Q^6 \beta^{11} + Q^5 \beta^{12} - 18 Q^6 \beta^{12} + 6 Q^7 \beta^{12} - 15 Q^6 \beta^{13} \\
	&+ 12 Q^7 \beta^{13} + 4 Q^6 \beta^{14} + 7 Q^7 \beta^{14} - 4 Q^8 \beta^{14} - Q^6 \beta^{15} - 
    3 Q^8 \beta^{15} + Q^9 \beta^{16}) \\
 &+\alpha^3 (Q \beta - 2 Q \beta^2 - 2 Q^2 \beta^2 + 3 Q \beta^3 + Q^3 \beta^3 - 4 Q \beta^4 + 2 Q^2 \beta^4 + 
    6 Q^3 \beta^4 \\
	&+ 4 Q^2 \beta^5 - 3 Q^4 \beta^5 - 20 Q^3 \beta^6 - 6 Q^4 \beta^6 + 3 Q^3 \beta^7 - 
    Q^4 \beta^7 + 2 Q^5 \beta^7 - 12 Q^3 \beta^8 \\
	&+ 40 Q^4 \beta^8 + 4 Q^5 \beta^8 + 3 Q^4 \beta^9 - 
    Q^5 \beta^9 + 2 Q^6 \beta^9 - 20 Q^5 \beta^{10} - 6 Q^6 \beta^{10} + 4 Q^5 \beta^{11}\\ 
    &-3 Q^7 \beta^{11} - 4 Q^5 \beta^{12} + 2 Q^6 \beta^{12} + 6 Q^7 \beta^{12} + 3 Q^6 \beta^{13} + 
    Q^8 \beta^{13} - 2 Q^7 \beta^{14} \\
	&- 2 Q^8 \beta^{14} + Q^8 \beta^{15}) \\
&+ \alpha^4 (Q^2 - \beta - 3 Q^2 \beta + 4 Q \beta^2 + 7 Q^2 \beta^2 - 4 Q^3 \beta^2 - 15 Q^2 \beta^3 + 
    12 Q^3 \beta^3\\
	&+ Q^2 \beta^4 - 18 Q^3 \beta^4 + 6 Q^4 \beta^4 - 8 Q^2 \beta^5 + 28 Q^3 \beta^5 - 
    18 Q^4 \beta^5 + 7 Q^3 \beta^6 + 20 Q^4 \beta^6 \\
	&- 4 Q^5 \beta^6 - 14 Q^4 \beta^7 + 
    12 Q^5 \beta^7 - 3 Q^4 \beta^8 - 14 Q^5 \beta^8 + Q^6 \beta^8 - 6 Q^4 \beta^9 + 4 Q^5 \beta^9 \\	 
    &-3 Q^6 \beta^9 + 6 Q^5 \beta^{10} + 5 Q^6 \beta^{10} - 3 Q^6 \beta^{11}) \\
&+ \alpha^5 (-Q + Q \beta + 2 Q^2 \beta - 2 Q \beta^2 - 2 Q^2 \beta^2 + 4 Q^2 \beta^3 - 4 Q^3 \beta^3 + 
    7 Q^3 \beta^4 \\
	& - 2 Q^3 \beta^5 + 2 Q^4 \beta^5 - 4 Q^3 \beta^6 - 4 Q^4 \beta^6 + 3 Q^4 \beta^7) \ .
\end{aligned}
\end{equation}

%%%%%%%%%%%%%
\newpage
\ifx\undefined\bysame
\newcommand{\bysame}{\leavevmode\hbox to3em{\hrulefill}\,}
\fi

\end{document}